# Évaluation cartographique du niveau de potentialités écologiques de sites
# CARPO
## Cadre méthodologique V0

Santiago FORERO

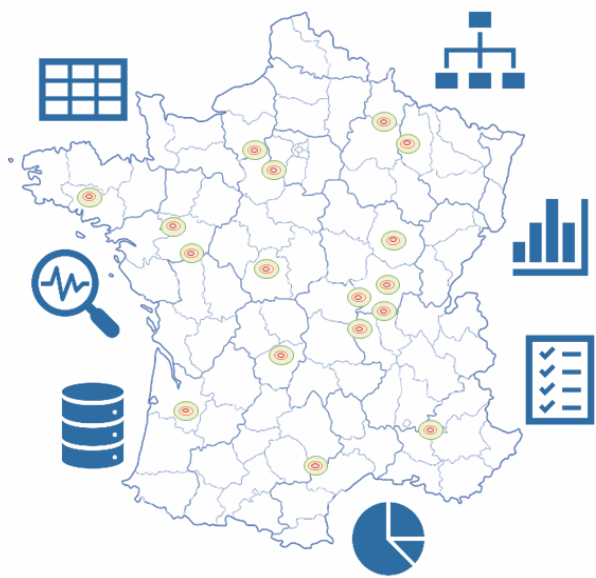

Juin 2021

# PatriNat

Centre d'expertise et de données
sur le patrimoine naturel

Un service commun
de l'Office français de la biodiversité,
du Muséum national d'Histoire naturelle,
du Centre national de la recherche scientifique
et de l'Institut pour la recherche et le développement

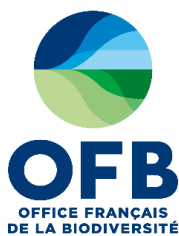 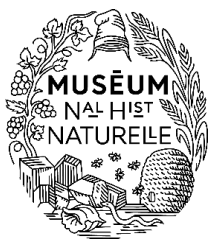 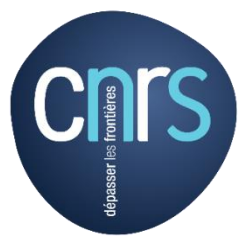 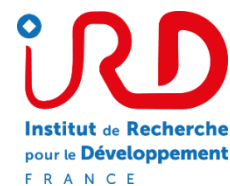

# PatriNat
Centre d'expertise et de données sur le patrimoine naturel

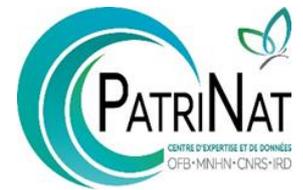

Dans une unité scientifique associant des ingénieurs, des experts et des spécialistes de la donnée, PatriNat rapproche les compétences et les moyens de ses quatre tutelles que sont l'OFB, le MNHN, le CNRS et l'IRD.

PatriNat coordonne des programmes nationaux d'acquisition de connaissance pour cartographier les écosystèmes, les espèces et les aires protégées, surveiller les tendances de la biodiversité terrestre et marine, répertorier les zones clefs pour la conservation de la nature (Znieff), et produire des référentiels scientifiques et techniques (TaxRef, HabRef, etc.). Ces programmes associent de nombreux partenaires et fédèrent les citoyens à travers des observatoires de sciences participatives (tels que Vigie-Nature, INPN espèces ou Vigie-terre).

PatriNat développe des systèmes d'information permettant de standardiser, partager, découvrir, synthétiser et archiver les données aussi bien pour les politiques publiques (SIB, SINP) que pour la recherche (PNDB) en assurant le lien avec les systèmes internationaux (GBIF, CDDA, etc.)

PatriNat apporte son expertise dans l'interprétation des données pour accompagner les acteurs et aider les décideurs à orienter leurs politiques : production d'indicateurs, notamment pour l'Observatoire national de la biodiversité (ONB) et des livrets de chiffres clés, élaboration des Listes rouges des espèces et écosystèmes menacés, revues systématiques, préparation des rapportages pour les directives européennes, élaboration d'outils de diagnostic de la biodiversité pour les acteurs des territoires, ou encore évaluation de l'efficacité des mesures de restauration. PatriNat organise également l'autorité scientifique CITES pour la France.

L'ensemble des informations (de la donnée brute à la donnée de synthèse) est rendu publique dans les portails NatureFrance, INPN et Compteur BIOM.

En savoir plus : www.patrinat.fr

Direction : Laurent PONCET et Julien TOUROULT

---

# Naturefrance
Le service public d'information sur la biodiversité

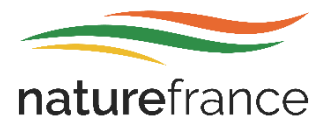

Naturefrance représente le service public d'information sur les politiques publiques de biodiversité en France. Il se décline dans plusieurs portails d'information, dont le portail général naturefrance.fr. Destiné à un public aussi large que possible, il propose des clés de lecture des grands enjeux liés à la biodiversité et à son évolution, aux pressions qu'elle subit, et aux réponses de la société. Naturefrance présente des chiffres clés, des indicateurs développés dans le cadre de l'ONB (Observatoire national de la biodiversité), des articles et des publications, issus de l'analyse scientifique des données provenant des politiques publiques de conservation ou d'activités socio-économiques favorables ou défavorables à la biodiversité.

Dans le cadre de cette mission confiée par l'OFB, PatriNat gère ce portail et participe au traitement, à l'analyse et à l'interprétation d'une partie des données versées sur Naturefrance : par exemple, celles provenant du Système d'information de l'inventaire du patrimoine naturel (SINP) ou encore du Système d'information de la Convention sur le commerce international des espèces de faune et de flore sauvages menacées d'extinction (SI CITES).

En savoir plus : naturefrance.fr

# Inventaire national du patrimoine naturel

Le portail de la biodiversité et de la géodiversité françaises,
de métropole et d'outre-mer

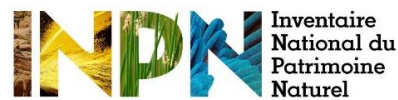

Dans le cadre de Naturefrance, l'Inventaire national du patrimoine naturel (INPN) est le portail de la biodiversité et de la géodiversité françaises, de métropole et d'outre-mer (www.inpn.fr). Il regroupe et diffuse les informations sur l'état et les tendances du patrimoine naturel français terrestre et marin (espèces animales, végétales, fongiques et microbiennes actuelles et anciennes, habitats naturels, espaces protégés et géologie) en France métropolitaine et ultramarine.

Les données proviennent du Système d'information de l'inventaire du patrimoine naturel (SINP) et de l'ensemble des réseaux associés. PatriNat organise au niveau national la gestion, la validation, la centralisation et la diffusion de ces informations. L'inventaire consolidé qui en résulte est l'aboutissement d'un travail associant scientifiques, collectivités territoriales, naturalistes et associations de protection de la nature, en vue d'établir une synthèse régulièrement mise à jour du patrimoine naturel en France.

L'INPN est un dispositif de référence français pour la connaissance naturaliste, l'expertise, la recherche en macroécologie et l'élaboration de stratégies de conservation efficaces du patrimoine naturel. L'ensemble de ces informations sont mises à la disposition de tous, professionnels, amateurs et citoyens.

En savoir plus : www.inpn.fr

---

# Compteur Biodiversité Outre-mer

Le portail des indicateurs, des enjeux et des initiatives
sur la biodiversité en outre-mer

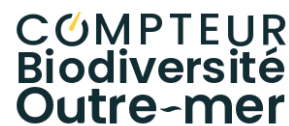

Dans le cadre de Naturefrance, le Compteur de la biodiversité Outre-mer (BiOM) développe une entrée dédiée aux territoires ultramarins français qui abritent une part importante de la biodiversité mondiale. Portail accessible, actualisé et pérenne, il favorise la rencontre des citoyens et des acteurs de la biodiversité, autour de trois objectifs : partager la connaissance scientifique, valoriser les actions des territoires ultramarins, et encourager chacun à agir. Cette démarche vise à relater les contextes culturels et mettre en avant des enjeux spécifiques de chaque territoire, pour répondre à un engagement du Livre bleu des Outre-mer.

Des études auprès des citoyens viennent compléter l'initiative : par exemple le premier panorama des programmes de sciences participatives dans les territoires, et une enquête sur la perception de la nature et l'utilisation des outils numériques.

PatriNat assure la mise en œuvre du projet et avec la participation des acteurs des outre-mer, suivant trois axes : production d'indicateurs de biodiversité (connaissances, espèces menacées, espaces protégés, etc.), relai des actions de mobilisation et de sciences participatives (écogestes, inventaires participatifs, etc.) et gestion technique du portail

En savoir plus : biodiversite-outre-mer.fr

# Table des matières

> **Avertissement :**
>
> Ce document présente une proposition d'une méthode d'évaluation des potentialités écologiques qui se base sur des premières réflexions relatives aux connaissances mobilisables. Cette version beta (V0) a été soumise à des tests sur deux groupes de sites nationaux en 2020, et les adaptations nécessaires suite à ces tests ont été intégrées à ce cadre méthodologique. Les futurs tests et le déploiement sur d'autres groupes diversifiés de sites pourraient faire évoluer cette V0 et conduiraient à son optimisation scientifique et opérationnelle. Dans ce cas, les éléments présentés dans ce document (notamment les indicateurs, les seuils et les règles décisionnelles) risquent de changer sensiblement en fonction des retours d'expérience.

# 1. Besoins et objectifs

Les partenaires publics et privés de PatriNat sont des acteurs majeurs de l'aménagement des territoires en raison de leurs nombreux sites d'implantation et des surfaces sur lesquelles ils sont amenés à intervenir. Pour ces partenaires, la conception et la mise en œuvre des stratégies permettant de conserver et de restaurer la biodiversité doit se baser sur une bonne connaissance du contexte écologique de leurs sites d'activité et de leur foncier. Cela permet de mieux appréhender les enjeux écologiques locaux et de dégager des pistes d'action en faveur de ce patrimoine naturel.

L'**objectif principal** de ce travail est d'évaluer les potentialités écologiques d'un groupe de sites à travers une caractérisation de leur contexte écologique et d'une identification des niveaux potentiels de responsabilité vis-à-vis de ces enjeux écologiques identifiés dans ce contexte. Ce travail se décline en deux étapes :

**Etape 1 : caractérisation du contexte écologique des sites**
Etudier le contexte naturel des sites d'implantation et de leurs environs, à partir des bases de données nationales standardisées, afin de mettre en évidence les éléments écologiques notables sur leur territoire et de rendre compte de la situation des sites par rapport à ces éléments de contexte.

**Etape 2 : évaluation d'un niveau de potentialités écologiques**
Analyser le positionnement des sites par rapport à différents enjeux écologiques sur la base de l'étude des éléments de contexte naturel. Pour ce faire, une batterie d'indicateurs sera calculée permettant ainsi d'évaluer les potentialités écologiques de chaque site et de regrouper les sites en fonction de leur niveau de potentialité.

# 2. Etape 1 : caractérisation du contexte écologique des sites

## 2.1. Démarche globale

Un **outil automatisé de croisement d'informations spatiales et de calcul** en cours de développement par PatriNat permet de caractériser le contexte écologique d'un groupe de sites. La Boîte à Outils Biodiversité (BOB), mobilise et compile la connaissance en synthétisant de manière automatisée, structurée, formalisée et reproductible les informations disponibles dans des **bases de données (BDD)** et de connaissance environnementales. Cet outil de synthèse des informations sur la biodiversité se base sur l'exploitation de couches cartographiques et de bases de données standardisées et publiques en France métropolitaine. Ces informations permettent de positionner un groupe de sites par rapport à différents éléments de son contexte écologique.

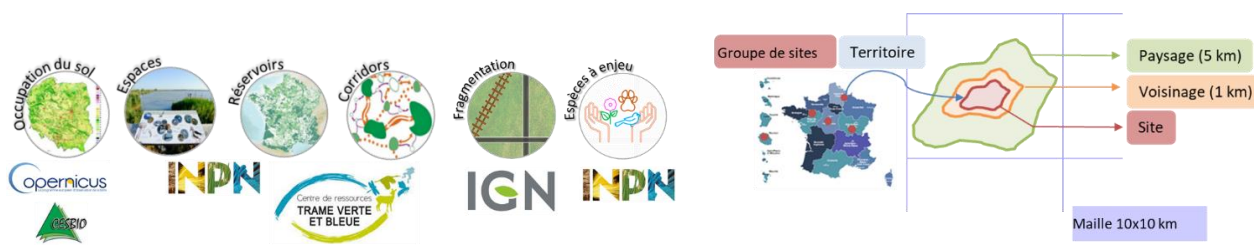

Figure 1 : Thématiques, bases de données et échelles d'étude traitées par l'outil automatisé

Cette analyse est (Figure 1) :

- **Multithématique** : car l'outil mobilise des bases de données relatives à plusieurs composants de écologiques tels que (1) les modes d'occupation du sol, (2) les zonages de protection, de gestion et d'inventaire de la biodiversité ; les éléments de connectivité écologique comme les (3) réservoirs de biodiversité et les (4) corridors écologiques ; (5) les éléments fragmentant du territoire ; (6) ainsi que les espèces à enjeu de conservation.

- **Multi-scalaire** : car elle se fait à trois échelles concentriques permettant d'explorer le territoire. Une première échelle relative au périmètre de l'implantation (appelée ici « **site** »), mais aussi au niveau de deux zones tampons appliquées au site, l'une dite "**voisinage**", qui couvre une surface définie par un rayon de 1 km autour du site, et l'autre dite "**paysage**", correspondant à la surface comprise à 5 km du site. Ces deux dernières échelles sont chacune exclusives de l'échelle d'étude « site ». Une quatrième échelle d'analyse correspondant à la **maille de 100 km²** (10 km x 10 km) est également prévue pour une des thématiques.

### 2.2. Principe méthodologique

Les étapes de la caractérisation du contexte écologique d'un groupe de sites se déclinent de cette façon (Figure 2) :

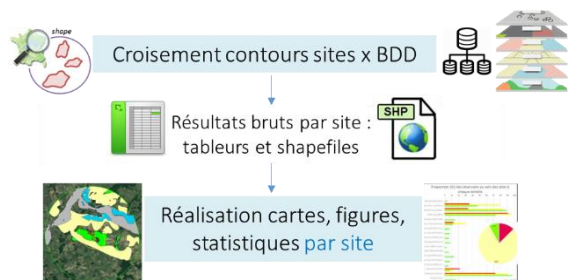

Figure 2 : Etapes de caractérisation du contexte écologique

1. Constitution d'une cartographie des sites (étape préalable à tout croisement) :
    a. Récupération et compilation des périmètres des sites à analyser sur un Système d'Information Géographique (SIG), via un logiciel tel qu'ArcGIS ou QGIS. Les polygones qui indiquent les limites des sites sont à privilégier par rapport aux points GPS des centroïdes.
    b. Obtention d'une couche cartographique unique (shapefile) contenant l'ensemble des périmètres des sites, et standardisation de la table attributaire associée (au moins 2 attributs par entité : ID_SITE et NOM_SITE).
2. Croisement avec des bases de données et de connaissance, calculs associés via l'outil automatisé :
    a. Intégration de la couche cartographique des sites dans l'outil.
    b. Croisement de la cartographie des sites avec les bases de données, calcul de différents indicateurs aux différentes échelles d'étude (site – voisinage 1km – paysage 5 km – maille).
    c. Sorties de l'outil de croisement et de calcul : obtention automatisée ou manuelle de données brutes (shapefiles avec des localisations d'éléments écologiques, tableurs avec des résultats numériques bruts) et de données de synthèse.
3. Traitement, analyse et représentation des données traitées :
    a. Cartographie (Figure 3) : intégration des shapefiles de sortie sur un logiciel SIG et élaboration de sorties cartographiques illustrant le placement des sites au sein des éléments écologiques mis en évidence.

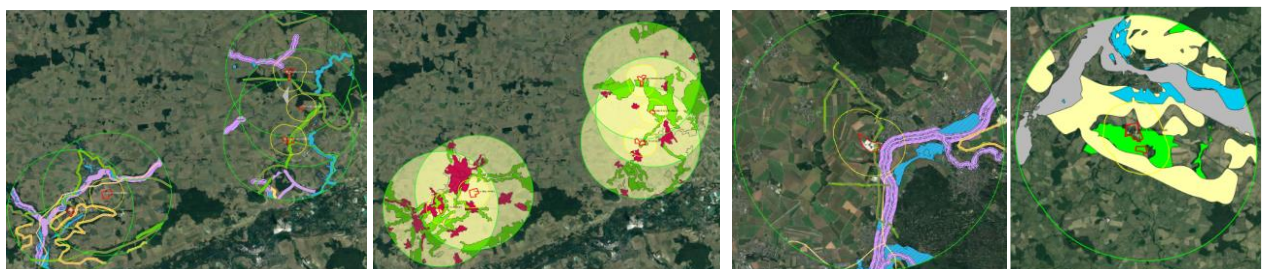

Figure 3 : Exemple de cartographie du groupe de sites pour les thématiques « corridors » et « occupation du sol » (gauche), zoom sur un site pour « corridors » et sur un autre site pour « réservoirs » (droite).

    b. Tri et sélection des informations d'intérêt dans les tableurs de données brutes et de synthèse pour l'obtention de représentations graphiques (tableaux de bord, diagrammes, histogrammes, etc.) qui répondent à des questions précises sur le contexte écologique des sites et de leur territoire : nombre de trames, proportion de la surface couverte par des zonages de protection, proportion de la surface couverte par différents types de milieux, etc. (Figure 4)

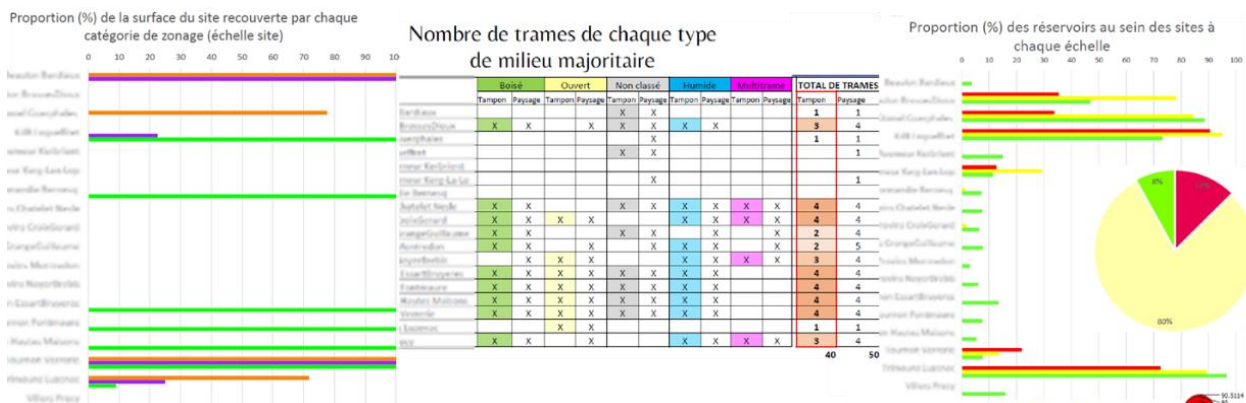
Figure 4 : Exemples de représentations graphiques obtenues suite aux traitements des sorties

## 3. Etape 2 : évaluation d'un niveau de potentialités écologiques

### 3.1. Démarche globale

En fonction des données mobilisables et des échelles d'étude considérées, ce travail se décline en deux phases avec des objectifs associés. Nous cherchons à répondre aux questions : comment caractériser 1) les potentialités écologiques et 2) les enjeux écologiques relatifs aux sites d'étude ?

La première phase porte sur l'**analyse du contexte écologique** de l'ensemble des sites d'implantation (à l'intérieur et à l'extérieur du site). Cette analyse est alimentée par des connaissances mobilisables issues des données cartographiques et naturalistes publiques. Ces données renseignent sur la capacité ou les **probabilités qu'un site présente des éléments de structure, de composition et de fonctionnalité favorables pour la biodiversité**. Par conséquent, on parle ici de « **potentialités écologiques** », en référence aux éléments qui donnent à un site une meilleure capacité d'accueillir de la biodiversité. Afin d'évaluer cette potentialité, plusieurs éléments seront pris en compte : la présence de zonages de biodiversité, des éléments de connectivité écologique, des modes d'occupation du sol liés à des grands types de milieux, des espèces à enjeux de conservation, etc.

Les potentialités écologiques à l'échelle d'un site se fondent sur une analyse des enjeux écologiques connus sur le territoire et décrits au travers des éléments précités. Néanmoins, il faut noter que **ces potentialités écologiques dépendent non seulement du contexte naturel, mais aussi de la taille du site, de la part d'habitats naturels ou semi-naturels, et de la gestion mise en place**.

Le terme « enjeux écologiques », pouvant être avérés ou potentiels, se base non seulement sur le contexte écologique (potentialités), mais intègre aussi des données sur la faune, la flore et les habitats retrouvés sur les sites (données d'observations suite à des inventaires sur site). Ces données d'inventaire étant seulement disponibles pour une partie des sites, ne sont pas prises en compte par la méthode CARPO.

Dans cette première phase, le but est donc de se baser sur les données environnementales (étape 1) pour donner un sens aux éléments d'intérêt écologique identifiés sur le territoire du groupe de sites. Pour ceci il convient de définir une **méthode d'évaluation**, reposant sur des **indicateurs chiffrés** qui rendront compte du niveau **de potentialité écologique** de chaque site vis-à-vis des éléments écologiques étudiés. Le niveau de potentialité écologique correspondrait alors à **l'importance qu'un élément écologique étudié prend sur une zone d'étude, et donc le degré de contribution et de responsabilité d'un site pour le maintien de cet élément** favorable pour la biodiversité. Ce niveau se traduit aussi par un **niveau d'alerte relatif à une biodiversité qui doit être prise en compte** dans un site.

Cette méthode permettra non seulement de mettre en évidence un niveau de potentialité écologique pour chaque indicateur évalué, mais aussi de s'appuyer sur l'ensemble des indicateurs au sein d'une thématique **pour les cumuler** et pour déterminer un niveau de potentialité globale d'un site pour une thématique donnée. Cette **évaluation cartographique du niveau de potentialités écologiques (CARPO)** des sites, permettra *in fine* d'effectuer des **mettre en parallèle et d'identifier des sites avec des potentialités écologiques définies** et pour **prioriser les actions au sein d'un groupe de sites**.

### 3.2. Principe méthodologique

La méthode d'évaluation du niveau de potentialités écologiques se fonde sur le calcul des valeurs de plusieurs indicateurs, sur leur renseignement dans une grille d'analyse, puis sur leur correspondance à des classes de niveau, définis selon certains seuils. Cette méthode a été construite sur une base empirique, puis testée sur deux groupes de sites en France métropolitaine de taille variable (un premier groupe de 254 sites et un deuxième groupe de 43 sites), et enfin affinée suite aux retours d'expérience. La méthode tient compte de ces ajustements introduits au cours des tests et constitue une version beta de l'outil CARPO.

#### 3.2.1. Indicateurs et thématiques

Les **indicateurs** sont des variables quantitatives ou semi-quantitatives auxquelles sont associées une valeur continue ou ordinale permettant de réaliser une évaluation. Il s'agit ici de métriques qui cherchent à répondre à une question scientifique précise et qui apporteraient des informations pertinentes pour estimer le niveau de potentialité écologique. Ces indicateurs ont été initialement choisis sur la base des informations pouvant être obtenues suite à l'analyse des sorties de l'outil automatisé (script R de BOB), puis complétés avec d'autres indicateurs calculés à partir des bases de données qui n'étaient pas encore traitées par cet outil, et dont leur calcul doit se réaliser de façon manuelle. Ils peuvent être regroupés par **thématiques** (occupation du sol relatif aux grands types de milieux, zonages/espaces classés de biodiversité, réseaux écologiques, espèces à enjeux de conservation) et par **sous-thématiques**, de façon à ce que pour chaque thématique il y ait au moins un indicateur.

La méthode d'évaluation présente donc **4 thématiques**, décomposées en **7 sous-thématiques** et examinées à travers **8 indicateurs**. Le Tableau 1 liste les thématiques, sous-thématiques et indicateurs retenus.

| Thématique | Sous thématique | Indicateur |
|---|---|---|
| Occupation du sol | Caractère naturel | % surface de la zone d'étude occupée par des milieux considérés comme naturels et semi-naturels |
| | Hétérogénéité | Nombre de milieux naturels différents dans la zone d'étude |
| | Perméabilité | % surface de la zone d'étude occupée par des modes d'occupation du sol considérés comme perméables |
| Réseaux écologiques | Corridors | Nombre de sous-trames différentes (milieux majoritaires) traversant la zone d'étude |
| | Réservoirs | % surface de la zone d'étude occupée par des réservoirs |
| | Unité du territoire (non-fragmentation) | Densité du linéaire des réseaux de transport routier ou ferroviaire (Rapport : longueur km totale linéaire/surface $km^2$ de l'échelle d'étude) |
| Zonages | Patrimonialité | % surface de la zone d'étude couverte par au moins un zonage biodiversité à haute patrimonialité (score 3) |
| Espèces | | Nombre d'espèces à enjeux de conservation dans la maille de l'INPN |

Tableau 1 : Thématiques, sous-thématiques et indicateurs proposés pour les tests de la V0

Les raisons justifiant le choix de ces indicateurs pour rendre compte des potentialités écologiques, ainsi que leurs intérêts majeurs sont présentés dans le Tableau 2 :

| Sous thématique | Indicateurs | Intérêt de la thématique : Mettre en évidence les sites... |
|---|---|---|

| Caractère naturel | % surface de la zone d'étude occupée par des milieux considérés comme naturels et semi-naturels | …Avec une bonne proportion de milieux naturels et semi-naturels ces sites posséderaient des potentialités importantes pour l'accueil de la biodiversité. Bien que les sites déjà très artificiels aient un faible intérêt de préservation (enjeux écologiques présumés faibles), ils pourraient avoir un fort intérêt pour la restauration. Prioriser les actions de préservation sur les sites plutôt naturels ou prioriser les actions de restauration/désartificialisation sur les sites artificiels. |
|---|---|---|
| Hétérogénéité | Nombre de milieux naturels différents dans la zone d'étude | …Avec une bonne diversité d'habitats naturels fonctionnels différents (sites hétérogènes) qui offriraient des niches écologiques pour une plus grande diversité d'espèces, par rapport à des sites dont la diversité d'habitats est faible (homogènes). Les mosaïques d'habitats et la diversité paysagère sont considérées comme favorables pour la biodiversité. Maintenir la diversité des habitats naturels fonctionnels lorsqu'elle est présente sur les sites ou, pour les sites plutôt homogènes, diversifier en créant de nouveaux habitats naturels fonctionnels et cohérents avec la matrice paysagère. |
| Perméabilité | % surface de la zone d'étude occupée par des modes d'occupation du sol considérés comme perméables | …Avec des sols non construits, non revêtus et non compacts (peu artificialisés), permettant des échanges physico-chimiques (eau-air) et entre les organismes vivants du sol. Prise en compte de la trame brune. Prioriser les actions de préservation sur les sites plutôt perméables ou prioriser les actions de désimperméabilisation sur sites plutôt imperméables. |
| Unité du territoire (non-fragmentation) | Densité du linéaire des réseaux de transport | …Peu fragmentés, avec de bonnes continuités écologiques car la densité d'obstacles aux déplacements (axes de transport fragmentant : routes, voies ferrées) y est faible. Prioriser les actions de réduction d'éléments fragmentants pour favoriser les réseaux écologiques |
| Corridors | Nombre de sous-trames différentes traversant la zone d'étude (nombre de milieux majoritaires des corridors) | …Avec une diversité et une quantité de sous-trames importantes. Ces zones d'étude comprendraient des corridors (linéaires et surfaciques) assurant des connexions entre des réservoirs de biodiversité, offrant aux espèces des conditions favorables à leur déplacement, aux échanges génétiques entre populations, et à l'accomplissement de leur cycle de vie. Une diversité de types de corridors, selon le milieu majoritaire (boisé, ouvert, humide, multi trame, non classé, etc.) suggère une présence de trames favorables à des cortèges faunistiques de plusieurs types d'habitats. Prioriser les actions de préservation de ces éléments de connectivité sur les sites favorables aux réseaux écologiques ou consolider ces réseaux via le renforcement des corridors. |
| Réservoirs | % surface de la zone d'étude occupée par des réservoirs | …Comportant un nombre élevé d'espaces dans lesquels la biodiversité est la plus riche ou la mieux représentée, où les espèces peuvent effectuer tout ou partie de leur cycle de vie (alimentation, reproduction, repos) et où les habitats naturels peuvent assurer leur fonctionnement (taille suffisante). Ces sites comprendraient des espaces pouvant abriter des noyaux de populations d'espèces (pour la dispersion d'individus ou l'accueil de nouvelles populations). Prioriser les actions de préservation de ces éléments de connectivité sur les sites favorables aux réseaux écologiques ou consolider ces réseaux via le renforcement des réservoirs. |
| Zonages : Patrimonialité | % surface de la zone d'étude couverte par au moins un zonage biodiversité à haute patrimonialité (score 3) | …Recoupés par des zonages présentant des enjeux de conservation : zones d'inventaire de biodiversité, zones de gestion concertée ou aires protégées. Ces zones ont une forte valeur écologique et patrimoniale (indépendamment des aspects réglementaires) et sont susceptibles d'accueillir une diversité de milieux naturels et d'espèces en bon état de conservation. |
| Espèces : Patrimonialité | Nombre d'espèces à enjeux de conservation dans la maille | …Dans des mailles présentant un nombre important d'espèces menacées (espèces CR, EN, VU des LR européennes et nationales), à enjeux régionaux (déterminantes de ZNIEFF), à enjeux européens (espèces de la DHFF+DO) et endémiques. La présence d'espèces qui justifient le classement d'aires protégés (SCAP : Stratégie de création d'aires protégées) et de forte valeur pour la biodiversité, avec une méthode nationale, est pertinente pour déterminer des implantations qui se trouvent dans des contextes écologiques patrimoniaux. |

Tableau 2 : Intérêt des indicateurs proposés pour l'analyse des potentialités écologiques

Les Annexes 1 à 4 présentent les spécificités et les éléments de base pour le calcul des indicateurs.

Parallèlement, des **descripteurs** ont été identifiés et sont décrits dans le Tableau 3. Ils correspondent à des variables qualitatives ou quantitatives qui, au contraire des indicateurs, ne sont pas évaluables par rapport à une référence ou à un seuil, et permettent uniquement de décrire une information. Ces descripteurs ne contribuent pas à une évaluation du niveau de potentialité. Cependant, ils sont importants dans l'analyse car ils permettent de relativiser certains résultats.

| Thématique et Sous thématique | Descripteur | Intérêt |
|---|---|---|
| Caractéristiques techniques : Dimensions | Surface du site d'implantation | Les sites d'une surface plus importante ont tendance à accueillir un plus grand nombre de potentialités écologiques et à présenter un plus grand nombre, proportion et diversité de milieux naturels favorables à l'accomplissement du cycle de vie des espèces. |
| Eco zones | Région biogéographique | Cette unité écologique renseigne sur les types de formations végétales et les conditions climatiques qui modèlent les dynamiques des écosystèmes locaux. Des potentialités plus importantes peuvent être attendues dans certaines régions biogéographiques par rapport à d'autres. |
| Espèces : Irremplaçabilité | Valeur de l'ICBG (Indice de Contribution à la Biodiversité Globalisée) de la maille (%) | L'irremplaçabilité d'une maille reflète son caractère unique et sa responsabilité régionale pour la conservation des espèces. Une maille est complètement irremplaçable si elle contient une ou plusieurs espèces qui n'existent nulle part ailleurs. La perte de cette maille entraînerait une perte importante pour cette espèce. L'Indice de Contribution à la Biodiversité Globalisée (ICBG) (Witté et Touroult, 2014) est un score sur 100 qui définit les points chauds de biodiversité, "irremplaçables" du fait de l'assemblage d'espèces qu'ils abritent. Les mailles avec un score de 100 contiennent un grand nombre de taxons ou des taxons rares ou endémiques (ou les deux), alors que les mailles à faible score contiennent des taxons plus répandus ou une richesse moins importante (Léonard et al., 2020). |
| Espèces : Méconnaissance | Valeur du taux méconnaissance (%) pour les taxons classiques dans la maille | Les résultats obtenus pour des indicateurs sur le nombre d'espèces patrimoniales dans une maille doivent être relativisés selon le niveau de connaissance de la biodiversité sur ladite maille. Le taux de méconnaissance est un indice établi en cumulant sur 27 groupes taxonomiques classiques le nombre de groupes méconnus par maille. L'indice correspond au pourcentage de groupes méconnus par rapport au total des 27 groupes étudiés (% élevés dans mailles méconnues). Il met en évidence les zones déficitaires en données naturalistes disponibles au niveau national (Witté & Touroult, 2017). |

Tableau 3 : Descripteurs proposés pour les tests de la V0

Tous les indicateurs des thématiques «occupation du sol», «réseaux écologiques» et «zonages» seront calculés sur les **3 échelles** d'analyse de la façon suivante (Figure 5) :

- **Site :** Calcul sur l'ensemble de la surface inclue à l'intérieur du périmètre du site.
- **Voisinage (1 km) :** Calcul sur la surface comprise entre le périmètre du site et son tampon de 1000 m autour de ce périmètre. La surface à l'intérieur des limites du site est donc exclue des calculs.
- **Paysage (5 km) :** Calcul sur la surface comprise entre le périmètre du site et son tampon de 5000 m autour de ce périmètre. La surface à l'intérieur des limites du site est donc exclue des calculs, mais la surface du voisinage (1 km) présentée auparavant est prise en compte.

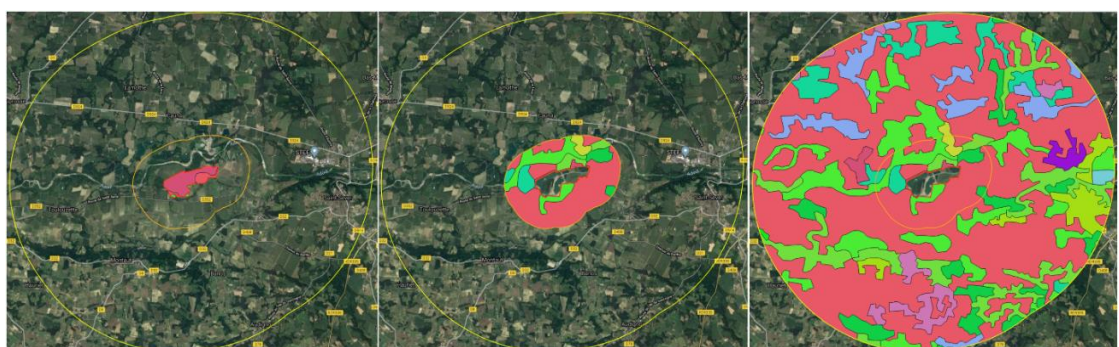

Figure 5 : Exemple de visualisation de l'occupation du sol (CLC18) pour un site selon l'échelle d'étude, (telle que) de gauche à droite figurent le site, le voisinage et le paysage.

Les indicateurs relatifs aux **espèces à enjeux de conservation** seront calculés uniquement à l'échelle d'une **maille de 100 km²** (10 km x 10 km) car c'est une des échelles géographiques les plus fines où des données publiques et standardisées au niveau national sont disponibles.

Le résultat de la majorité des indicateurs de la méthode CARPO (à chaque échelle géographique) peut être obtenu à partir des **fichiers de sortie** (SIG et tableurs). L'**Annexe 5** présente le contenu de ces fichiers et indique le champ contenant la valeur d'intérêt pour chaque indicateur CARPO. Ces valeurs issues de ces fichiers doivent être renseignées dans la **grille d'évaluation** (Annexe 8) dans la colonne « Résultat » comme indiqué dans le Tableau 4. Elles permettront de déterminer par la suite le niveau de potentialité associé.

| Thématique | Sous thématique | Indicateur | Résultat : |
|---|---|---|---|
| Thém. 1 | Sous-thém. 1.1 | Indicateur 1.1 | Site : _______ <br> Voisinage : _____ <br> Paysage : _______ |
|  | Sous-thém. 1.2 | Indicateur 1.2 | Site : _______ <br> Voisinage : _____ <br> Paysage : _______ |
| Thém. 2 | Sous-thém. 2 | Indicateur 2 | Site : _______ <br> Voisinage : _____ <br> Paysage : _______ |
| Thém. 3 | Sous-thém. 3 | Indicateur 3 | Maille : _______ |
| … | … | … | … |

Tableau 4 : Colonne de renseignement des valeurs des indicateurs au sein de la grille d'évaluation

### 3.2.2. Notation de chaque indicateur : seuils et classes de niveau de potentialité écologique

Une fois l'indicateur calculé aux trois échelles d'étude classiques (site-voisinage-paysage) ou à l'échelle de la maille, ces valeurs numériques seront traduites en un niveau de potentialité écologique.

Ainsi, pour chaque indicateur et pour chaque échelle d'étude, une évaluation du résultat obtenu sera faite, en se référant a de **seuils déterminés (délimitant des intervalles de valeurs)**. Ces seuils correspondent à des valeurs repères qui définissent les bornes de chacune des classes de niveau. Ainsi, les résultats des calculs pour un site étudié sont comparés à ces seuils pour retrouver le niveau de potentialité associé.

La **définition des valeurs seuils doit se faire au cas par cas pour chaque groupe de sites**. Il n'y a donc **pas de seuils de référence nationaux fixé dans la méthode CARPO, mais la définition des seuils et des niveaux doit se faire préalablement à toute évaluation du groupe de sites**. Cette définition de seuils se base donc sur une étude rapide de la distribution de valeurs observées dans la série pour chaque indicateur. Les seuils vont diviser la répartition des résultats en classes, de sorte que n seuils produiront n+1 classes (un seuil crée deux classes, deux seuils créent trois classes, et ainsi de suite).

Par conséquent, **CARPO propose une méthode pour le calcul des seuils** et des classes de niveaux pour chaque groupe de sites, **mais il ne propose pas de valeurs nationales de référence** à appliquer à n'importe quel groupe de sites. Néanmoins, **une exception est faite pour l'indicateur sur les espèces à enjeux de conservation**, pour lequel les différents seuils et niveaux sont définis sur des références nationales et donc applicables à tout groupe de sites (cf. Annexe 4).

Les seuils sont définis selon **3 classes de niveau de potentialité écologique (moyenne - forte - très forte) par indicateur et par échelle d'étude**. Afin d'aboutir à ces 3 classes, il s'agit de **retrouver deux seuils sur la distribution de valeurs de chaque indicateur** comme illustré dans le Tableau 5.

| Suite à l'analyse de la distribution des résultats au sein d'un groupe de sites, il est possible de définir 2 seuils pour chaque indicateur et à chaque échelle. Les deux seuils définis pour l'indicateur à l'échelle du site sont notés ici par b % et c % qui permettront de définir trois intervalles de valeurs chacun associé à un niveau de potentialité. Ceci est également le cas pour les seuils f % et g % à l'échelle du voisinage et finalement j % et k % à l'échelle du paysage | | | | |
|---|---|---|---|---|
| Indicateur | Moyen d'appréciation | Fourchette/seuil | Conclusion | |
| Indicateur X : Proportion de la zone d'étude recouverte par… | Si % du **SITE** recouvert se trouve dans la fourchette | [a % - **b** %[ | Potentialités moyennes | Pour l'échelle SITE |
|  |  | ]**b** % – **c** %] | Potentialités fortes |  |
|  |  | ]**c** % - d %] | Potentialités très fortes |  |
|  | Si % du **VOISINAGE** recouvert se trouve dans la fourchette | [e % - **f** %[ | Potentialités moyennes | Pour l'échelle Voisinage |
|  |  | ]**f** % – **g** %] | Potentialités fortes |  |
|  |  | ]**g** % - h %] | Potentialités très fortes |  |
|  |  | [i % - **j** %[ | Potentialités moyennes |  |

| | Si % du **PAYSAGE** recouvert se trouve dans la fourchette | ]j % – k %] | Potentialités fortes | Pour l'échelle Paysage |
| | | ]k % - l %] | Potentialités très fortes | |

Tableau 5 : Principe de notation d'un indicateur avec des seuils et classes de niveau

Cette définition de classes est appelée « **discrétisation** » et consiste à découper en classes (ou groupe de valeurs) une série de variables quantitatives ou qualitatives en vue de sa représentation graphique ou cartographique.

Il existe **plusieurs méthodes de discrétisation** (regroupement de valeurs en classes) telles que l'équidistance (intervalles égaux), la progression arithmétique, l'équi-fréquence (quantiles), la méthode de Jenks, la méthode manuelle (seuils observés), les classes standardisées, les moyennes emboîtées, etc. En fonction de la méthode choisie, les seuils vont varier et donc les classes de niveau en résultant aussi (Figure 6).

Il n'y a pas une théorie pour le choix de la méthode de discrétisation, mais il est possible de guider cette sélection via l'observation de la distribution de la série. Ainsi, **la forme de la distribution des valeurs** (uniforme, normale, symétrique, asymétrique, uni-modale, multimodale, etc.) oriente **le choix du mode de discrétisation**, comme présenté dans l'Annexe 6a.

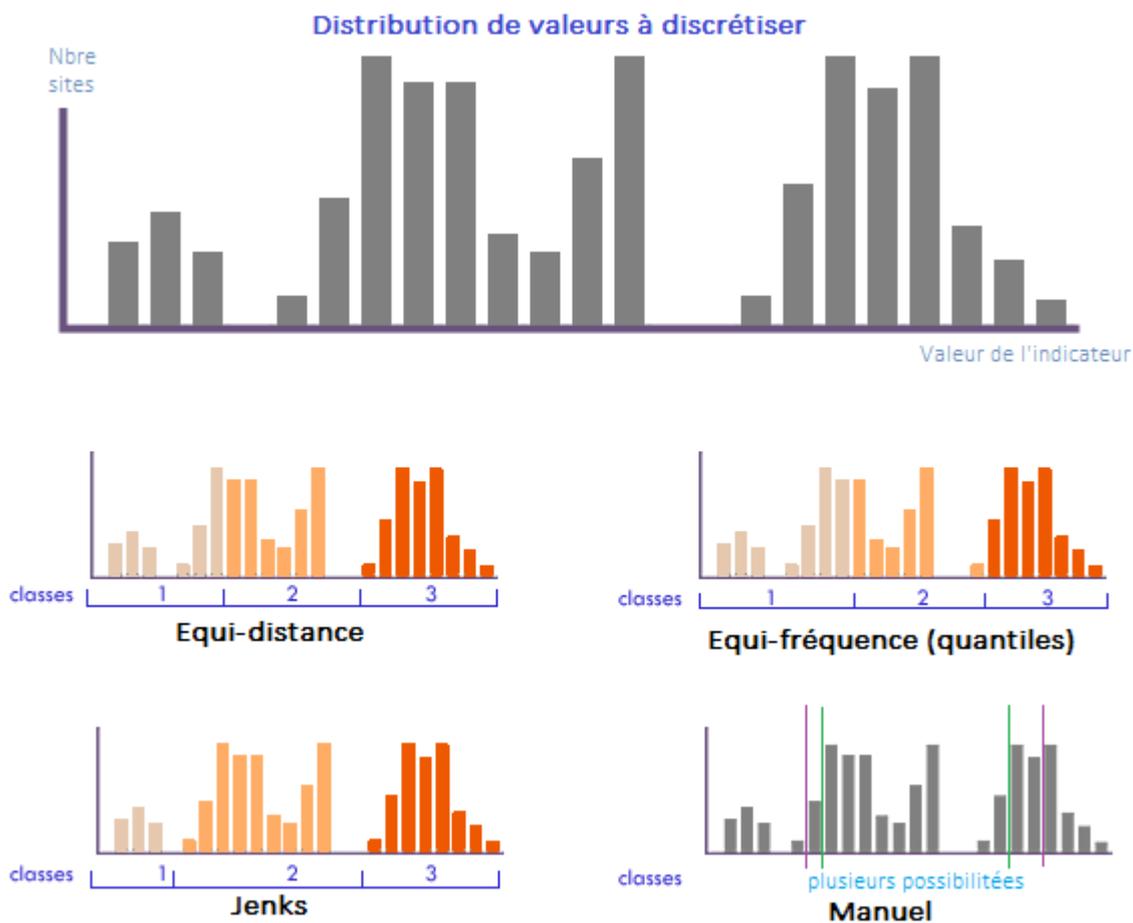

Figure 6 : Différentes possibilités de découpage d'une distribution de valeurs en fonction de la méthode de discrétisation choisie. Adaptée de AxisMaps.

Dans le cas des variables étudiées au travers des 8 indicateurs de CARPO, les distributions présentent majoritairement des **formes asymétriques et multimodales**, pouvant être discrétisées préférentiellement au moyen des méthodes d'équi-fréquence ou de Jenks. Les premiers tests de la méthode CARPO ont permis de définir laquelle de ces deux méthodes constitue l'approche la plus précise et adaptée pour l'évaluation de potentialités écologiques. Ils ont donc mis en évidence plusieurs problématiques avec la méthode de l'équi-fréquence et ont permis de confirmer la pertinence de la méthode de Jenks.

**La discrétisation de classes doit donc se faire à travers la méthode de Jenks**, aussi appelée « **seuils naturels** ». Elle se base « sur le principe de ressemblance/dissemblance en calculant la distance paramétrique entre toutes les valeurs de la série. La méthode minimise la variance intraclasse et maximise la variance interclasse. Il s'agit d'un algorithme itératif formant, dans un premier temps, autant de couples qu'il y a de combinaisons de valeurs (triées par ordre croissant) pour un nombre de classes donné. On calcule alors la variance intraclasse et interclasse. » (Jegou *et al.* s. d.). Ainsi, **les plus fortes discontinuités observées dans la distribution constituent des paliers** marquants, qu'il convient donc de séparer en classes (Magrit, 2014). Le fonctionnement de l'algorithme de Jenks est présenté dans l'Annexe 6 b. La méthode permettrait donc de définir de façon optimale les 2 paliers bornant les 3 classes de niveau de potentialité. Le nombre de sites dans chaque classe peut être homogène ou hétérogène, en fonction de chaque indicateur. Cette méthode, qui respecte l'allure de la série et qui est une des méthodes statistiques les plus puissantes, **nécessite l'emploi d'un logiciel statistique** (comme QGIS).

L'Annexe 7 présente la procédure pour **calculer facilement les seuils et donc les classes de niveau pour la méthode de Jenks via QGIS**, ainsi qu'un exemple pour illustrer ce calcul de seuils et de classes de niveau.

Au sein des 7 indicateurs relatifs aux thématiques « occupation du sol », « réseaux écologiques », « zonages », **3 intervalles de valeurs sont définis pour chacune des trois échelles d'étude (site-voisinage-paysage)** afin de retrouver les 3 niveaux de potentialités écologiques dans chacune des échelles. Ainsi, un site peut présenter, pour un même indicateur, des niveaux de potentialité différents en fonction de l'échelle d'étude (cf. exemple dans le Tableau 6).

| Indicateur | Résultat | Moyen d'appréciation | Fourchette/seuil | Conclusion | |
|---|---|---|---|---|---|
| Indicateur X : Proportion de la zone d'étude recouverte par des milieux naturels et semi-naturels | SITE : 1,8 % | Si % du SITE recouvert se trouve dans la fourchette | 0 - 2 % | Potentialités moyennes | Pour l'échelle SITE |
| | | | 2,1 – 15 % | Potentialités fortes | |
| | | | 15,1 - 100 % | Potentialités très fortes | |
| | VOISINAGE : 13,4 % | Si % du VOISINAGE recouvert se trouve dans la fourchette | 0 % - 9 % | Potentialités moyennes | Pour l'échelle Voisinage |
| | | | 9,1 - 22 % | Potentialités fortes | |
| | | | 22,1 % - 100 % | Potentialités très fortes | |
| | PAYSAGE : 40,3 % | Si % du PAYSAGE recouvert se trouve dans la fourchette | 0 - 20 % | Potentialités moyennes | Pour l'échelle Paysage |
| | | | 20,1 - 39 % | Potentialités fortes | |
| | | | 39,1 % - 100 % | Potentialités très fortes | |
| *Le site en question présente des potentialités moyennes à l'échelle du site quant à son caractère naturel, mais ces potentialités augmentent aux échelles supérieures : elles sont fortes dans son voisinage et très fortes dans son paysage. Ceci suggère que le site peut être un élément artificialisant dans son environnement et des pistes d'action de désartificialisation sont envisageables.* | | | | | |

Tableau 6 : Exemple de notation d'un indicateur avec des résultats et des intervalles de valeurs fictifs

**Pour l'indicateur relatif aux espèces à enjeux de conservation (patrimoniales), les 3 classes de niveau de potentialité** sont également définies, mais elles ne sont plus rattachées à une échelle d'étude (site - voisinage - paysage) comme les autres indicateurs, car l'analyse se fait uniquement **à l'échelle d'une maille** (10 x 10km). Autrement dit, seulement le périmètre du site d'étude est croisé avec le maillage INPN, et non pas son voisinage et son paysage. Les 3 classes de niveau relatives au nombre d'espèces patrimoniales ont été définies en fonction de la distribution de valeurs de toutes les mailles en France métropolitaine (Annexe 4). En conséquence, des fourchettes de valeurs constituant un référentiel national ont été établies ; ce qui implique qu'**un calcul de seuils et de classes de niveau pour cet indicateur et pour chaque groupe de sites n'est pas nécessaire, mais qu'il suffit d'appliquer les classes nationales définies dans la méthode CARPO**.

> *Pour l'indicateur « nombre d'espèces à enjeux de conservation », le niveau de potentialité écologique est considéré moyen si le site se trouve dans une maille avec 0-38 espèces patrimoniales. Il est considéré fort si la maille présente 39-62 espèces patrimoniales et très fort si la maille présente 63-132 espèces patrimoniales.*

La **grille d'évaluation** (Annexe 8) permet de renseigner pour chaque indicateur les **valeurs du calcul de l'indicateur**, de les **confronter aux intervalles de valeurs décisionnelles** correspondantes, pour **définir le niveau de potentialité écologique** de chaque indicateur et chaque échelle, comme illustré dans l'exemple sur le Tableau 7 (cas d'un site fictif au sein d'un groupe de 200 sites). En plus de la valeur calculée pour chaque indicateur et les intervalles des classes de niveau, l'utilisateur peut également enregistrer les **moyennes** observées au sein du groupe de sites évalués pour chaque indicateur, ainsi que le **nombre de sites** présents dans chaque classe. Ces moyennes et nombres de sites fourniront, **à titre informatif, un curseur des tendances observées au sein du groupe**, sans que ces valeurs deviennent des références à atteindre pour les sites se trouvant en dessous de ces valeurs moyennes.

| Thématique | Indicateur | Résultat | | | Fourchette décisionnelle | | | Niveau/ échelle | Niveau échelle retenue Préciser échelle : **SITE** |
|---|---|---|---|---|---|---|---|---|---|
| | | Moyenne | Echelle | Résultat | Fourchettes | Nb sites | Niveau | | |
| Thématique1 | Indicateur 1.1 | µ=85,37 % | SITE | 54,2 % | 0 - 43,70 % | 18 sites | Moyen | ☐ Moyennes ■ Fortes ☐ Très fortes | ■ Fortes |
| | | | | | 43,71 - 80,90 % | 42 sites | Fort | | |
| | | | | | 80,91 - 100 % | 140 sites | Très fort | | |
| | | µ=91,48% | VOISINAGE | 84,05 % | 6,20 - 41,80 % | 10 sites | Moyen | ☐ Moyennes ☐ Fortes ■ Très fortes | |
| | | | | | 41,81 - 83,60 % | 30 sites | Fort | | |
| | | | | | 83,61 - 100 % | 160 sites | Très fort | | |
| | | µ=92,06% | PAYSAGE | 11,83 % | 5,10 - 12,70 % | 8 sites | Moyen | ■ Moyennes ☐ Fortes ☐ Très fortes | |
| | | | | | 12,71 - 85,3 % | 30 sites | Fort | | |
| | | | | | 85,31 - 100 % | 162 sites | Très fort | | |
| | Indicateur 1.2 | µ=3 | SITE | 0 | 0 – 2 | 50 sites | Moyen | ■ Moyennes ☐ Fortes ☐ Très fortes | SITE : ■ Moyennes |
| | | | | | 3 – 4 | 70 sites | Fort | | |
| | | | | | 5 – 7 | 80 sites | Très fort | | |
| | | µ=5 | VOISINAGE | 5 | 1 – 4 | 68 sites | Moyen | ☐ Moyennes ■ Fortes ☐ Très fortes | |
| | | | | | 5 | 90 sites | Fort | | |
| | | | | | 6 – 7 | 42 sites | Très fort | | |
| | | µ=6 | PAYSAGE | 6 | 4 – 5 | 85 sites | Moyen | ☐ Moyennes ■ Fortes ☐ Très fortes | |
| | | | | | 6 | 60 sites | Fort | | |
| | | | | | 7 – 8 | 55 sites | Très fort | | |
| Thématique 2 | Indicateur 2.1 | µ=88.76 | MAILLE | 152 | 1 - 76 | | Moyen | ☐ Moyennes ☐ Fortes ■ Très fortes | MAILLE : ■ Très fortes |
| | | | | | 77 - 114 | | Fort | | |
| | | | | | 115 - 226 | | Très fort | | |

Tableau 7 : Exemple des résultats renseignés dans un extrait de la grille d'évaluation pour un site fictif et des intervalles fictifs, suite à l'étude de la distribution des valeurs des 200 sites, et où l'analyse se concentrera sur les résultats de l'échelle du site

Comme expliqué précédemment, l'indicateur sur les espèces à enjeux aboutit à un résultat unique de niveau de potentialité (celui de la maille). En revanche, les 7 autres indicateurs auront un niveau de potentialité par échelle d'étude (site-étude-paysage), donc trois résultats de niveaux pour chaque indicateur. Il est possible, via la grille d'évaluation, de **se concentrer uniquement sur une seule échelle pour chaque indicateur, afin d'obtenir un bilan des potentialités et une visualisation synthétique des résultats à l'échelle souhaitée, et de pouvoir faire ensuite un cumul de potentialités à ladite échelle** (cf. 3.2.3). Ceci implique de faire une **sélection de l'échelle d'intérêt**.

Il est donc nécessaire de **sélectionner une même échelle d'analyse pour l'ensemble d'indicateurs :** Chaque site évalué fera l'objet d'une grille d'évaluation qui sera renseignée avec les résultats obtenus pour chaque indicateur et à chaque échelle, permettant de déterminer les niveaux de potentialité associés. L'utilisateur décide ensuite de se concentrer sur une seule échelle d'étude commune pour laquelle il souhaite visualiser le bilan des niveaux des indicateurs. Il fera donc une seule évaluation s'il souhaite obtenir un bilan pour une seule échelle (par exemple pour apercevoir toutes les potentialités uniquement à l'intérieur des sites), ou bien deux ou trois évaluations s'il souhaite avoir un bilan des potentialités de son site à chaque échelle d'étude (par exemple pour comparer les niveaux de potentialité à l'intérieur de son site, avec ceux à l'extérieur de celui-ci).
Ainsi, **pour chaque site évalué, il y aura autant de grilles d'évaluation à remplir qu'il y aura d'échelles (parmi les 3 échelles) pour lesquelles un bilan des potentialités et une visualisation synthétique des résultats seraient souhaités**. La grille d'évaluation est structurée de façon à ce que la première moitié des colonnes (recensant les résultats/niveaux pour chaque indicateur et à chaque échelle) sera commune à toutes les grilles ; alors que seulement les dernières colonnes (vouées à reprendre les résultats à une échelle choisie et à cumuler les niveaux des indicateurs) changeront d'une grille à l'autre ; ce qui rend la démarche assez rapide. Pour cela, **l'évaluateur notera, dans la colonne « Niveau de potentialité de l'échelle retenue », l'intitulé de l'échelle** (site-voisinage-paysage) retenue pour l'analyse et les résultats associés, afin d'obtenir le bilan pour le site d'implantation évalué.

L'Annexe 8 présente un exemple d'évaluation des niveaux de potentialité aux trois échelles pour un même site (3 grilles d'évaluation).

Suite à ce choix d'échelle, le renseignement de la grille d'évaluation peut être finalisé, une fois que les résultats à l'échelle retenue y seront enregistrés. Ceci aboutit à un diagnostic synthétique des niveaux de potentialités écologiques des indicateurs individuels d'un site d'implantation, en fonction de l'échelle(s) choisie(s).

### 3.2.3. Cumul des niveaux de potentialité des indicateurs

L'évaluation d'un site via la grille, permet de retrouver le niveau des potentialités écologiques de celui-ci pour chacun des 8 indicateurs relatifs aux 4 thématiques. Bien que ces 8 résultats offrent un bilan du site d'implantation, il est possible **d'obtenir un bilan plus synthétique, via le calcul du niveau de potentialité de chacune des 4 thématiques** (occupation du sol, réseaux écologiques, zonages et espèces).

Ainsi, **les niveaux de potentialités individuels des indicateurs d'une même thématique peuvent être cumulés pour avoir une information synthétique sur le niveau de potentialité écologique de ladite thématique**. Ceci est notamment le cas de deux thématiques présentant chacune trois indicateurs :
- Les niveaux retenus à une échelle donnée pour les indicateurs « caractère naturel », « hétérogénéité » et « perméabilité », appartenant tous à la même thématique, peuvent être cumulés pour retrouver le niveau de potentialité de la thématique « occupation du sol ».
- Les niveaux retenus à une échelle donnée pour les indicateurs « corridors », « réservoirs » et « unité du territoire » appartenant tous à la même thématique, peuvent être cumulés pour retrouver le niveau de potentialité de la thématique « réseaux écologiques ».

Ce cumul se base sur l'hypothèse selon laquelle **tous les indicateurs et toutes les thématiques ont la même importance :** tous les indicateurs ont le même poids entre eux et toutes les thématiques ont le même poids entre elles. **Il n'y a pas de pondération pour les indicateurs ni les thématiques.**

Une fois les niveaux de potentialités définis pour chaque indicateur individuel au sein d'une thématique, ils peuvent être cumulés entre eux, tout en suivant **la règle de décision pour le cumul de niveaux de potentialités** présentée dans le Tableau 8. Elle indique lorsqu'une thématique englobe plusieurs indicateurs avec des niveaux de potentialité différents, la façon pour agréger ces niveaux **selon le nombre d'indicateurs de chaque niveau dans la thématique évaluée** :

| Légende | | Critère | Si le nombre d'indicateurs et leurs niveaux sont : | | …alors le niveau global est |
|---|---|---|---|---|---|
| M | Potentialité moyenne | A) | 3 TF | 100 % TF | **TF** |
| F | Forte potentialité | B) | 2 TF et (1 F ou 1 M) | ≥ 66 % TF et (≤ 34 % F ou ≤ 34 % M) | |
| TF | Très forte potentialité | C) | 2 M et 1 F | ≥ 66 % M et ≤ 34 % F | **M** |
| | | D) | 3 M | 100 % M | |
| | | E) | Les autres cas | Les autres cas | **F** |

Tableau 8 : Règle décisionnelle pour le cumul de niveaux de potentialités

Cette règle de décision est inspirée des méthodes d'Évaluation de l'Etat de Conservation (EVAL) des habitats d'intérêt communautaire (Natura 2000). Elle a été calibrée via des tests de différents cas d'agrégation de niveaux individuels de potentialité lorsqu'un cumul de 2 à 5 éléments doit être réalisé (Annexe 9). Le test de la V0 de cette méthode sur d'autres groupes de sites, permettra d'affiner ces seuils. Un exemple théorique de notation du cumul de niveaux d'indicateurs est présenté au Tableau 9, et des exemples pratiques sont illustrés dans l'Annexe 8.

Il faut noter que **cette règle décisionnelle a été choisie de façon à faire « remonter » le niveau cumulé (« tirer vers le haut »)** pour mettre en avant les très fortes potentialités écologiques par rapport aux niveaux plus faibles. Cela permettra de diminuer le risque de juger un site comme ayant des potentialités moyennes (risques d'activités humaines qui découlent de ce jugement), alors que ses potentialités seraient réellement plus importantes. De la même façon que pour les indicateurs et les seuils, cette règle décisionnelle pourra évoluer suite au déploiement de CARPO sur d'autres sites et sera affinée par la suite, si besoin.

| Thémat. | Indicateur | Niveau échelle retenue | Critère décisionnel | Potentialité thém. |
|---|---|---|---|---|
| Thémat.1 | Indicateur 1.1 | ■ Très fortes<br>□ Fortes<br>□ Moyennes | B) 2 TF et (1 F ou 1 M) | Très fortes pour la thématique 1 |
| | Indicateur 1.2 | ■ Très fortes<br>□ Fortes<br>□ Moyennes | | |
| | Indicateur 1.3 | □ Très fortes<br>□ Fortes<br>■ Moyennes) | | |
| Thémat.2 | Indicateur 2.1 | ■ Fortes<br>□ Fortes<br>□ Moyennes | C) 2M et 1 F | Moyennes pour la thématique 3 |
| | Indicateur 2.2 | □ Très fortes<br>□ Fortes<br>■ Moyennes | | |
| | Indicateur 2.3 | □ Très fortes<br>□ Fortes<br>■ Moyennes | | |

Ce cumul d'indicateurs permettrait de constituer un tableau de bord offrant un aperçu global des sites et du niveau de potentialité global associé à chaque thématique évaluée.

Il est fortement déconseillé de cumuler les niveaux individuels des thématiques pour **définir un niveau global de potentialité de l'implantation des sites,** car cela conduirait à combiner et superposer des enjeux peu comparables, ce qui apporterait un biais considérable.

Tableau 9 : Exemple de notation du cumul des niveaux de potentialités pour une thématique

### 3.2.4. Visualisation des résultats

Les résultats obtenus pour chaque site lors de l'évaluation des 8 indicateurs peuvent être représentés sous forme de la **grille de diagnostic** du niveau de potentialité de chaque site, ainsi que via un **radar** (Figure 7). Ce dernier précise le niveau de potentialité obtenu pour chaque indicateur (8 bras du radar) et à chacune des échelles évaluées. Ceci fournit un bilan des points forts de chaque site sur une thématique donnée, ainsi que des points d'amélioration qui pourraient être adressés via un plan d'action, afin d'augmenter les potentialités écologiques du site pour ledit indicateur.

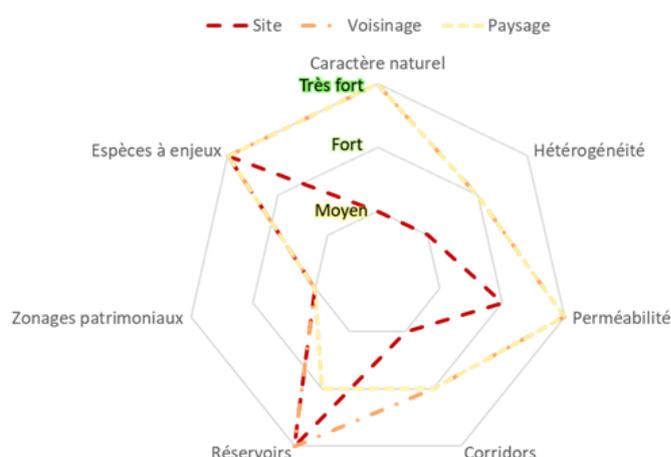

Figure 7 : Exemple de diagnostic pour un site via la grille, et d'un radar synthétique multi échelle et multithématique

Par exemple, un site présentant des potentialités moyennes à l'échelle du site pour un indicateur, et des potentialités très fortes à l'échelle du voisinage/paysage, met en évidence un décalage entre le contexte à l'intérieur du site et celui de son environnement proche. Il sera donc possible d'identifier des pistes d'actions prioritaires afin que le site puisse répondre aux enjeux de son territoire d'implantation.

Cette représentation des résultats individuels de chaque site peut être complétée par une présentation globale des résultats du groupe, via un **atlas** : un recueil ordonné de cartes permettant de représenter un espace donné (un site, un groupe de sites ou un sous-ensemble) et donc d'exposer une ou plusieurs thématiques via des planches, des figures (tableaux de synthèse, histogrammes, camemberts, radars), des commentaires, etc. Un exemple de contenu d'un tel atlas est présenté dans la section suivante (cf. 4. Livrables potentiels).

Il est également possible de produire un **tableau de bord du groupe de sites** (un tableur par échelle étudiée) avec des nuances de couleurs où chaque ligne correspond à un site et chaque colonne est un des 8 indicateurs (et des descripteurs). Les cellules sont les résultats de l'indicateur et l'intensité de leur couleur représente la magnitude de la valeur du site au sein du groupe (Figure 8)

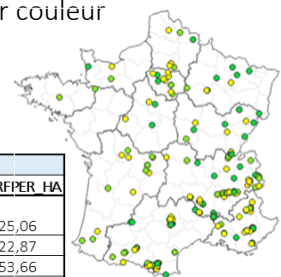

| NOM_SITE | Indicateurs CARPO | | | | | | | | Descripteurs CARPO | | | |
|---|---|---|---|---|---|---|---|---|---|---|---|---|
| | CARACT_NATUREL | HETEROGENEITE | PERMEABILITE | CORRIDORS | RESERV | TRSP_FRGM | ZONAGE_PATRIMO | ESP_SCAP20 | MECON. | CBG | REG_BIOGEO | SURFPER_HA |
| Site A | 100,0% | 3 | 100,0% | 0 | 100,0% | 0,0000 | 99,4% | 79 | 63 | 100 | Alpin + Méditerranéen | 25,06 |
| Site B | 100,0% | 0 | 100,0% | 1 | 0,0% | 0,0000 | 0,0% | 39 | 78 | 100 | Continental | 22,87 |
| Site C | 24,8% | 1 | 100,0% | 1 | 35,4% | 0,0000 | 0,0% | 42 | 74 | 100 | Continental | 53,66 |
| Site D | 54,5% | 1 | 100,0% | 1 | 0,0% | 0,7325 | 0,0% | 42 | 74 | 100 | Continental | 112,82 |
| Site E | 50,4% | 3 | 100,0% | 1 | 35,4% | 0,0883 | 0,0% | 37 | 70 | 0 | Continental | 111,14 |
| Site F | 2,8% | 1 | 100,0% | 1 | 100,0% | 0,0000 | 0,0% | 52 | 74 | 2 | Atlantique | 25,39 |
| Site G | 25,5% | 2 | 100,0% | 1 | 47,4% | 2,0660 | 0,0% | 54 | 70 | 9 | Atlantique | 57,11 |
| Site H | 41,6% | 3 | 100,0% | 0 | 0,0% | 0,5000 | 0,0% | 48 | 70 | 100 | Méditerranéen | 187,83 |
| Site I | 100,0% | 1 | 100,0% | 0 | 97,3% | 2,3369 | 95,1% | 74 | 67 | 100 | Méditerranéen | 61,08 |
| Site J | 51,6% | 1 | 100,0% | 2 | 13,4% | 1,1111 | 66,5% | 39 | 81 | 0 | Continental | 46,77 |
| Site K | 90,4% | 2 | 100,0% | 1 | 33,8% | 0,4566 | 0,0% | 56 | 52 | 100 | Atlantique | 267,11 |
| Site L | 100,0% | 1 | 100,0% | 1 | 77,6% | 0,0000 | 0,0% | 54 | 70 | 9 | Atlantique | 14,26 |
| Site M | 100,0% | 2 | 100,0% | 0 | 90,5% | 1,2492 | 0,0% | 55 | 70 | 99 | Atlantique | 40,81 |
| Site N | 25,5% | 1 | 86,6% | 0 | 0,0% | 0,6666 | 0,0% | 56 | 67 | 100 | Atlantique | 57,06 |
| Site O | 25,6% | 2 | 96,8% | 0 | 12,7% | 0,2020 | 0,0% | 56 | 67 | 100 | Atlantique | 201,44 |
| Site P | 30,5% | 1 | 30,5% | 2 | 76,4% | 0,1784 | 0,0% | 56 | 70 | 100 | Atlantique | 8,31 |
| Site Q | 10,1% | 1 | 100,0% | 1 | 4,4% | 3,2604 | 0,0% | 52 | 70 | 0 | Continental | 46,86 |
| Site R | 32,0% | 1 | 100,0% | 1 | 100,0% | 0,0000 | 0,0% | 52 | 74 | 2 | Atlantique | 41,95 |
| Site S | 34,6% | 1 | 100,0% | 1 | 84,8% | 0,3764 | 85,1% | 67 | 67 | 100 | Continental | 58,55 |
| Site T | 76,4% | 4 | 100,0% | 3 | 0,0% | 0,0000 | 0,0% | 53 | 30 | 100 | Atlantique | 111,44 |
| Site U | 94,3% | 0 | 100,0% | 0 | 0,0% | 0,0000 | 0,0% | 41 | 81 | 0 | Atlantique | 4,33 |
| Site V | 14,7% | 1 | 100,0% | 1 | 0,0% | 0,0000 | 0,0% | 33 | 78 | 0 | Atlantique | 34,83 |
| Site W | 1,3% | 1 | 100,0% | 0 | 0,0% | 1,1008 | 0,0% | 33 | 78 | 0 | Atlantique | 12,25 |
| Site X | 18,9% | 1 | 100,0% | 1 | 0,0% | 0,7903 | 0,0% | 33 | 78 | 0 | Atlantique | 34,57 |
| Site Y | 28,2% | 3 | 100,0% | 0 | 88,1% | 0,6948 | 14,7% | 34 | 63 | 3 | Atlantique | 74,64 |
| Site Z | 68,8% | 4 | 100,0% | 1 | 72,4% | 0,3359 | 71,8% | 67 | 63 | 100 | Alpin | 1100,01 |

Figure 8 : Tableau de bord de nuances des couleurs pour 27 sites et carte des niveaux (pastilles) pour un indicateur donné

## 4. Productions et résultats potentiels

Les travaux de développement de cet outil, ainsi que sa mise en application sur un groupe de sites ont permis de produire les livrables suivants :

- **Guide** cadre méthodologique CARPO pour l'évaluation des potentialités écologiques (ce document).

- **Sorties** de l'outil automatisé et calculs manuels additionnels : Données brutes et données de synthèse (tableurs et SIG pour une ré exploitation ultérieure des résultats).

- **Atlas cartographique du contexte écologique** des sites d'implantation avec :
    - Des cartes descriptives pour chaque thématique (occupation du sol, zonages, réservoirs et corridors) sur la base des résultats bruts : cartes pour chaque site à différentes échelles, pour le groupe de sites, etc.
    - Des cartes analytiques sur la base des résultats CARPO :
        - Pour chaque thématique (occupation du sol, zonages et espèces patrimoniaux, réseaux écologiques) : cartes pour chaque site à différentes échelles, pour le groupe de sites
        - Pour chaque indicateur CARPO (voir chaque thématique) : une carte illustrant le niveau de potentialité écologique (pastille de couleur) pour le groupe de sites et pour l'indicateur en question.
        - Pour chaque thématique : une carte illustrant le niveau de potentialité écologique (pastille de couleur) pour le groupe de sites au sein de la thématique (cumul d'indicateurs d'une thématique).
    - Des représentations graphiques (tableaux synthétiques, diagrammes, notes commentées) sur la base des résultats bruts. Elles accompagnement les cartes.
    - Note technique sur la définition de seuils et classes de niveau : visualisation des résultats de la discrétisation Jenks via QGIS (tableaux de seuils/classes, histogrammes de distribution de valeurs, aperçu cartographique des sites dans chaque classe) (cf. Annexe 7 : exemple).
    - Grille d'indicateurs et de notation pour l'évaluation du niveau d'enjeu/potentialité : grille vierge, ainsi que les grilles renseignées pour les sites évalués.
    - Radars illustrant le niveau de potentialité écologique pour chaque site : un radar avec l'ensemble d'indicateurs et/ou un radar avec l'ensemble des thématiques. Ils accompagnent les grilles d'évaluation
    - Tableau de bord de nuances de couleurs reprenant les résultats de tous les sites pour chacun des indicateurs et descripteurs.

La Figure 9 synthétise le principe méthodologique de CARPO, depuis le croisement des BDD jusqu'à l'élaboration d'un atlas cartographique. Elle peut être complétée par l'Annexe 7 sur la définition des seuils de niveau de potentialité.

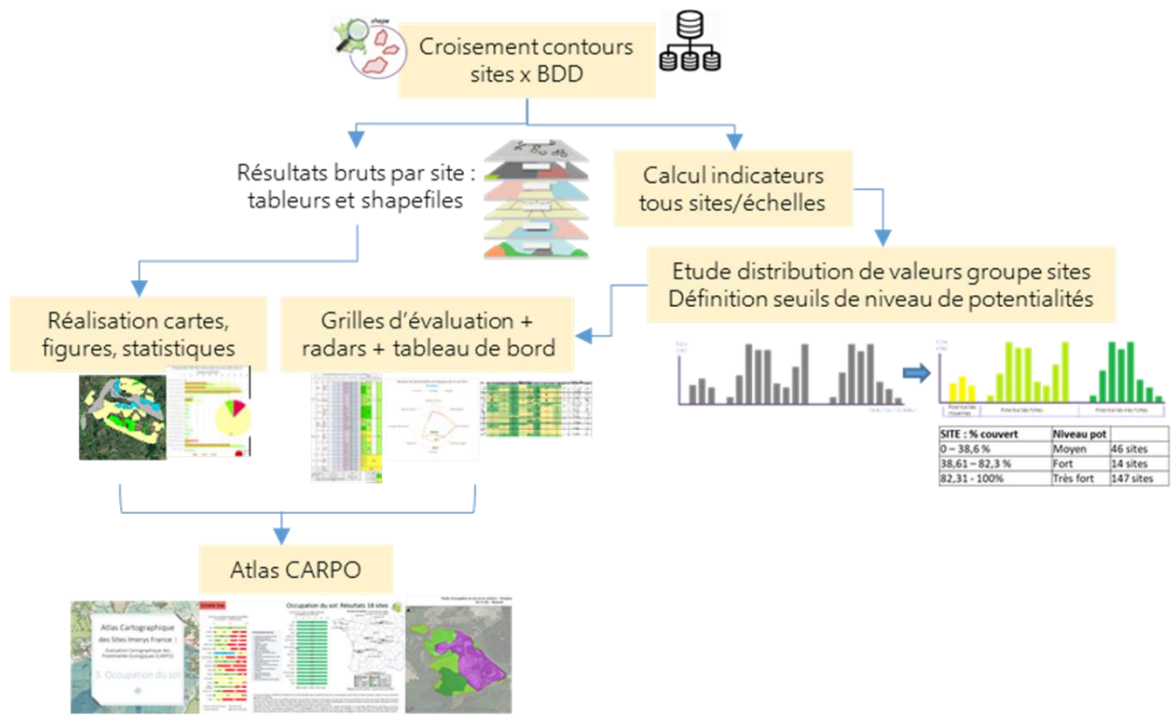

Figure 9 : Principe méthodologique de la mise en application de CARPO

## 5. Limites et perspectives

Démarche globale

- Le bon fonctionnement des outils de croisement et de calcul, ainsi que la valorisation des données de l'utilisateur dépendent de la bonne structure et l'homogénéisation de ces dernières. **L'utilisateur doit veiller à une bonne mise en forme du SIG** des sites à analyser avant toute analyse.

- Le calcul des résultats bruts et des indicateurs se fait sur trois échelles d'étude concentriques (site-voisinage-paysage) avec des surfaces qui sont inclues ou exclues dans l'analyse lorsque l'échelle d'étude varie (Figure 5). Ceci permet, entre autres, de mettre en évidence les potentialités écologiques à l'intérieur du site, en parallèle des potentialités extra-site. La pertinence du **choix de surfaces analysées selon l'échelle (inclusion/exclusion du périmètre intrasite aux échelles supérieures)** sera vérifiée lors des futurs tests et adaptée si besoin.

- Applicabilité de l'outil CARPO et conditions du groupe de sites à évaluer :
  - **Méthode nationale** : Cette méthode est applicable initialement sur le territoire terrestre de la France métropolitaine. En fonction du groupe de sites étudié, certains d'entre eux peuvent se trouver aux frontières françaises terrestres ou maritimes avec d'autres pays européens, ou à proximité de celles-ci. Etant donné que les bases de données cartographiques utilisées sont bornées au territoire français, le calcul de certains éléments sur une ou plusieurs échelles d'étude peut se voir biaisé lorsque l'échelle d'étude dépasse le territoire français. Ceci peut aboutir actuellement à des données incomplètes pour le périmètre non français. La question du traitement des données pour ces **sites limitrophes** doit être réfléchie pour une prochaine version de l'outil (par exemple une déformation des tampons (voisinage, paysage) sur le tracé des frontières françaises). Ces sites présenteraient donc des surfaces inférieures à celles normalement attendues, et seront dûment identifiés au niveau des sorties. Dans l'attente de ces optimisations, il a été décidé de **supprimer ces sites limitrophes lors des calculs des seuils/niveaux à appliquer, puis de les réintégrer pour leur appliquer ces seuils, une fois définis,** comme pour tous les autres sites non-problématiques.
  - **Taille du groupe de sites (nombre de sites)** : Cette méthode a été testée sur des réseaux de 43 à 254 sites, avec des résultats satisfaisants. Les seuils et les classes de niveau étant définis sur la base de la distribution des valeurs au sein du groupe de sites à analyser, il faudrait que l'échantillon (nombre de sites) soit significatif. En théorie, étant donné que trois classes de niveaux sont définies pour chaque indicateur, un échantillon de 3 sites pourrait déjà permettre de définir des seuils et de discrétiser les classes. Cependant, les formules de Brooks-Carruthers (k = 5 × log (N)) et de Sturges-Huntsberger (k = 1 + (10 ⁄ 3) × log (N)), où k est le nombre de classes souhaité et N la population (Paegelow, 2000), indiquent qu'à fin d'obtenir 3 classes, il faudrait au moins 4 valeurs. En conclusion, **CARPO est uniquement applicable pour les groupes de sites qui comptent au moins 4 sites**. Pour un groupe de sites de taille réduite, il est recommandé d'avoir recours à l'avis d'expert et des analyses de terrain pour identifier les potentialités et prioriser les actions.
  - **Taille des sites (surface)** : Les tests ont mis en évidence que des sites d'une petite surface peuvent présenter des problèmes de croisements des BDD et donc des manques ou des erreurs de calcul des indicateurs. Par conséquent, **CARPO est uniquement applicable aux sites d'une surface d'au moins 1 ha**. Pour de très petits sites (moins de 1 ha), il faudra ajouter un tampon de 100 m autour du périmètre avant tout croisement. Cette règle pourrait être réadaptée suite aux réflexions dans le cadre des futurs développements de l'outil.

- L'évaluation du niveau de potentialité se base uniquement pour l'instant sur **l'analyse du contexte écologique du territoire des sites d'implantation (BDD publiques et mobilisables) et non pas sur les enjeux réels au sein des sites**, qui nécessitent une mise à disposition et une intégration de données d'inventaire des sites. Des réflexions seront engagées sur la façon d'intégrer ces enjeux propres aux sites dans une future version de la méthode.

- Des réflexions pourraient également être engagées sur la manière de traduire les potentialités écologiques en enjeux d'actions prioritaires : enjeux de conservation, enjeux de connaissance (si lacunes identifiées), enjeux de gestion, etc. Si les atlas cartographiques fournissent une information du degré des caractéristiques écologiques du site, il sera nécessaire de travailler avec le gestionnaire des sites sur des pistes d'actions à mettre en place, suite à cette identification des potentialités.

- Ce document présente un prototype de la méthode qui a été soumis à quelques tests, mais qui nécessite un déploiement sur d'autres groupes de sites, afin d'étendre ces tests à d'autres contextes. Suite aux retours d'expérience, **cette méthode sera amenée à évoluer, ce qui entraînerait des modifications sur les indicateurs, les seuils et les règles décisionnelles définis dans cette V0**.

### Thématiques et indicateurs

Le Tableau 11 liste les limites et faiblesses identifiées pour les indicateurs actuellement considérés dans cette V0, ainsi que quelques perspectives :

| Thématiques et sous-thématiques indicateurs | Limites ou faiblesses | Perspectives |
|---|---|---|
| Occupation du sol : <br> - Hétérogénéité <br> - Caractère naturel <br> - Perméabilité | L'indicateur sur l'hétérogénéité (diversité) de milieux naturels part du principe que les potentialités écologiques d'un site augmentent lorsque la diversité des milieux augmente. Cependant se pose la question d'une possible préférence pour des secteurs avec une moindre diversité des milieux, lorsque ceux-ci permettraient d'avoir une matrice paysagère cohérente et fonctionnelle et qui serait plus importante qu'un site avec un grand nombre de milieux, mais des surfaces très faibles pour chacun et donc peu favorables à la biodiversité. Au-delà de la diversité, il serait intéressant de considérer la fonctionnalité et l'originalité des milieux recensés. | Les choix réalisés pour cette V0 correspondent à un consensus d'experts. La pertinence des choix réalisés sera examinée lors des futurs tests de la V0 sur d'autres groupes de sites, pour définir si des modifications doivent être apportées lors d'une prochaine V1. |
| | La définition des postes d'occupation du sol considérés comme naturels, semi-naturels, artificiels, perméables, imperméables ou mixtes, se fait sur la base du dire d'expert et est donc sujette à une part de subjectivité. Le classement des postes sur la base de Corine Land Cover et OSO n'est toujours pas aisée, et certains postes peuvent regrouper différents contextes plus complexes. Ceci est à considérer dans l'interprétation des résultats. | |
| | Les indicateurs sur le caractère naturel, l'hétérogénéité et la perméabilité, peuvent fournir des résultats très similaires en fonction des postes d'occupation du sol choisis pour les trois indicateurs. | La redondance entre indicateurs pourra être vérifiée lors des futurs tests, afin de faire un tri pertinent d'indicateurs pour une prochaine version de l'outil. |
| Réseaux écologiques : corridors et réservoirs | La couche utilisée correspond à une compilation de SRCE régionaux. Par conséquent, il existe une certaine hétérogénéité et une précision variable des corridors et des réservoirs en fonction de la région (contours précis, pixels, flèches, etc.). En outre, la localisation et la largeur de ces corridors sont données de façon approximative, voire sommaire, dans certaines occasions, ce qui implique une incertitude pour le calcul de sous-trames interceptées par une zone d'étude et qui suscite des doutes sur un éventuel besoin d'inclure un tampon autour des corridors. | |
| Réseaux écologiques : | Cet indicateur n'est actuellement pas calculable de manière automatique par l'outil atomatisé car les couches de la BDD TOPO | Dans le cas où les croisements et calculs seraient difficiles à traiter, cet indicateur |
| Unité du territoire (non-fragmentation) | ne sont pas intégrées à l'outil. **Le croisement des couches de la BD TOPO et du groupe de sites et le calcul de cet indicateur doit donc se faire « à la main ».** | pourrait être exclu ou remplacé dans une future version de CARPO. |
| | Selon la précision du tracé des périmètres des sites, certains axes de transport qui pourraient ne pas appartenir au site d'étude peuvent être inclus dans le calcul si le découpage dudit périmètre n'est pas suffisamment précis. | S'assurer avant tout croisement des données cartographiques que les périmètres des sites sont les plus précis possibles et corriger si besoin pour exclure des axes de transport hors-site. |
| | Le choix du type de linéaire de transport considéré comme fragmentant (axes larges, imperméabilisés, actifs, au niveau du sol) par rapport à ce qui constituerait seulement des obstacles mineurs (axes réduits, perméables, inactifs, surélevés/souterrains), se fait sur la base du dire d'expert et est donc sujet à une part de subjectivité. Ceci est à considérer dans l'interprétation des résultats. | Les choix réalisés pour cette V0 correspondent à un consensus d'experts. La pertinence des choix réalisés sera examinée lors des futurs tests sur d'autres groupes de sites, pour définir si des modifications doivent être apportées lors d'une prochaine V1. |
| Zonages : patrimonialité | La notation des différents zonages selon un score de patrimonialité (de 1 pour le moins patrimonial à 3 pour le plus patrimonial), se fait sur la base du dire d'expert et est donc sujette à une part de subjectivité. Ceci est à considérer dans l'interprétation. | |
| | Cet indicateur tient compte actuellement de 9 types de zonages hautement patrimoniaux (score 3). La question se pose sur la pertinence de considérer toutes ces zones à périmètres souvent superposés entre eux, et des fois redondantes, ou de ne choisir que certains zonages qui suffiraient à eux seuls à mettre en évidence les zones à plus forte patrimonialité. | Vérifier si un nombre plus réduit de zonages pourrait fournir une information plus pertinente et moins redondante. p. e. les périmètres des ZNIEFF de type I suffiraient-ils à mettre en évidence les zonages à haute patrimonialité ? |
| | Cette V0 propose un indicateur sur la proportion de la surface de la zone d'étude occupée par des zonages et un autre indicateur pour la proportion des réservoirs. Cependant, un grand nombre des zonages d'inventaire ou de protection de la biodiversité en France sont classés sur la base des périmètres des réservoirs. Ceci pourrait entraîner des résultats très similaires, voire redondants, pour ces deux indicateurs. | Les tests de cette méthode permettront de mettre en évidence s'il existe une corrélation entre ces deux indicateurs suggérant une redondance, et s'il est pertinent de ne considérer qu'un seul indicateur. |
| Espèces : patrimonialité | Les données mobilisables se retrouvent à une échelle plus large (maille 10 km) que celle du site, de son voisinage et son paysage. Ce ne sont donc pas des observations réelles sur les sites évalués, mais un aperçu des espèces potentielles et connues dans le territoire de la maille à laquelle le site évalué appartient. Cet aspect prédictif doit être pris en compte dans l'interprétation. | Réflexion à une intégration des données des espèces issues des inventaires au sein des sites, dans une V1. |
| | Cet indicateur croise la localisation de la zone d'étude avec les mailles de 10 km afin d'associer un site à une maille et d'afficher la valeur de la maille relative au nombre d'espèces à enjeux de conservation (SCAP20) et une valeur de méconnaissance naturaliste. Cependant, certains sites/voisinages/paysages se trouvent à cheval entre plusieurs mailles, ce qui pose la question de quelle valeur (quelle maille) afficher pour ces sites. | En attendant la définition de règles de décision au sein du service, la solution temporaire consiste à afficher la valeur de la maille avec la valeur de richesse la plus élevée et de signaler ces sites, pour tenir compte de ces particularités dans l'interprétation. |
| | La définition des critères permettant de juger si une espèce est patrimoniale se fait sur la base d'un choix des statuts les plus pertinents et communément utilisés dans plusieurs études. Cependant ce choix est issu de l'avis de plusieurs experts et est donc sujet à une part de subjectivité. | Ce choix est le fruit des travaux d'une équipe d'experts et d'un programme national (SCAP). Cependant, dans le cas où ces choix s'avéreraient inadaptés lors des tests de la V0, des modifications seraient apportées dans une V1. |
| | Pas de prise en compte des espèces menacées au niveau régional (listes rouges régionales) dans la V0. | Réflexion à une intégration des espèces des listes rouges régionales dans une V1. |
| | Les espèces à enjeux régionaux (déterminantes des ZNIEFF) ressortent comme à enjeu dans toutes les régions. Autrement dit, une espèce qui est déterminante de ZNIEFF dans une région sera tout de même comptabilisée dans la richesse d'une maille d'une autre région où elle n'est pas déterminante de ZNIEFF. Si ces espèces peuvent constituer une alerte à l'échelle nationale, cet aspect est à prendre en compte pour relativiser les résultats. | Réflexion à une possibilité de régionaliser les enjeux d'espèces déterminantes de ZNIEFF dans une future version. |
| | La présence d'un nombre réduit d'espèces à enjeux de conservation dans une maille ne traduit pas systématiquement un niveau réduit de patrimonialité et de potentialités écologiques de celle-ci. Ce résultat peut être dû à un nombre réduit des données | Relativiser les résultats de l'indicateur d'espèces à enjeux de conservation par rapport à la valeur du descripteur sur la méconnaissance naturaliste au sein de |

|  | d'espèces qui remonteraient à l'INPN dans cette maille et donc à un taux de connaissance faible des espèces dans celle-ci. | chaque maille et de la contribution à la biodiversité globalisée (CBG). |
|---|---|---|

Tableau 11 : Limites et perspectives pour les indicateurs proposés

- Le choix des indicateurs retenus actuellement est lié aux BDD et aux couches dont la connaissance et l'utilisation sont maîtrisées, et pour lesquelles les calculs d'une partie des indicateurs sont automatisés via l'outil de croisement et de calcul. Il est également fondamental que ces couches géographiques mobilisables soient **disponibles au format vecteur** (et non pas uniquement au format raster), afin de garantir une exploitation complète des entités et de leurs attributs. Cependant, il serait éventuellement intéressant d'intégrer d'autres BDD et d'autres couches si une analyse sur leur fiabilité, complétude, homogénéité et donc exploitabilité le démontre. Dans ce cas, il faudrait considérer les possibilités d'intégrer d'autres indicateurs qui sont présentés dans l'Annexe 10.

- Ces indicateurs prennent en compte toutes les entités des couches géographiques intersectées par les zones d'étude, peu importe la représentativité de cette entité dans la zone d'étude. Ainsi, des éléments occupant moins de 0,01 % de la zone d'étude pourraient être comptabilisés dans le calcul des indicateurs. Ceci soulève la question de choisir un **seuil en dessous duquel des éléments sous-représentés seront omis**.

- **Prise en compte incomplète des «enjeux réglementaires»** : Pour cette V0 de la méthode **sont uniquement considérés les aspects patrimoniaux** et non pas les aspects réglementaires. Les espèces protégées, ainsi que les zonages de protection ne sont pas considérés dans cette V0 par leur statut réglementaire, mais uniquement par leur aspect patrimonial : il n'y a pas d'indicateur sur les espèces protégées, et les zonages de protection sont considérés au même titre que les zonages d'inventaire ou le réseau Natura2000. Les aspects réglementaires qui sont complexes à noter/pondérer et qui ne traduisent pas toujours un intérêt écologique fort, pourront être reconsidérés dans une V1 de la méthode ou pourraient être évaluées au travers d'autres outils (p.e. la notation des zonages via ŒIL - Outil d'Évaluation de la biodiversité en contexte d'Infrastructure Linéaire). Les aspects réglementaires qui sont complexes à noter/pondérer et qui ne traduisent pas toujours un intérêt écologique fort, pourront être reconsidérés dans une V1 de la méthode ou pourraient être évaluées au travers d'autres outils (p.e. la notation des zonages via ŒIL - Outil d'Évaluation de la biodiversité

- Les valeurs obtenues pour chaque indicateur et la classe de niveau de potentialité résultante sont le fruit des croisements de bases de données mobilisées et ces **résultats dépendent donc de la précision et la finesse de ces bases**. En fonction de celles-ci, il peut y avoir :
  - Des décalages mineurs entre les périmètres de certains éléments, ce qui peut entraîner des surestimations ou sous-estimations de certains indicateurs; p. e. un axe routier qui ressortirait comme faisant partie d'un site, à cause d'un tracé imprécis du site, alors que cet axe borderait uniquement le site. Un autre exemple serait le cas des corridors ou réservoirs dont la géométrie n'est pas exacte et qui pourrait influencer leur croisement.
  - Quelques différences entre la nature attribuée à un objet dans une couche cartographique et la réalité sur le terrain, p. e. un secteur noté avec un mode d'occupation du sol qui n'est pas exactement celui qui est réellement présent sur site (par erreur, par changement, etc.).

Il est fortement conseillé de **toujours nuancer les résultats obtenus avec les descripteurs** (taille des sites, région biogéographique, méconnaissance naturaliste) et le contexte naturel/artificiel des sites.

Il est aussi recommandé de **privilégier une analyse aux échelles du voisinage et du paysage pour les indicateurs sur les corridors, les réservoirs et l'unité du territoire, ainsi que pour les indicateurs sur l'occupation du sol lorsque Corine Land Cover est utilisée** comme source des données. Ceci en raison de la précision et la finesse de ces bases dont leur prise en compte est plus pertinent à une échelle plus large que celle du site.

### Seuils, fourchettes de valeur et niveaux

- Comme indiqué précédemment, des intervalles de valeurs nationales de référence ne sont pas proposés pour chaque classe de niveau de potentialité ; mais les seuils et les classes de niveaux doivent être calculés au **préalable à toute évaluation du groupe de sites, sur la base d'une étude de la distribution de valeurs de la série**. Ceci a été choisi afin d'étudier du contexte écologique du groupe de sites, d'analyser les niveaux de potentialités au sein de ce groupe, et non pas d'une simplification d'un contexte national très hétérogène qui complexifierait la définition de seuils applicables sur tout le territoire. De plus, cela implique que l'étude de la distribution de valeurs doit se faire **sur l'ensemble des sites en une seule fois**, et qu'il est impossible de définir des seuils sur une première sub-série qui serait complété par une seconde.

- Parallèlement, si le choix des seuils se fait initialement sur la base de la distribution de valeurs de la série de chaque groupe de sites à étudier, il serait possible de réaliser des tests statistiques avec des tirages aléatoires. Ainsi, une future version de la méthode pourrait éventuellement proposer des seuils/intervalles régionaux. Dans ce cas, il faudrait étudier la distribution des valeurs d'un groupe fictif de sites qui soit représentatif de chaque région (p. e. sur un ensemble de >300 sites fictifs de tailles variables) afin de **définir des tendances à l'échelle de chaque région biogéographique**. Ceci permettrait de tenir compte des différences sur les connaissances mobilisables et l'hétérogénéité des données remontées selon les régions, ainsi que des différences écologiques au sein des régions biogéographiques.

- Suite aux tests réalisés, il a été établi que le calcul de seuils et de classes de niveau se fait via la méthode de **discrétisation de données (regroupement en classes) de l'algorithme de Jenks**. Les avantages et limites de cette méthode, ainsi que d'autres méthodes de discrétisation existantes sont présentées dans l'Annexe 6a. Ce choix de méthode de discrétisation à appliquer, détermine les valeurs retenues comme seuils pour chaque classe. Bien que cette méthode semble être la plus adaptée, si de futurs tests et le déploiement de CARPO suggèrent qu'une autre méthode doit être adoptée, ces modifications seraient apportées lors d'une future version.

- La V0 de CARPO propose une démarche de classement des sites en 3 niveaux de potentialités (moyen - fort - très fort) afin de réduire le nombre de possibilités, et d'offrir une vision synthétique et opérationnelle d'un classement intra-groupe de sites. Cependant les tests de cette V0 pourraient amener les réflexions vers une augmentation du nombre de niveaux de potentialités (et donc du nombre de seuils).

### Cumul des niveaux de potentialité

- Ce type de démarche d'agrégation de niveaux offre l'avantage de donner des informations de synthèse pour avoir une vision globale d'une thématique et d'un groupe de sites. Cependant, il présente la faiblesse de l'incertitude liée à la fiabilité d'un groupement et à la superposition d'indicateurs différents et d'une homogénéité des BDD qui n'est pas toujours optimale. Par conséquent, la robustesse de ces choix méthodologiques sera mise à l'épreuve afin de déterminer si le cumul des niveaux de potentialité est faisable, pertinent et fiable sur la base de la méthode proposée. **Dans le cas où cette approche de cumul s'avérerait inadéquate, l'analyse s'arrêtera à la notation d'un niveau de potentialité de chaque indicateur** (pas de cumul pour la thématique).

# Bibliographie

# Annexe 1 :
## Spécificités sur les indicateurs relatifs à l'occupation du sol

L'occupation du sol aux différentes échelles d'étude rend compte de l'utilisation des espaces au sein d'un territoire, et donc d'un degré de préservation ou de perturbation des milieux naturels qui seraient favorables à l'accueil d'une biodiversité fonctionnelle.

Les 3 indicateurs liés aux modes d'occupation du sol (MOS) sont :

| Sous thématique | Indicateurs | Intérêt de la thématique : Mettre en évidence les sites… |
|---|---|---|
| Caractère naturel | % surface de la zone d'étude occupée par des milieux naturels et semi-naturels | Avec une bonne proportion de milieux naturels et semi-naturels car ils posséderaient de potentialités importantes pour l'accueil de la biodiversité. Bien que les sites déjà très artificiels aient un faible intérêt de préservation (enjeux écologiques soupçonnés faibles), ils pourraient avoir un fort intérêt pour la restauration. Prioriser les actions de préservation sur sites plutôt naturels ou prioriser les actions de restauration/désartificialisation sur sites plutôt artificiels. |
| Hétérogénéité | Nombre de milieux naturels différents dans la zone d'étude | Avec une bonne diversité d'habitats naturels différents (sites hétérogènes) qui offriraient des niches écologiques pour une plus grande diversité d'espèces, par rapport à des sites avec une faible diversité d'habitats (homogènes). Les mosaïques d'habitats et la diversité paysagère est considérée comme favorable pour la biodiversité. |
| Perméabilité | % surface de la zone d'étude occupée par des MOS perméables | Avec des sols non construits, non revêtus et non compacts (peu artificialisés), permettant des échanges physicochimiques (eau-air) et entre les organismes vivants du sol. Prise en compte de la trame brune. Prioriser les actions de préservation sur sites perméables ou prioriser les actions de désimperméabilisation sur sites imperméables. |

Ces indicateurs sont calculés sur la base d'une de couches cartographique publique, nationale, homogène, standardisée et mobilisable. Le tableau suivant illustre et compare les deux couches cartographiques existantes et qui ont testées dans cette méthode : la couche européenne de la typologie Corine Land Cover (CLC) dans sa version 2018, et la couche française Occupation du Sol (OSO) du Centre d'Etudes Spatiales de la Biosphère (Cesbio), dans sa version 2017 ou 2018.

|  | Corine Land Cover | OSO |
|---|---|---|
|  | 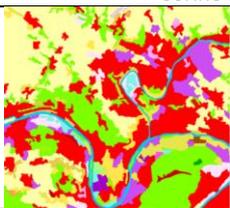 | 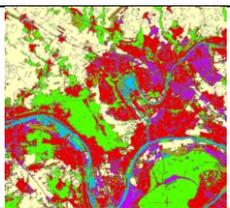 |
| Producteur (France) | Institut national de l'information géographique et forestière (IGN) (depuis 2018) | CES OSO (centre d'expertise scientifique) & CESBIO (centre d'étude Spatiale de la BIOsphère) |
| Première édition | 1985 | 2016 |
| Mise à jour | 2018 - Tous les 6 ans | 2018- Annuelle |
| Méthode d'acquisition | Photo-interprétation humaine d'images satellites (Landsat, SPOT, IRS, …) | Classification automatique d'images satellites (Sentinel,…) |
| Echelle d'utilisation | 1/100 000 | 1/10 000 – 1/25 000 |
| Unité min. de collecte | 25 ha | 0,1 ha (résolution entre 10 m et 20 m) |
| Postes | 44 classes (au niveau 3) | 23 postes (17 en 2017) |
| Couverture | France entière (dont l'outre-mer) et 39 pays en Europe | France métropolitaine |

La couche utilisée lors de l'évaluation d'un groupe de sites doit être la même pour les trois indicateurs. Il est donc indispensable de faire un choix. Chaque couche présente ses avantages et inconvénients et suite aux tests effectués, **il est fortement recommandé d'utiliser provisoirement la couche vecteur OSO17 (version** 2017), étant donné que la version 2018 d'OSO n'est pas encore disponible en format vecteur. Bien que OSO17 compte un nombre de postes plus réduit que celui de CLC18, elle présente une meilleure résolution cartographique (échelle d'utilisation plus fine), moins de problèmes au moment du choix des étiquettes « caractère naturel » et « perméabilité » de chacun des postes, et malgré l'existence d'une version plus récente (OSO18), la version OSO17 reste assez récente. **Au moment où des fichiers vecteur d'OSO18 (version 2018) seront disponibles** (actuellement seulement des fichiers raster le sont), **cette dernière deviendra la source de données d'entrée à privilégier**. Parallèlement, pour les groupes de sites nationaux qui font partie intégrante d'un plus vaste ensemble européen, **il est possible d'utiliser CLC18 comme couche d'entrée, afin d'assurer une comparabilité des résultats avec le contexte européen**.

Il faut noter qu'en raison de la résolution de ces couches, il est difficile d'obtenir une information très précise à l'échelle du site sans des relevés de terrain. Ces couches donneront des informations nettement descriptives des grands milieux présents, mais leur interprétation devra tenir compte de ces limites de précision.

Les indicateurs CARPO en lien avec l'occupation du sol englobent deux concepts clefs : celui de l'artificialisation, et celui de l'imperméabilisation des sols :

> **L'artificialisation** « *consiste à transformer un sol naturel, agricole ou forestier, par des opérations d'aménagement pouvant entraîner une imperméabilisation partielle ou totale, afin de les affecter notamment à des fonctions urbaines ou de transport (habitat, activités, commerces, infrastructures, équipements publics…)* » (Cerema, 2019).
> Un espace peut être considéré comme artificialisé « *lorsque celui-ci ne présente plus les caractéristiques d'un écosystème fonctionnel évoluant librement ou géré de manière à favoriser le maintien des populations d'espèces qui y accomplissent tout ou partie de leur cycle de vie, les habitats naturels présents et les fonctions biotiques et abiotiques qu'il remplit* » (Padilla *et al.*, 2020).

> **L'imperméabilisation** des sols *« est le recouvrement permanent d'une parcelle de terre et de son sol par un matériau artificiel imperméable tel que l'asphalte ou le béton »* (Commission Européenne, 2012).

Si ces deux termes sont souvent liés, des nuances existent et certains sols peuvent être artificiels et perméables en même temps. « *Un sol est imperméabilisé ou minéralisé lorsqu'il est recouvert d'un matériau imperméable à l'eau et l'air, tel que l'asphalte ou le béton (routes, voies ferrées, parkings, constructions…) de manière irréversible. Un sol artificialisé (pelouse, gravillons, chantiers, chemin…) perd tout ou partie de ses fonctions écologiques, mais de manière réversible* » (Magdelaine, 2019).

Afin de calculer ces proportions de surface ou ces nombres de modes/postes d'occupation du sol pour chaque indicateur, il faut **déterminer pour les deux couches et pour tous les modes d'occupation du sol** :

- Les postes considérés comme étant liés à des milieux « naturels », « semi-naturels » et « artificiels »
- Les postes considérés comme étant liés à des sols « perméables », « imperméables », « mixtes »

La définition de ces termes est sujette à de débats, et plusieurs auteurs s'accordent à privilégier une notion de degré d'artificialisation/naturalité et de perméabilité, au lieu d'une série d'états fixes et clairement définies. Cependant, tout en restant dans le principe de ces degrés fluides et dynamiques, il a été choisi de définir les classes relatives au caractère naturel et à la perméabilité de la façon suivante :

| Naturel | Semi-naturel | Artificiel |
|---|---|---|
| Milieux intacts ou non bouleversés régulièrement par les activités humaines, en considérant que la mise en assec d'étangs ou la sylviculture correspondent à un niveau de perturbation soutenable par les écosystèmes. | Milieux peu perturbés (perturbations réduites dans le temps) ou en état d'équilibre entre les processus naturels et les activités humaines maintenant ces milieux (Garcin, 2018). Inclut des pratiques de gestion comme la fauche, le pâturage ou le traitement des prairies aux produits phytosanitaires | Milieux présentant des perturbations anthropiques d'un niveau supérieur et répétées dans le temps (labour régulier, mise à nu des sols), d'une durée plus ou moins permanente, qui entrainent des conséquences majoritairement délétères pour les espèces. |
| Milieux résultant de l'expression à un temps T de déterminismes écologiques, en dehors de toute gestion humaine, abstraction faite des conditions initiales de mise en place de ce système. Cette notion est détachée du concept de climax, ce qui permet de considérer comme naturel tout type de végétation, quel que soit son stade au sein d'une succession (Bota-phytoso-flo, 2014). | Milieux où l'action de l'homme a pour seul effet de bloquer durablement un ou quelques processus participant au déterminisme écologique de ce système (p. e. via la fauche annuelle, pâturage extensif et débroussaillage). La gestion se limite en un événement perturbateur cycliquement répété permettant de bloquer une dynamique naturelle (Bota-phytoso-flo, 2014). | Milieux où l'homme prend pleinement part à son organisation en l'adaptant et en l'ajustant à ces besoins. Cette gestion forte et récurrente se fait sur la structuration du système (travail du sol, choix actif des espèces vivantes) et sur le contrôle des conditions écologiques (irrigation, drainage, utilisation d'intrants). (Bota-phytoso-flo, 2014) |

| Perméable | Mixte | Imperméable |
|---|---|---|
| Milieux associées à des sols composés de surfaces permettant la circulation et pénétration des fluides (eau), de l'air et des organismes vivants. Ces surfaces absorbantes sont souvent d'origine végétale et / ou minérale (surfaces rocheuses naturelles comprises). | Milieux composés souvent par des sols perméables, mais pouvant comprendre ponctuellement des infrastructures rendant les sols partiellement imperméables (cas de certaines décharges, terrains sportifs, etc.) | Milieux associés à des sols compactés, recouverts de matériaux plus ou moins imperméables ou simplement ceux sur lesquels des bâtiments ont été érigés. Ces sols sont étanches par l'action humaine (à différence de surfaces rocheuses naturelles) ce qui empêche l'infiltration de l'eau dans le sol et augmente le ruissellement (Cobali, 2016). Les milieux imperméables sont compris ici comme « imperméabilisés » (rendus imperméables par l'action humaine). |

Les tableaux suivants indiquent le **statut de chaque poste d'occupation du sol et pour chaque couche, vis-à-vis de ces questions d'artificialisation et d'imperméabilisation**. Cette typologie a été testée dans le cadre de la V0 de cette méthode et pourrait évoluer lors d'une future V1.

| Occupation du Sol (OSO) 2018 - Cesbio ||||
|---|---|---|---|
| Code | Nom_Poste | Caractère naturel | Perméabilité |
| 1 | bâtis denses | Artificiel | Imperméable |
| 2 | bâtis diffus | Artificiel | Mixte |
| 3 | zones ind et com | Artificiel | Imperméable |
| 4 | surfaces routes | Artificiel | Imperméable |
| 5 | colza | Artificiel | Perméable |
| 6 | céréales et pailles | Artificiel | Perméable |
| 7 | protéagineux | Artificiel | Perméable |
| 8 | soja | Artificiel | Perméable |
| 9 | tournesol | Artificiel | Perméable |
| 10 | maïs | Artificiel | Perméable |
| 11 | riz | Artificiel | Perméable |
| 12 | tubercules/racines | Artificiel | Perméable |
| 13 | prairies | Semi-naturel | Perméable |
| 14 | vergers | Semi-naturel | Perméable |
| 15 | vignes | Semi-naturel | Perméable |
| 16 | forêts de feuillus | Naturel | Perméable |
| 17 | forêts de conifères | Naturel | Perméable |
| 18 | pelouses | Naturel | Perméable |
| 19 | landes ligneuses | Naturel | Perméable |
| 20 | surfaces minérales | Naturel | Perméable |
| 21 | plages et dunes | Naturel | Perméable |
| 22 | glaciers ou neiges | Naturel | Perméable |
| 23 | eau | Naturel | Perméable |

| Occupation du Sol (OSO) 2017 - Cesbio ||||
|---|---|---|---|
| Code | Nom_Poste | Caractère naturel | Perméabilité |
| 11 | culture été | Artificiel | Perméable |
| 12 | culture hiver | Artificiel | Perméable |
| 31 | foret feuillus | Naturel | Perméable |
| 32 | foret conifères | Naturel | Perméable |
| 34 | pelouses | Naturel | Perméable |
| 36 | landes ligneuses | Naturel | Perméable |
| 41 | urbain dense | Artificiel | Imperméable |
| 42 | urbain diffus | Artificiel | Mixte |
| 43 | zones ind et com | Artificiel | Imperméable |
| 44 | surfaces routes | Artificiel | Imperméable |
| 45 | surfaces minérales | Naturel | Perméable |
| 46 | plages et dunes | Naturel | Perméable |
| 51 | eau | Naturel | Perméable |
| 53 | glaciers ou neige | Naturel | Perméable |
| 211 | prairies | Semi-naturel | Perméable |
| 221 | vergers | Semi-naturel | Perméable |
| 222 | vignes | Semi-naturel | Perméable |

| | | | | | Corine Land Cover (CLC) | | | |
|---|---|---|---|---|---|---|---|---|
| code niv1 | libellé | code niv2 | libellé | code niv3 | libellé | | Caractère naturel | Perméabilité |
| 1 | Territoires artificialisés | 11 | Zones urbanisées | 111 | Tissu urbain continu | | Artificiel | Imperméable |
| | | | | 112 | Tissu urbain discontinu | | Artificiel | Mixte |
| | | 12 | Zones industrielles ou commerciales et réseaux de communication | 121 | Zones industrielles ou co0mmerciales et installations publiques | | Artificiel | Imperméable |
| | | | | 122 | Réseaux routier et ferroviaire et espaces associés | | Artificiel | Imperméable |
| | | | | 123 | Zones portuaires | | Artificiel | Imperméable |
| | | | | 124 | Aéroports | | Artificiel | Imperméable |
| | | 13 | Mines, décharges et chantiers | 131 | Extraction de matériaux | | Artificiel | Perméable |
| | | | | 132 | Décharges | | Artificiel | Mixte |
| | | | | 133 | Chantiers | | Artificiel | Mixte |
| | | 14 | Espaces verts artificialisés, non agricoles | 141 | Espaces verts urbains | | Semi-naturel | Mixte |
| | | | | 142 | Equipements sportifs et de loisirs | | Artificiel | Mixte |
| 2 | Territoires agricoles | 21 | Terres arables | 211 | Terres arables hors périmètres d'irrigation | | Artificiel | Perméable |
| | | | | 212 | Périmètres irrigués en permanence | | Artificiel | Perméable |
| | | | | 213 | Rizières | | Artificiel | Perméable |
| | | 22 | Cultures permanentes | 221 | Vignobles | | Semi-naturel | Perméable |
| | | | | 222 | Vergers et petits fruits | | Semi-naturel | Perméable |
| | | | | 223 | Oliveraies | | Semi-naturel | Perméable |
| | | 23 | Prairies | 231 | Prairies et autres surfaces toujours en herbe à usage agricole | | Semi-naturel | Perméable |
| | | 24 | Zones agricoles hétérogènes | 241 | Cultures annuelles associées à des cultures permanentes | | Artificiel | Perméable |
| | | | | 242 | Systèmes culturaux et parcellaires complexes | | Semi-naturel | Perméable |
| | | | | 243 | Surfaces essentiellement agricoles, interrompues par des espaces naturels importants | | Semi-naturel | Perméable |
| | | | | 244 | Territoires agroforestiers | | Semi-naturel | Perméable |
| 3 | Forêts et milieux semi-naturels | 31 | Forêts | 311 | Forêts de feuillus | | Naturel | Perméable |
| | | | | 312 | Forêts de conifères | | Naturel | Perméable |
| | | | | 313 | Forêts mélangées | | Naturel | Perméable |
| | | 32 | Milieux à végétation arbustive et/ou herbacée | 321 | Pelouses et pâturages naturels | | Naturel | Perméable |
| | | | | 322 | Landes et broussailles | | Naturel | Perméable |
| | | | | 323 | Végétation sclérophylle | | Naturel | Perméable |
| | | | | 324 | Forêt et végétation arbustive en mutation | | Naturel | Perméable |
| | | 33 | Espaces ouverts, sans ou avec peu de végétation | 331 | Plages, dunes et sable | | Naturel | Perméable |
| | | | | 332 | Roches nues | | Naturel | Perméable |
| | | | | 333 | Végétation clairsemée | | Naturel | Perméable |
| | | | | 334 | Zones incendiées | | Naturel | Perméable |
| | | | | 335 | Glaciers et neiges éternelles | | Naturel | Perméable |
| 4 | Zones humides | 41 | Zones humides intérieures | 411 | Marais intérieurs | | Naturel | Perméable |
| | | | | 412 | Tourbières | | Naturel | Perméable |
| | | 42 | Zones humides côtières | 421 | Marais maritimes | | Naturel | Perméable |
| | | | | 422 | Marais salants | | Semi-naturel | Perméable |
| | | | | 423 | Zones intertidales | | Naturel | Perméable |
| 5 | Surfaces en eau | 51 | Eaux continentales | 511 | Cours et voies d'eau | | Naturel | Perméable |
| | | | | 512 | Plans d'eau | | Naturel | Perméable |
| | | 52 | Eaux maritimes | 521 | Lagunes littorales | | Naturel | Perméable |
| | | | | 522 | Estuaires | | Naturel | Perméable |
| | | | | 523 | Mers et océans | | Naturel | Perméable |

Evaluation cartographique du niveau de potentialités écologiques (CARPO)

L'exemple suivant illustre les différentes représentations cartographiques pouvant être produites à partir des fichiers de sortie pour un site étudié à l'échelle de son voisinage (1000 m) : une visualisation des différents MOS, du caractère naturel associé à ces MOS, ainsi que de leur perméabilité. Les encadrés précisent les résultats attribués à ce site (à l'échelle du voisinage) pour les trois indicateurs évalués dans cette thématique.

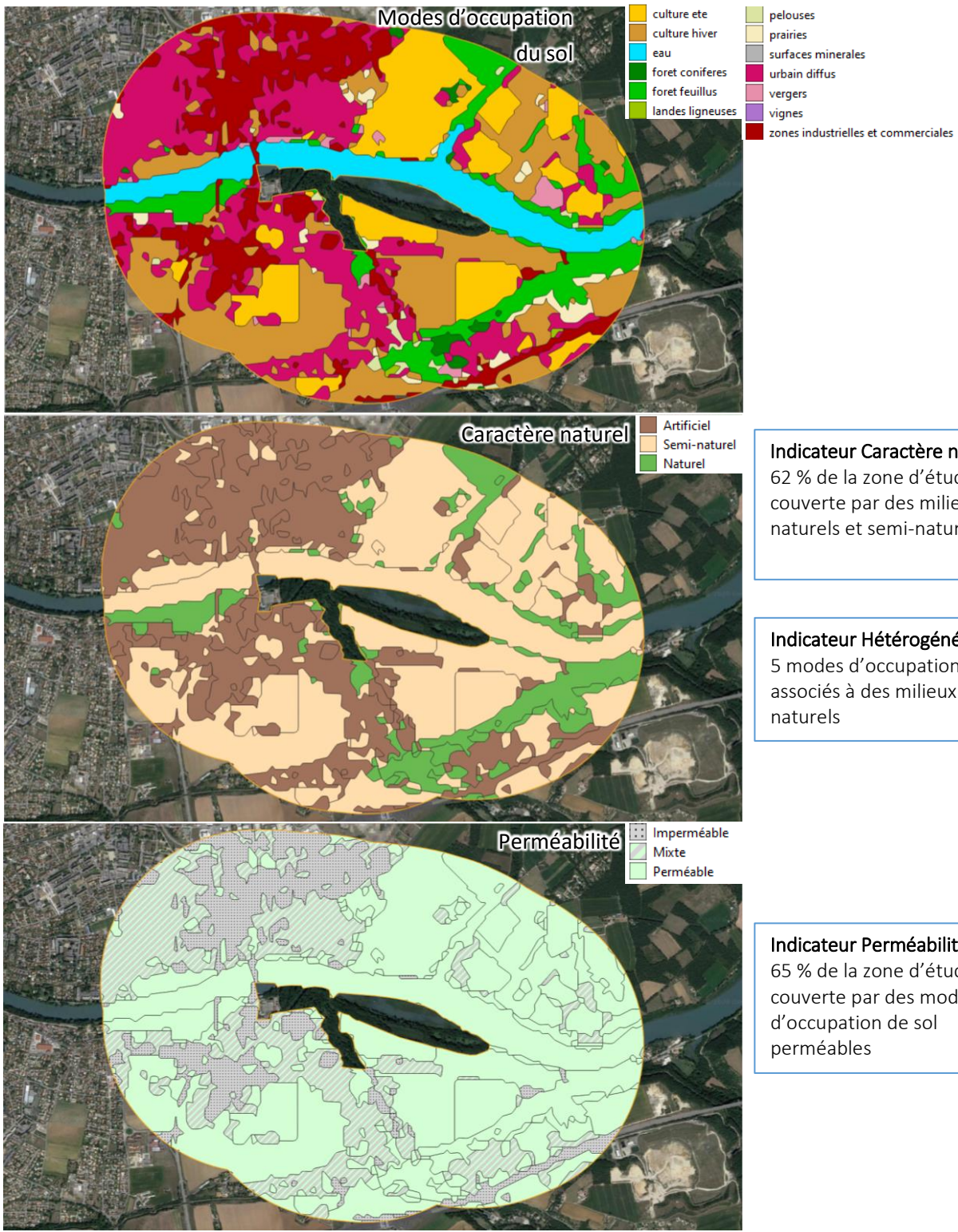

**Indicateur Caractère naturel :**
62 % de la zone d'étude couverte par des milieux naturels et semi-naturels

**Indicateur Hétérogénéité:**
5 modes d'occupation de sol associés à des milieux naturels

**Indicateur Perméabilité :**
65 % de la zone d'étude couverte par des modes d'occupation de sol perméables

## Annexe 2 :
## Spécificités sur les indicateurs relatifs aux réseaux écologiques

Les 3 indicateurs proposés et liés aux réseaux écologiques dans les trois échelles d'étude sont :

| Sous thématique | Indicateurs | Intérêt de la thématique : Mettre en évidence les sites… |
|---|---|---|
| Corridors | Nombre de sous-trames différentes traversant la zone d'étude (nombre de milieux majoritaires des corridors) | Une diversité et une quantité de sous-trames importantes. Ces zones d'étude comprendraient des corridors (linéaires et surfaciques) assurant des connexions entre des réservoirs de biodiversité, offrant aux espèces des conditions favorables à leur déplacement, aux échanges génétiques entre populations, et à l'accomplissement de leur cycle de vie. Une diversité de types de corridors, selon le milieu majoritaire (boisé, ouvert, humide, multi trame, non classé, etc.) suggère une présence de trames favorables à des cortèges faunistiques de plusieurs types d'habitats. |
| Réservoirs | % surface de la zone d'étude occupée par des réservoirs | Comportant un nombre élevé d'espaces dans lesquels la biodiversité est la plus riche ou la mieux représentée, où les espèces peuvent effectuer tout ou partie de leur cycle de vie (alimentation, reproduction, repos) et où les habitats naturels peuvent assurer leur fonctionnement (taille suffisante). Ces sites comprendraient des espaces pouvant abriter des noyaux de populations d'espèces (pour la dispersion d'individus ou l'accueil de nouvelles populations). |
| Unité du territoire (non-fragmentation) | Densité du linéaire des réseaux de transport routier ou ferroviaire (Rapport : longueur totale linéaires / surface de l'échelle d'étude) | Peu fragmentés, avec de bonnes continuités écologiques car la densité d'obstacles aux déplacements (axes de transport fragmentant : routes, voies ferrées) y est faible. Prioriser les actions de préservation sur les sites favorables aux réseaux ou prioriser les actions de restauration de continuités sur les sites plutôt fragmentés. |

### A) Indicateurs sur les corridors et les réservoirs

La localisation des sites au sein des éléments de connectivité écologique peut être appréciée à travers la caractérisation de leur localisation au sein d'un réseau de corridors et des réservoirs de biodiversité :
- « Les **réservoirs** de biodiversité sont des espaces dans lesquels la biodiversité, rare ou commune, menacée ou non menacée, est la plus riche ou la mieux représentée, où les espèces peuvent effectuer tout ou partie de leur cycle de vie (alimentation, reproduction, repos) et où les habitats naturels peuvent assurer leur fonctionnement, en ayant notamment une taille suffisante. Ce sont des espaces pouvant abriter des noyaux de populations d'espèces à partir desquels les individus se dispersent, ou susceptibles de permettre l'accueil de nouvelles populations d'espèces ».
- « Les **corridors** écologiques assurent des connexions entre des réservoirs de biodiversité, offrant aux espèces des conditions favorables à leur déplacement et à l'accomplissement de leur cycle de vie… Les corridors écologiques peuvent prendre plusieurs formes et n'impliquent pas nécessairement une continuité physique ou des espaces contigus. On distingue ainsi trois types de corridors écologiques : les corridors linéaires (haies, chemins et bords de chemins, ripisylves, bandes enherbées le long des cours d'eau) ; les corridors discontinus (ponctuation d'espaces-relais ou d'îlots-refuges, mares permanentes ou temporaires, bosquets) ; les corridors paysagers (mosaïque de structures paysagères variées) » (Centre de ressources TVB, 2020).

Ces réservoirs et ces corridors peuvent être également classés en fonction du **milieu majoritaire** les composant, à savoir, des milieux de type boisé, ouvert, humide ou multi-trame. De plus les corridors peuvent être associés à un objectif attribué dans le cadre du SRCE : préservation, restauration ou à préciser.

La présence, la qualité et la diversité de réservoirs et de corridors dans le territoire d'un groupe de sites, rend compte de la place de ces secteurs dans l'accueil de la biodiversité, de leur contribution aux échanges des populations des espèces et donc du rôle stratégique de leur préservation et de leur reconnexion pour le maintien de ces dispersions. Ces échanges génétiques entre populations sont fondamentaux pour le maintien des espèces.

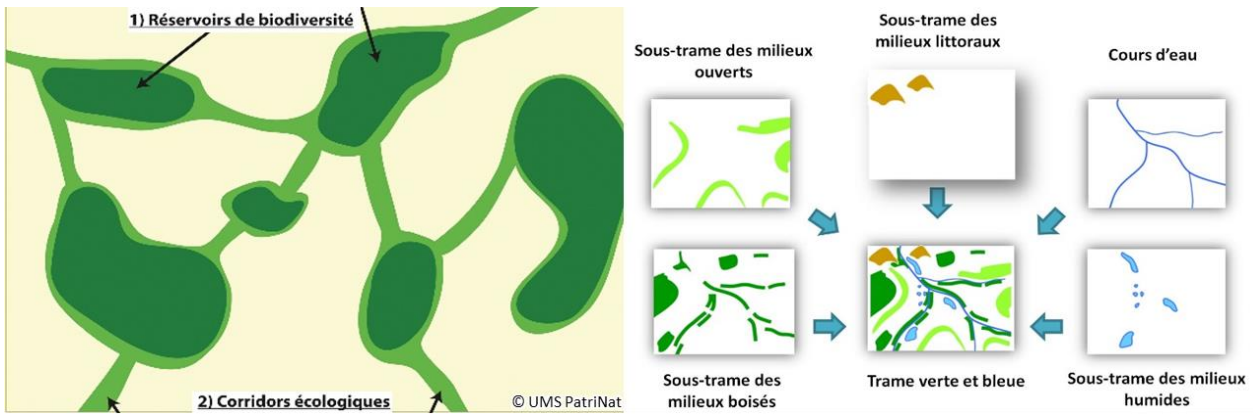

Ces éléments de connectivité sont déterminés et pris en considération par chaque région en France, au travers les Schémas Régionaux de Cohérence Ecologique (SRCE), inclus dans les SRADDET. L'analyse de ces réseaux écologiques proposés dans cette méthode se base sur **le travail de compilation de ces différentes cartographies régionales**. En effet, la base de données nationale Trame verte et bleue constitue une compilation et une standardisation des données des différents Schémas Régionaux de Cohérence Ecologique (SRCE) (Billon et al., 2016). Or, les bilans techniques et scientifiques réalisés indiquent une **grande hétérogénéité** dans les méthodes d'identification des composantes de la TVB (Sordello et al., 2017), les représentations cartographiques de la TVB (Billon et al., 2017) ainsi que dans les méthodes d'identification des obstacles et d'attribution des objectifs (Vanpeene et al., 2017). La constitution de la base de données nationale a nécessité la mise en conformité des bases régionales, tout en restant au plus proche des données produites par les régions. Les représentations géographiques des différentes composantes de la TVB peuvent donc être différentes d'une région à l'autre (Fournier et al., 2019).

L'exemple suivant illustre les différentes représentations cartographiques possibles pour deux sites étudiés à l'échelle du voisinage (1000 m) : une visualisation des corridors (gauche) et des réservoirs (droite). Les encadrés précisent les résultats de ces sites (voisinage) pour 2 indicateurs évalués dans cette thématique.

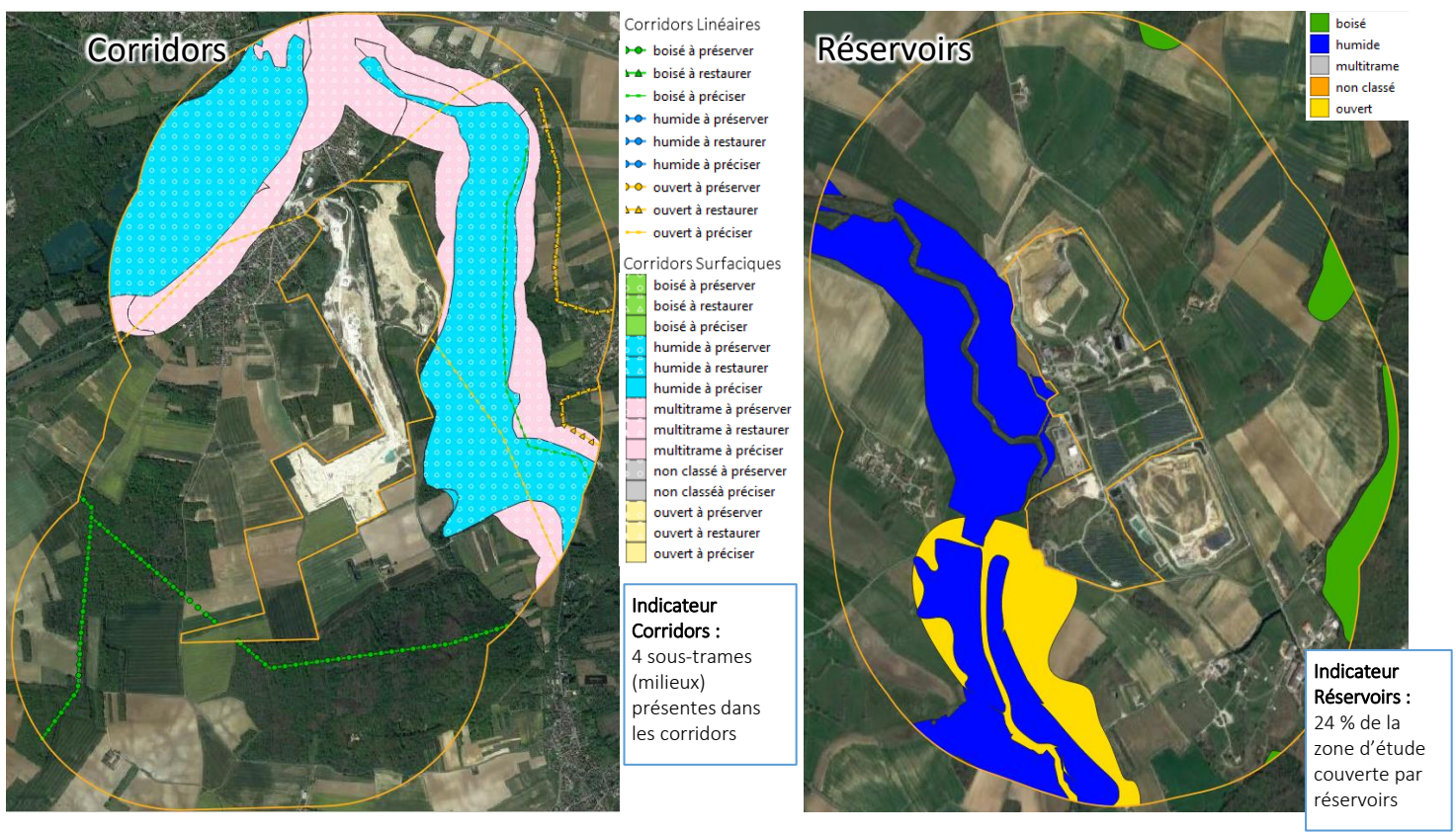

Evaluation cartographique du niveau de potentialités écologiques (CARPO)

### B) Indicateur sur la l'unité du territoire (non-fragmentation)

Bien qu'un site puisse avoir un rôle favorisant les réseaux écologiques, ils peuvent aussi présenter des barrières et des obstacles fragmentant le territoire.

La **fragmentation** décrit un ensemble de processus qui transforme une surface continue d'habitat naturel en un nombre plus ou moins important de fragments de taille variable (Thompson et al., 2010), suite à un changement d'usage des terres ou à la création d'infrastructures de transport. Ces îlots d'habitats se trouvent ainsi isolés, séparés : on parle de **perte de connectivité** (OFB, 2020). Ce morcellement peut empêcher un ou plusieurs individus, espèces, population ou association de ces entités vivantes de se déplacer comme elles le devraient et le pourraient en l'absence de facteur de fragmentation (Norpac, 2011).

Les **infrastructures linéaires de transport** sont un des principaux éléments de fragmentation écologique. Ils peuvent induire un morcellement significatif, non seulement en raison du découpage du territoire et de l'artificialisation entrainé par ces aménagements, mais aussi par le trafic associé. En outre, même sans trafic, de nombreuses espèces d'invertébrés refusent de traverser ces axes. L'aire écologiquement impactée par la route dépasse largement la superficie de la route elle-même. En divisant les milieux naturels, les réseaux de transport peuvent avoir plusieurs **effets** sur les espèces :
- Contraints au cycle biologique des espèces
- Mortalité directe par collision
- Réduction de la taille des habitats de telle sorte que les populations d'espèces à domaine vital important ne peuvent plus y vivre
- Isolement des tâches d'habitats restantes de telle sorte que les individus ont peu de chances de se déplacer de l'une à l'autre (ESRI France, 2020).

L'**isolement** des populations peut même entraîner leur **extinction** par **limitation de la dispersion et des échanges méta populationnels**. En effet, la faible taille des populations les rend très **sensibles aux aléas** de survie et de reproduction tout au long de leur cycle de vie. La recolonisation de ces fragments est alors difficile du fait de leur isolement. Le succès de reproduction dans les fragments de petite taille peut également être limité. La **perte de diversité génétique** associée aux faibles échanges peut compromettre leur **viabilité** car la consanguinité accrue peut conduire à un **déclin des performances** des individus (Thompson et al., 2010).

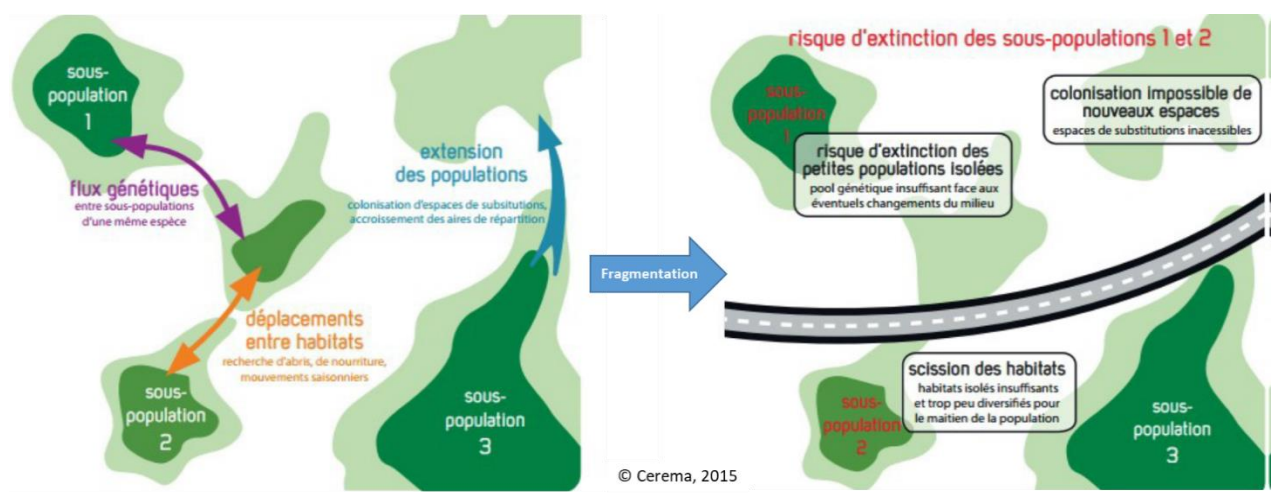

Afin d'évaluer le **degré d'unité (non-fragmentation)** du territoire, il est proposé inversement d'observer son **degré de fragmentation**. Pour cela, de nombreux indicateurs existent, dont un des plus simples qui consiste à étudier la **densité du linéaire des réseaux de transport routier ou ferroviaire**. Autrement dit, il s'agit de calculer le **rapport entre la longueur totale des linéaires de transport et la surface de l'échelle d'étude** (site/voisinage/paysage). En effet, la densité du réseau associée à la zone d'étude permet d'avoir une idée

de la surface du territoire sous l'influence des routes (Forman et al. 1995) (SETRA, 2000). Des études en 2000, montrent que la fragmentation routière moyenne de la forêt française était de 0,24 km/km² et que près de 230 000 ha de territoires boisés présentaient une densité routière de >1 km/km² (SETRA, 2000).

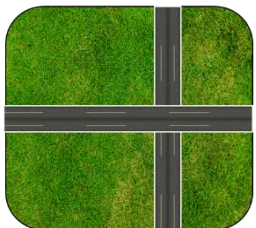
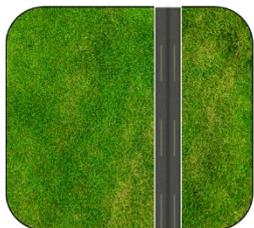

Etant donné que l'aspect étudié ici est l'unité du territoire, **le niveau de potentialités écologique « très fort » correspondra aux valeurs de densité minimales**. Autrement dit, les sites d'étude avec des potentialités écologiques les plus fortes seront ceux qui ont une fragmentation la plus réduite **(densité la plus proche de 0)** et donc une unité de territoire plus importante.

Afin d'étudier cette fragmentation entraînée par les réseaux de transport, la **BD TOPO 3.0® de l'IGN** (Institut National de l'Information Géographique et Forestière) est utilisée. Elle contient une description des éléments du paysage sous forme de vecteurs de précision métrique, classés selon des thématiques adaptées, et elle couvre de manière cohérente l'ensemble des entités géographiques et administratives du territoire national. Au sein de cette BD, les couches géographiques d'intérêt pour le linéaire de transport correspondent aux **« tronçon de route » et « tronçon de voie ferrée »**. En outre, **un tri doit être effectué, afin de ne considérer que les éléments les plus fragmentant** (les voies actives, larges, au sol et avec matériaux imperméables) par rapport aux éléments peu fragmentant (voies inactives, étroites, hors-sol et avec des matériaux perméables). Ce tri se fait sur la base des attributs présentés sur le tableau suivant.

|                      | Tronçon de route | Tronçon de voie ferrée |
|---|---|---|
| **Attributs retenus** |  | **Etat de l'objet** : En construction \| En service |
|  | **Nature** : Type autoroutier \| Bretelle \| Route à 2 chaussées \| Route à 1 chaussée | Nature : Funiculaire/crémaillère \| LGV \| Métro Tramway \| Voie de service \| Voie ferrée ppale |
|  | **Position par rapport au sol** : 0 (au sol) | |
| **Attributs exclus** |  | **Etat de l'objet** : En projet \| Non exploité |
|  | **Nature** : Route empierrée \| Chemin \| Escalier \| Bac / liaison maritime \| Rond-point \| Piste cyclable \| Sentier. | **Nature** : Sans objet (souvent non exploité) |
|  | **Position par rapport au sol** : \|-4 \| -3 \| -2 \| -1 \| (souterrain)  & \| 1 \| 2 \| 3 \| 4 \| (surélevé) | |

L'exemple suivant illustre une représentation cartographique possible pour cet indicateur sur l'unité du territoire (non-fragmentation) pour un site analysé à l'échelle du paysage (5000 m), ainsi que la logique du calcul de la densité de fragmentation par les infrastructures linéaires de transport.

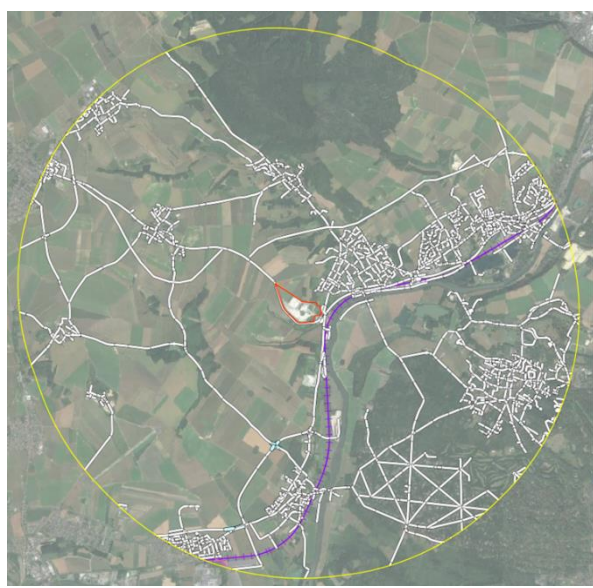

L'intérêt écologique de certains secteurs géographiques se traduit notamment par des mesures de protection, d'inventaire et de gestion. Ces espaces géographiques sont clairement définis afin d'assurer à long terme la conservation de la nature, ainsi que les services écosystémiques et les valeurs culturelles qui leur sont associés. Les zonages de biodiversité se trouvent donc au niveau des endroits présentant des milieux naturels fonctionnels et dynamiques, des habitats ou des populations d'espèces patrimoniaux. Plusieurs statuts existent selon des modalités réglementaires, des objectifs d'action (gestion ou non-gestion), d'intégration des activités socio-économiques, etc.

Les projets d'aménagement du territoire doivent faire attention à la présence de ces zonages, notamment dans les phases les plus en amont de la séquence Eviter-Réduire-Compenser (ERC) les impacts.

Dans le cadre de cette méthode, l'indicateur lié à la présence des zonages de biodiversité dans les trois échelles d'étude pour chaque site d'implantation est :

| Sous thématique | Indicateurs | Intérêt de la thématique : Mettre en évidence les sites… |
|---|---|---|
| Zonages : Patrimonialité | % surface de la zone d'étude couverte par au moins un zonage biodiversité à haute patrimonialité (score 3) | Recoupés par des zonages présentant des enjeux de conservation : zones d'inventaire de biodiversité, zones de gestion concertée, aires protégées. Ces zones ont une forte valeur écologique et patrimoniale (inhérente aux aspects réglementaires) et sont susceptibles d'accueillir une diversité des milieux naturels et d'espèces en bon état de conservation. |

Ces zonages peuvent être hiérarchisés :

a) Selon leurs enjeux réglementaires : les zones protégées au titre de la réglementation ressortiraient au-dessus des zonages d'inventaires qui ne posséderaient pas des contraintes réglementaires.
b) Selon leur patrimonialité et donc leur contribution au maintien d'espaces à fort intérêt écologique, indépendamment de leur statut réglementaire.

La notion de **patrimonialité** « *est une construction sociale, qui accorde de la valeur à ce qui est rare ou risque de ne plus exister. Elle ne doit pas occulter la valeur intrinsèque de la biodiversité. La valeur patrimoniale d'un site (si l'on s'en tient au seul patrimoine naturel) peut être estimée à partir des listes d'habitats et d'espèces répertoriés sur le site. La valeur patrimoniale de ce site sera alors appréciée en fonction de la valeur patrimoniale cumulée de chacune de ces composantes* » (Delzons, 2015).

Dans cette V0 de la méthode d'évaluation, **l'intérêt porte uniquement sur les aspects patrimoniaux** des zonages de biodiversité. Les aspects réglementaires sont volontairement mis à l'écart par soucis de cohérence globale de la méthode qui cherche à mettre en évidence des potentialités écologiques en raison d'une bonne structure, fonctionnalité et diversité des milieux naturels, des espèces et du paysage, et non pas à la présence de zones ou d'espèces avec des contraintes réglementaires. Il n'est donc pas pertinent de mélanger dans un système de notation des enjeux patrimoniaux avec des enjeux réglementaires. La possible intégration des enjeux réglementaires sera étudiée pour une future version de la méthode d'évaluation.

Les croisements des bases de données et les analyses via l'outil automatisé permettent de faire ressortir **20 types de zonages de biodiversité**. Ces zonages sont :

- Zones Naturelles d'Intérêt Ecologique, Faunistique et Floristique (ZNIEFF) :
    o ZNIEFF de type 1 : espaces homogènes écologiquement, définis par la présence d'espèces, d'associations d'espèces ou d'habitats rares, remarquables ou caractéristiques du patrimoine naturel régional. Ce sont les zones les plus remarquables du territoire.

- ZNIEFF de type 2 : espaces qui intègrent des ensembles naturels fonctionnels et paysagers, possédant une cohésion élevée et plus riches que les milieux alentours. Elles peuvent inclure des zones de type I.
- Sites Natura 2000
    - SIC : Site d'Importance Communautaire participant au réseau européen Natura 2000, et visant la conservation des types d'habitats et des espèces animales et végétales figurant aux annexes I et II de la Directive "Habitats"
    - ZPS : Zone de Protection Spéciale participant au réseau européen Natura 2000, et visant la conservation des espèces d'oiseaux sauvages figurant à l'annexe I de la Directive "Oiseaux" ou qui servent d'aires de reproduction, de mue, d'hivernage ou de zones de relais à des oiseaux migrateurs. (INPN, 2019)
- Espaces protégés (avec un statut juridique) :
    - APPB : les arrêtés préfectoraux de protection de biotope
    - ASPIM : les aires spécialement protégées d'importance méditerranéenne
    - BIOS : les réserves de biosphère
    - BPM : les biens inscrits sur la liste du patrimoine mondial de l'UNESCO
    - CDL : les sites du conservatoire de l'espace littoral et des rivages lacustres
    - CEN : les sites des conservatoires d'espaces naturels
    - OSPAR : Convention pour la protection du milieu marin de l'Atlantique du Nord-Est. Ici, cela concerne les 37 Aires Marines Protégées (AMP) en France
    - PN : les parcs nationaux
    - PNM : les parcs naturels marins
    - RAMSAR : les sites de la Convention Ramsar (Zones humides d'importance internationale)
    - RB : les réserves biologiques de l'Office National des Forêts (ONF) : Réserves biologiques dirigées (RBD) ou intégrales (RBI), selon le mode de gestion et d'activité sylvicole.
    - RIPN : les réserves intégrales de parc national
    - RNC : les réserves naturelles de Corse
    - RNCFS : les réserves de la chasse et de la faune sauvage
    - RNN : les réserves naturelles nationales
    - RNR : les réserves naturelles régionales

Cependant, certains de ces zonages sont redondants ou n'offrent pas une vrai plus-value écologique par rapport aux réservoirs de biodiversité étudiés dans les indicateurs relatifs aux réseaux écologiques. C'est pourquoi il a été choisi de s'intéresser uniquement aux zonages présentant la plus forte **patrimonialité**.

Pour cela, un **coefficient/score de patrimonialité** est attribué à chacun des grands types de zonage renseignés dans l'INPN, en fonction de la probabilité qu'un tel zonage abrite des enjeux de patrimonialité (habitats/espèces), et indépendamment du statut règlementaire des espaces. Le coefficient varie de :

- Score 1 : enjeu réduit de patrimonialité
- Score 2 : enjeu modéré de patrimonialité
- **Score 3 : enjeu élevé de patrimonialité**

**Il se base sur le dire d'expert et reflète la situation moyenne observée, à l'échelle nationale, et peut dans certains cas être sujet à caution**, en fonction de particularismes locaux. En effet, certaines exceptions existent pour des zonages considérés comme à haute patrimonialité (score 3) et peuvent présenter des enjeux écologiques plus réduits dans certains secteurs géographiques. Il permet cependant d'offrir une vision synthétique de ces enjeux, à l'échelle de la France. Il faut noter ici que le cumul n'est pas pris en compte (deux zonages empilés ne valent pas forcément plus qu'un), mais seulement l'aspect patrimonial.

Les coefficients proposés et le cumul de caractéristiques associées à chaque score sont résumés dans le tableau suivant :

| Type de zonage | Score | Caractéristiques du score |
|---|---|---|
| APPB | 3 | • Ensemble de la surface du zonage présentant une forte valeur patrimoniale<br>• Vocation principalement écologique avec peu d'intervention (mais pouvant être parfois couplée avec d'autres vocations agricoles, sylvicoles, etc.) : préservation d'espèces et de milieux, prévention de dégradations, protéger et assurer la gestion conservatoire d'habitats<br>• Catégorie I II et III de l'UICN qui visent en premier lieu à protéger l'intégrité écologique des écosystèmes et des processus naturels. Dans une moindre mesure des zonages de catégorie IV qui s'appliquent à des sites dans lesquels des interventions de gestion régulières sont nécessaires pour conserver et, le cas échéant, restaurer des espèces ou des habitats.<br>• Zonages de biodiversité définis à l'échelle régionale voire nationale et donc en cohérence avec les enjeux de patrimonialité locale, qui sont souvent le plus forts.<br>• Espaces souvent associées aux réservoirs de biodiversité (APPB, RNR, RNN, ZNIEFF1, etc.) dans les réseaux écologiques |
| PN (zone cœur du parc) | 3 | |
| RBD | 3 | |
| RBI | 3 | |
| RIPN | 3 | |
| RN Corse | 3 | |
| RNN | 3 | |
| RNR | 3 | |
| ZNIEFF 1 (Terre et Mer) | 3 | |
| Sites des CEN | 2 | • Seulement une partie de la surface des zonages présentant une forte valeur patrimoniale, diluée dans une matrice paysagère plus ou moins naturelle.<br>• Vocation principalement écologique avec une intervention ou gestion plus marquée : restaurer des fonctionnalités de milieux, mener une politique foncière, mettre en place des projets de démonstrations et d'expérimentation, contribuer au développement durable<br>• Catégorie IV de l'UICN (sites dans lesquels des interventions de gestion régulières sont nécessaires pour conserver et, le cas échéant, restaurer des espèces ou des habitats), catégorie V (zonages de paysages culturels habités) ou sans catégorie clairement identifiée<br>• Inclut des zonages associés à des directives européennes cherchant une cohérence de la préservation et la restauration d'une patrimonialité européenne, voire des conventions et traités internationaux sans un cadre d'application strict. |
| SIC | 2 | |
| Site du conservatoire du littoral | 2 | |
| ZNIEFF 2 (Terre et mer) | 2 | |
| ZPS | 2 | |
| RNCFS | 2 | |
| OSPAR | 2 | |
| Réserve de biosphère | 2 | |
| ASPIM | 1 | • Seulement une partie de la surface des zonages présentant une forte valeur patrimoniale, diluée dans une matrice paysagère plus ou moins naturelle.<br>• Vocation souvent associée à des dimensions socio-économiques et culturelles qui doivent cohabiter avec le volet écologique : protection de patrimoine culturel ; contribution à l'aménagement du territoire ; contribution au développement économique, social et culturel et qualité de vie ; accueil, éducation et information du public<br>• Catégorie V de l'UICN (zonages de paysages culturels habités), catégorie VI (aires d'utilisation durable des ressources naturelles, essentiellement au profit des populations locales) ou sans catégorie clairement identifiée<br>• Inclus des zonages souvent issus de conventions et traités internationaux sans un cadre d'application strict |
| PN (zone d'adhésion) | 1 | |
| PNM | 1 | |
| PNR | 1 | |
| Sites RAMSAR | 1 | |

L'indicateur choisi dans cette thématique ne tiendra compte uniquement que de la **proportion totale de la surface de la zone d'étude recouverte par des zonages avec un score de 3**. Par conséquent 9 types de zonages seront comptabilisés dans le calcul de cet indicateur, à savoir : APPB, PN (cœur), RIPN, RNC, RNN, RNR, RBI, RBD, et ZNIEFF 1. Pour cela, le principe du calcul se base sur le rapport entre la surface (en hectare) de la zone d'étude recouverte par des zonages patrimoniaux (score 3) et la surface totale de la zone d'étude (en hectare). L'exemple suivant illustre la logique du calcul de l'indicateur dans un cas fictif.

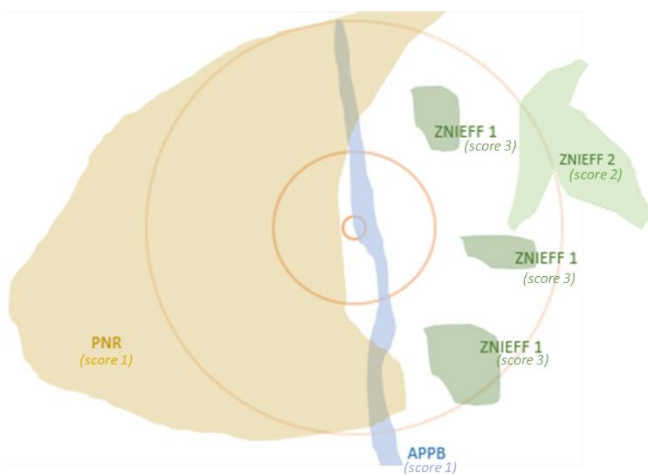

| | % surface cumulée en | | |
|---|---|---|---|
| | Enjeu élevé (3) | Enjeu modéré (2) | Enjeu réduit (1) |
| Site | 50 % | 0 | 0 |
| Voisinage | 10 % | 0 | 35 % |
| Paysage | 20 % | 5 % | 40 % |

| Indicateurs | Valeur |
|---|---|
| Proportion de zonages patrimoniaux dans la zone d'étude | Site : 50 %<br>Voisinage : 10 %<br>Paysage : 20 % |

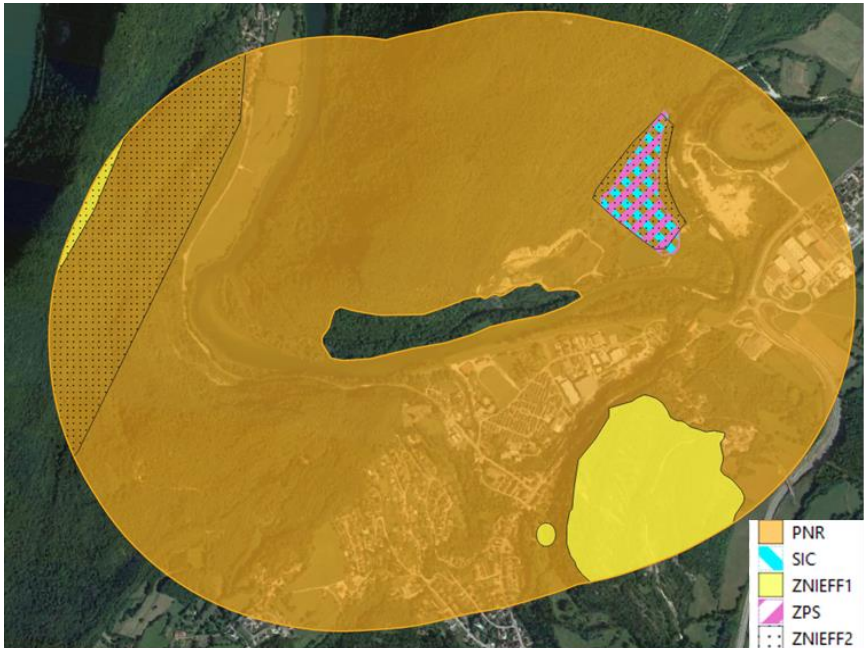

L'exemple suivant illustre les différentes représentations cartographiques possibles pour un site étudié à l'échelle du voisinage (1000 m) : une visualisation des types des zonages intersectés (haut), des catégories des zonages (milieu) et des scores de patrimonialité (bas). L'encadré précise le résultat de ce site (échelle du voisinage) pour l'indicateur évalué dans cette thématique.

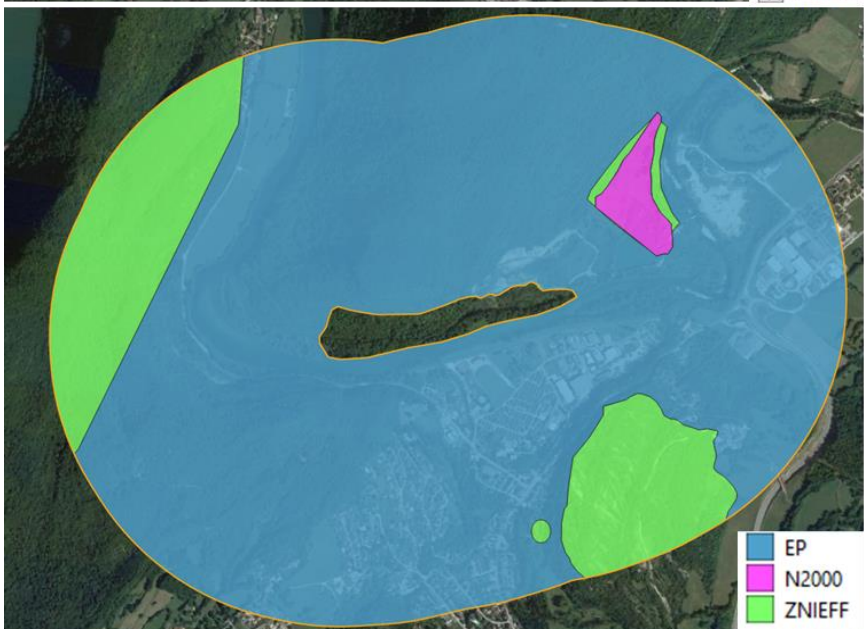

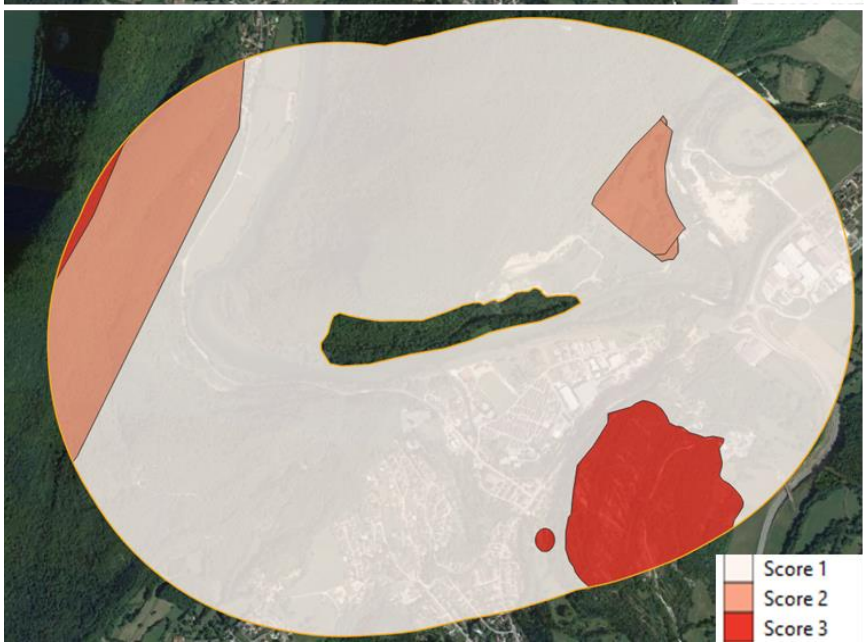

**Indicateur Zonages :**
6 % de la zone d'étude couverte par zonages patrimoniaux (score 3)

Evaluation cartographique du niveau de potentialités écologiques (CARPO)

# Annexe 4 :
# Spécificités sur les indicateurs relatifs aux espèces à enjeux (à la maille de l'INPN)

L'intérêt écologique de certains secteurs géographiques peut être mis en évidence par leur capacité à offrir des habitats et des interactions favorables à l'accueil d'espèces typiques du territoire. Ces espèces peuvent être :

- **Communes** : Espèce dont les populations sont largement distribuées et/ou dont les individus sont souvent observés par l'Homme, par opposition à espèce rare (INPN, 2019). On parle souvent de nature ordinaire.
- **Patrimoniales** (ou remarquables) : Notion subjective qui attribue une valeur d'existence forte aux espèces qui sont plus rares que les autres et qui sont bien connues (INPN, 2019). La patrimonialité ou valeur patrimoniale d'une espèce peut être définie selon sa rareté, le degré de menace pesant sur elle ou selon l'importance relative d'un site (avec la plus grosse colonie pour une espèce par exemple). Les espèces patrimoniales peuvent être protégées (ou pas). Par exemple, les espèces déterminantes de ZNIEFF sont considérées comme patrimoniales, mais ce seul statut n'a pas de valeur juridique (Delzons, 2015).

Dans le cadre de cette méthode, l'indicateur lié à la présence des espèces dans le secteur d'étude correspond à la **richesse d'espèces à enjeux de conservation dans la maille**, comme indiqué dans le tableau suivant :

| Sous thématique | Indicateurs | Intérêt de la thématique : Mettre en évidence les sites... |
|---|---|---|
| Patrimonialité | Nombre d'espèces à enjeux de conservation dans la maille de l'INPN | Dans des mailles présentant un nombre important d'espèces menacées (LR nationales et européennes), à enjeux régionaux (déterminantes de ZNIEFF), à enjeux européens (espèces de la DHFF+DO), endémiques et de la SCAP (années précédentes). La présence d'espèces qui justifient le classement d'aires protégés (SCAP : Stratégie de création d'aires protégées) et de forte valeur pour la biodiversité, avec une méthode nationale (Léonard *et al.*, 2019), est pertinente pour déterminer des implantations qui se trouvent dans des contextes écologiques patrimoniaux. |

De la même manière que pour l'indicateur sur les zonages patrimoniaux, **l'intérêt porte uniquement sur les aspects patrimoniaux** des espèces. Les aspects réglementaires sont volontairement mis à l'écart par souci de cohérence globale de la méthode qui cherche à mettre en évidence des potentialités écologiques en raison d'une bonne structure, fonctionnalité et diversité, et non pas à la présence de zones ou d'espèces avec des contraintes réglementaires. Il n'est donc pas pertinent de mélanger dans un système de notation des enjeux patrimoniaux avec des enjeux réglementaires. La possible intégration des enjeux réglementaires sera étudiée pour une future version de la méthode d'évaluation.

**Sont considérées comme des espèces patrimoniales et donc à enjeux de conservation** dans cette étude, les espèces relevant des **5 critères de patrimonialité** suivants :

- **Déterminantes de ZNIEFF** dans la région concernée par la maille, et remplissant les conditions d'éligibilité. Ce sont des espèces qui peuvent justifier l'établissement d'une ZNIEFF dans une région (en fonction de leur nombre et coefficients associés).
  *« Dans le cadre des ZNIEFF, sont qualifiées de déterminantes :*
  *1) les espèces en danger, vulnérables, rares ou remarquables répondant aux cotations mises en place par l'UICN ou extraites des livres rouges publiés nationalement ou régionalement ;*
  *2) les espèces protégées nationalement, régionalement, ou faisant l'objet de réglementations européennes ou internationales lorsqu'elles présentent un intérêt patrimonial réel au regard du contexte national ou régional ;*

*3) les espèces ne bénéficiant pas d'un statut de protection ou n'étant pas inscrites dans des listes rouges, mais se trouvant dans des conditions écologiques ou biogéographiques particulières, en limite d'aire ou dont la population est particulièrement exceptionnelle (effectifs remarquables, limite d'aire, endémismes...) »* (INPN, 2019).

- Des espèces **d'intérêt communautaire** figurant en annexe II et IV de la Directive Habitats Faune Flore (DHFF) ou en annexe 1 de la Directive Oiseaux (DO).

    *« Une espèce d'intérêt communautaire est une espèce en danger ou vulnérable ou rare ou endémique (c'est-à-dire propres à un territoire bien délimité ou à un habitat spécifique) énumérée… pour lesquelles doivent être désignées des Zones Spéciales de Conservation… et pour lesquelles des mesures de protection doivent être mises en place sur l'ensemble du territoire »* (INPN, 2019).

- Des espèces **menacées** d'extinction et donc figurant sur les **listes rouges** mondiale, européenne, et/ou nationales (validées par l'UICN). Sont considérées uniquement les espèces des catégories CR « en danger critique d'extinction », EN « en danger d'extinction », VU « vulnérables », et dans certaines occasions NT « quasi-menacées », en fonction des cas et de la liste rouge considérée (cf. clé décisionnelle ci-dessous).

- Des espèces **endémiques** : Une espèce endémique est un « *taxon naturellement restreint à la zone géographique considérée. Cette notion est donc dépendante de la zone considérée : endémique d'un continent, endémique d'un pays, endémique d'une zone biogéographique, endémique d'une île… Dans l'INPN, l'endémisme est considéré par rapport aux limites administratives de la France (métropole, DOM, COM)* » (INPN, 2019). C'est en raison de cette localisation géographique restreinte et son absence dans d'autres pays du monde que l'espèce présente des enjeux de conservation.

- Des espèces de la **SCAP 2015** : La Stratégie Nationale de Création d'Aires Protégées (SCAP) « *est une stratégie nationale visant à améliorer la cohérence, la représentativité et l'efficacité du réseau métropolitain des aires protégées terrestres en contribuant au maintien de la biodiversité, au bon fonctionnement des écosystèmes et à l'amélioration de la trame écologique… avec l'objectif de placer au minimum 2 % du territoire terrestre métropolitain sous protection forte d'ici l'horizon 2019… des projets de création ou d'extension d'aires protégées, désignés pour la présence d'espèce ou d'habitats de la liste nationale SCAP et concourant à l'atteinte de l'objectif des 2 %* » (INPN, 2019). Un diagnostic des espèces pour la SCAP avait été réalisé en 2009 puis en 2015. Les espèces retenues lors de ces campagnes précédentes sont également prises en compte (sauf exceptions).

Les statuts d'espèces sont issus de la base de connaissances « statuts » (Gargominy et Régnier, 2018) de l'INPN en lien avec le référentiel taxonomique TAXREF. Ces éléments sont consultables sur http://inpn.mnhn.fr/isb/accueil/index.

Sont exclues de l'analyse les espèces éteintes, introduites, accidentelles ou marines, ainsi que les espèces non nicheuses pour les oiseaux (sélection des oiseaux nicheurs faite selon l'Atlas des oiseaux nicheurs de France métropolitaine de 2015 (Issa et Muller, 2015) ou à dire d'expert) (Léonard *et al.*, 2019).

**Le critère de protection nationale ou régionale des espèces n'est pas pris en compte directement** (bien qu'il puisse des fois se retrouver dans les espèces déterminantes de ZNIEFF ou des directives européennes). En effet, les espèces de certains groupes sont très largement protégés nationalement, bien que leur rareté et le degré de menace les concernant soient très variables d'une espèce à l'autre (Delzons, 2015). Si cette V0 se base uniquement sur les enjeux patrimoniaux, la possible intégration des enjeux réglementaires sera étudiée pour une éventuelle V1 de la méthode d'évaluation (comme pour les zonages).

Le choix des espèces retenues comme des espèces à enjeux de conservation se base sur la **clé décisionnelle** suivante (Léonard *et al.*, 2019) qui détermine la combinaison des 5 critères de patrimonialité nécessaires pour qu'un taxon soit éligible. Cette clé varie en fonction du groupe taxonomique étudié (vertébrés, invertébrés, flore et fonge), comme indiqué dans la figure suivante.

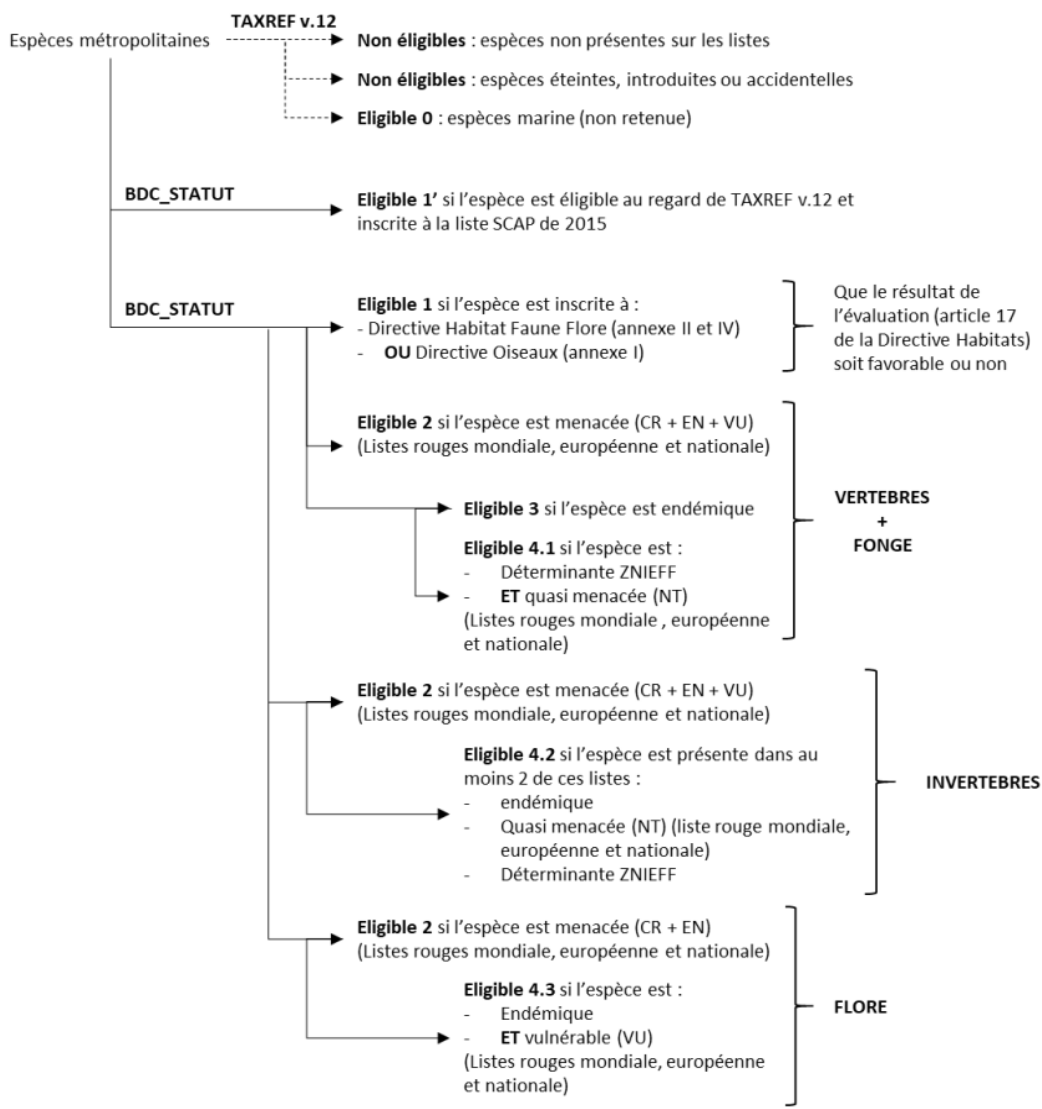

L'application de la clé décisionnelle permet d'aboutir à une liste **de 2 218 taxons à enjeux de conservation en France** (comprenant des espèces et des sous espèces) dans le cadre de la **SCAP 2020**. La liste a été mise à jour avec la version 13 du référentiel taxonomique TAXREF, et les taxons sont traités au rang de l'espèce ou de la sous-espèce le cas échéant (Léonard *et al.*, 2019). Le nombre d'espèces à enjeux (richesse) varie d'une maille à l'autre et a permis de calculer le nombre total d'espèces par maille, d'étudier leur distribution de valeurs à niveau national et de dresser la cartographie présentée à continuation qui sera intégrée dans cette V0 de la méthode d'évaluation

Ces résultats obtenus sur la **richesse d'une maille en espèces à enjeux de conservation (nombre d'espèces de la SCAP20 dans la maille)** doivent être relativisés via les deux descripteurs mentionnées dans le tableau ci-dessous. En effet, l'existence d'une maille présentant un nombre réduit d'espèces à enjeu de conservation peut être liée au fait que cette maille présente un nombre réduit de données d'espèces qui remontent à l'INPN.

| Sous thématique | Descripteur | Intérêt |
|---|---|---|
| Irremplaçabilité | Valeur de l'ICBG (Indice de Contribution à la Biodiversité Globalisée) de la maille (%) | L'irremplaçabilité d'une maille reflète son caractère unique et sa responsabilité régionale pour la conservation des espèces. Une maille est complètement irremplaçable si elle contient une ou plusieurs espèces qui n'existent nulle part ailleurs. La perte de cette maille entraînerait une perte importante pour cette espèce. L'Indice de Contribution à la Biodiversité Globalisée (ICBG) (Witté et Touroult, 2014) est un score sur 100 qui définit les points chauds de biodiversité, "irremplaçables" du fait de l'assemblage d'espèces qu'ils abritent. Les mailles avec un score de 100 contiennent un grand nombre de taxons ou des taxons rares ou endémiques (ou les deux), alors que les mailles à faible score contiennent des taxons plus répandus ou une richesse moins importante (Léonard et al., 2020) |
| Méconnaissance | Valeur du taux méconnaissance (%) pour les taxons classiques dans la maille | Les résultats obtenus pour des indicateurs sur le nombre d'espèces patrimoniales dans une maille doivent être relativisés selon le niveau de connaissance de la biodiversité sur ladite maille. Le taux de méconnaissance est un indice établi en cumulant sur 27 groupes taxonomiques classiques le nombre de groupes méconnus par maille. L'indice correspond au pourcentage de groupes méconnus par rapport au total des 27 groupes étudiés (% élevés dans mailles méconnues). Il met en évidence les zones déficitaires en données naturalistes disponibles au niveau national (Witté & Touroult, 2017). |

Plusieurs différences sont à noter dans le traitement de cet indicateur, par rapport aux indicateurs des thématiques liées à l'occupation du sol, aux réseaux écologiques et aux zonages de biodiversité :

- L'indicateur se base sur des **données tabulaires** correspondant aux listes d'espèces trouvées au sein d'une maille et remontées à l'**INPN**. Ce ne sont donc pas les espèces retrouvées au sein d'un périmètre d'un site d'implantation, mais un répertoire d'espèces potentielles observées par des acteurs de l'environnementaux (et sciences participatives) dans le territoire de la maille.

- Il n'est pas calculé aux trois échelles d'étude habituelles (site – voisinage 1 km – paysage 5 km), mais uniquement à une échelle plus large : celle de la **maille de l'INPN**. La maille est une unité géométrique normalisée utilisée comme unité d'échantillonnage dans un inventaire et/ou comme mode de restitution synthétique de données de distribution. À l'échelle française, le maillage standard est la maille de 10 km de côté Lambert 93. Cela constituerait donc à une division imaginaire du territoire français en au moins **5868 mailles carrées de 10 km x 10 km**.

- En raison du calcul de l'indicateur uniquement à l'échelle de la maille, le choix des 3 seuils pour les 3 classes de niveau de potentialité ne se fait pas de façon à avoir 1 seuil par chaque échelle (pour les autres thématiques), mais 3 seuils à la même échelle de la maille. Autrement dit, pour une maille sont définis 3 fourchettes de valeurs relatives à la richesse d'espèces à enjeux dans la maille, à partir desquelles 3 seuils (nombre d'espèces) permettront de déterminer si les potentialités écologiques par rapport aux espèces à enjeux sont moyennes, fortes ou très fortes.

- Cet indicateur croise la localisation de la zone d'étude avec les mailles de 10 km de l'INPN afin d'associer un site à une maille et d'afficher la valeur de la maille relative au nombre d'espèces à enjeux de conservation et une valeur de méconnaissance naturaliste. L'indicateur sera calculé **uniquement via le croisement des périmètres des sites avec les mailles INPN (pas de croisement des échelles « voisinage » et « paysage »)**. Ceci permettra de réduire le risque des sites intersectant plusieurs mailles.

- Cependant, certains sites se trouvent à cheval entre plusieurs mailles, ce qui pose la question de quelle valeur (quelle maille) afficher pour ces sites. En attendant la définition de règles de croisement au sein de PatriNat, la solution temporaire consiste à afficher la valeur de la maille avec la valeur de richesse la plus élevée et de signaler ces sites, pour tenir compte de ces particularités dans l'interprétation. La couche SIG produite lors des calculs automatisés permet de mettre en évidence ces cas en affichant les limites des mailles à l'intérieur des périmètres des sites.

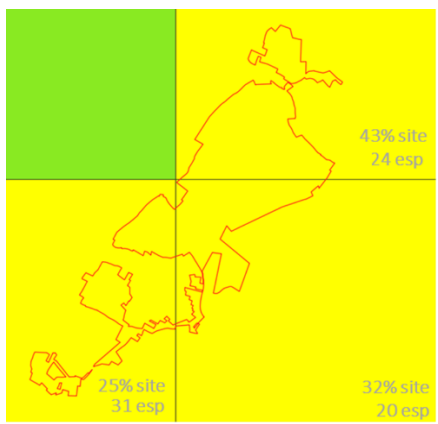
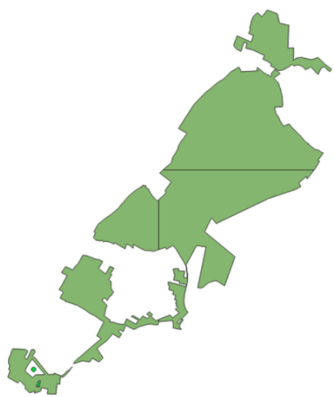

- **Les seuils (et classes de niveau) ne doivent pas être calculés pour le groupe de sites évalués, mais ce sont des seuils nationaux applicables à toute groupe de sites**. Ainsi, ils ont été calculés sur la base des données de l'ensemble des mailles françaises (et non pas seulement les mailles intersectées par un groupe en particulier). En effet, ces valeurs de richesse proviennent d'un programme national de création d'aires protégées nourrit par des données de tout le territoire métropolitain, ce qui permet de constituer des seuils nationaux (et donc des classes de niveaux), applicables à tout groupe de sites évalué. Il suffit donc d'appliquer ces seuils à tout groupe de sites à analyser. Les fourchettes des valeurs nationales de chaque classe de niveau de potentialité, et la répartition géographique des mailles au sein de ces classes sont présentées ci-dessous.

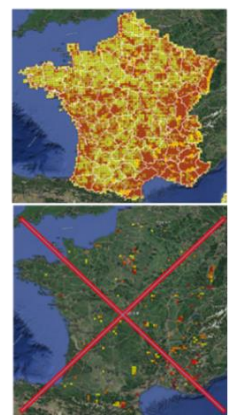
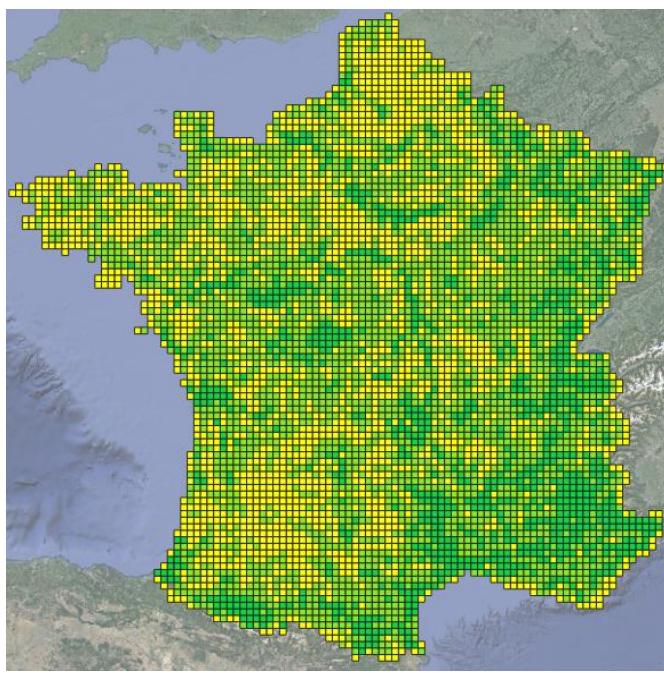
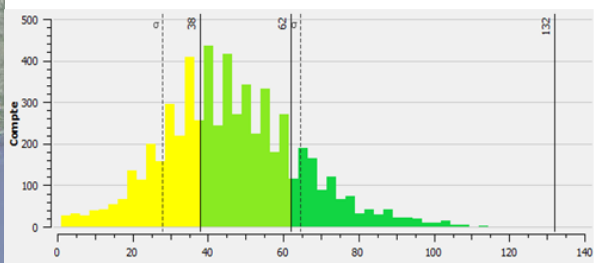

| Site : Richesse (nb sp) | Niveau pot | Mailles |
|---|---|---|
| 0 - 38 | Moyen | 2070 |
| 39 – 62 | Fort | 2780 |
| 63 – 132 | Très fort | 1018 |

Le maillage 10km x 10km est disponible pour téléchargement sur :
https://inpn.mnhn.fr/telechargement/cartes-et-information-geographique/ref/referentiels

## Annexe 5 :
## Informations mobilisables et indicateurs dont le calcul est automatisé : nomenclature

Le tableau de nomenclature suivant présente les différents fichiers de sortie (couche SIG et tableurs) produits via le croisement des BDD et les calculs automatisés. Il explique le contenu de chaque fichier de sortie, ainsi que celui de chaque champ des tables attributaires et des tableurs.

Les champs notés en bleu correspondent aux champs propres au module CARPO (par rapport aux champs en noir, intrinsèques à l'outil automatisé d'origine).

De plus, les cases notées en jaune indiquent les champs contenant les valeurs à utiliser pour renseigner les indicateurs CARPO, sur le SIG synthèse (Annexe 7) et sur la grille d'évaluation (Annexe 8).

### Thématique : Occupation du sol

| Nom du fichier | Description du Fichier |
|---|---|
| OS01_*X*_INTERSECT_*Y* | SIG (Fichiers Shape) : Données brutes de la couche X (où X = CLC18, OSO17 ou OSO18), présentant les différents modes d'occupation du sol (MOS) aux différentes échelles d'analyses Y (Site, 1000 m ou 5000 m)<br>3 shapes en sortie pour chacune des 3 couches X utilisées : OS01_X_INTERSECT_Site, OS01_X_INTERSECT_1000, OS01_X_INTERSECT_5000<br>Table récapitulative des tables attributaires des 3 shapes |
| **CHAMPS** | **DESCRIPTION** |
| PERIMET | Echelle / Périmètre d'analyse |
| SURFPER | Superficie du polygone de zone d'étude considérée (Site, tampon 1 km, tampon 5km) en hectare |
| ID_SITE | Identifiant du site d'étude |
| NOM_SIT | Nom du site d'étude |
| CD_OS | Code Corine/OSO de la classe d'occupation des sols |
| LB_OS | Libellé associé au code Corine/OSO indiqué dans le champ CD_OS |
| SURFINT_H | Surface intersectée en ha |
| SURFINT_M | Surface intersectée en m² |
| PERMEAB | Classe de perméabilité associée au mode d'occupation du sol : 3 attributs possibles : « perméable », « imperméable » et « mixte » |
| CARACT_NAT | Classe de caractère naturel associée au mode d'occupation du sol : 3 attributs possibles : « naturel », « semi-naturel » et « artificiel » |
| OS02_PROP_OS_*X* | Tableur : Surfaces de chacun des MOS aux différentes échelles d'analyses, en fonction de la couche X (CLC18, OSO17, OSO18) d'entrée. Chaque échelle d'analyse est traitée dans une feuille Excel distincte.<br>Dans le cas de CLC18 : Il y a 3 tableurs en sortie, un pour chaque niveau CLC18 différent (1 à 3). |
| **CHAMPS** | **DESCRIPTION** |
| PERIMETRE | Echelle / Périmètre d'analyse |
| SURFPERI_HA | Superficie du polygone de zone d'étude considérée (Site, tampon 1 km, tampon 5km) en hectare |
| ID_SITE | Identifiant unique du site d'étude |
| NOM_SITE | Nom du site d'étude |
| CD_OS | Code Corine/OSO de la classe d'occupation des sols |
| LB_OS | Libellé associé au code Corine/OSO indiqué dans le champ CD_OS |
| SURFINT_HA | Surface cumulée du MOS concernée (Niveau X nomenclature CLC) en ha |
| PROP_AREA | Proportion du périmètre d'analyse concerner par le MOS. |

| OS03_CARACT_NAT_X | Tableur : Surfaces et proportions de chaque classe de caractère naturel (naturel, seminaturel ou artificiel), ainsi que le nombre de MOS naturels et seminaturels, aux différentes échelles d'analyses, en fonction de la couche X (CLC18, OSO17, OSO18) d'entrée. Chaque échelle d'analyse est traitée dans une feuille Excel distincte. | |
|---|---|---|
| CHAMPS | DESCRIPTION | |
| PERIMETRE | Echelle / Périmètre d'analyse | |
| SURFPERI_HA | Superficie du polygone de zone d'étude considérée (Site, tampon 1 km, tampon 5km) en hectare | |
| ID_SITE | Identifiant unique du site d'étude | |
| NOM_SITE | Nom du site d'étude | |
| SURF_NATUR | Surface totale de l'ensemble de MOS associés à des milieux « naturels » (Niveau 3 CLC18 ou OSO) dans l'échelle d'étude, en ha | |
| SURF_SEMINAT | Surface totale de l'ensemble de MOS associés à des milieux « semi-naturels » (Niveau 3 CLC18 ou OSO) dans l'échelle d'étude, en ha | |
| SURF_ARTIF | Surface totale de l'ensemble de MOS associés à des milieux « artificiels » (Niveau 3 CLC18 ou OSO) dans l'échelle d'étude, en ha | |
| PROP_NATUR | Proportion (% de la surface totale) de la zone d'étude concernée par des MOS associés à des milieux « naturels » (Niveau 3 CLC18 ou OSO) | La somme des 2 valeurs |
| PROP_SEMINAT | Proportion (% de la surface totale) de la zone d'étude concernée par des MOS des milieux « semi-naturels » (Niveau 3 CLC18 ou OSO) | = Indicateur Caract. Natur |
| PROP_ARTIF | Proportion (% de la surface totale) de la zone d'étude concernée par des MOS associés à des milieux « artificiels » (Niveau 3 CLC18 ou OSO) | |
| NB_NATUREL | Nombre des différents MOS associés à des milieux « naturels » (Niveau 3 CLC18 ou OSO) dans la zone d'étude | Indicateur Hétérogénéité |
| NB_SEMINAT | Nombre des différents MOS associés à des milieux « semi-naturels » (Niveau 3 CLC18 ou OSO) dans la zone d'étude | |
| OS03_PERMEABILITE_CLC18 | Tableur : Surfaces et proportions de chaque classe de perméabilité (perméable, mixte, imperméable), aux différentes échelles d'analyses, en fonction de la couche X (CLC18, OSO17, OSO18) d'entrée. Chaque échelle d'analyse est traitée dans une feuille Excel distincte. | |
| CHAMPS | DESCRIPTION | |
| PERIMETRE | Echelle / Périmètre d'analyse | |
| SURFPERI_HA | Superficie du polygone de zone d'étude considérée (Site, tampon 1 km, tampon 5km) en hectare | |
| ID_SITE | Identifiant unique du site d'étude | |
| NOM_SITE | Nom du site d'étude | |
| SURF_PERM | Surface totale de l'ensemble de MOS associés à des milieux « perméables » (Niveau 3 CLC18 ou OSO) dans l'échelle d'étude, en ha | |
| SURF_MIXTE | Surface totale de l'ensemble de MOS associés à des milieux « mixtes » (Niveau 3 CLC18 ou OSO) dans l'échelle d'étude, en ha | |
| SURF_IMPER | Surface totale de l'ensemble de MOS associés à des milieux « imperméables » (Niveau 3 CLC18 ou OSO) dans l'échelle d'étude, en ha | |
| PROP_PERM | Proportion (% de la surface totale) de la zone d'étude concernée par des MOS associés à des milieux « perméables » (Niveau 3 CLC18 ou OSO) | Indicateur Perméabilité |
| PROP_MIXTE | Proportion (% de la surface totale) de la zone d'étude concernée par des MOS associés à des milieux « mixtes » (Niveau 3 CLC18 ou OSO) | |
| PROP_IMPER | Proportion (% de la surface totale) de la zone d'étude concernée par des MOS des milieux « imperméables » (Niveau 3 CLC18 ou OSO) | |

## Thématique : Réseaux écologiques : Réservoirs

| RES01_INTERSECT_Y | SIG (Fishier Shape) : Données brutes de croisement entre les sites et la couche N_SRCE_RESERVOIR_S_000 (décembre 2017) aux différentes échelles d'analyses Y (Site, 1000 m ou 5000 m).<br>3 shapes en sortie : RES01_INTERSECT_Site, RES01_INTERSECT_1000, RES01_INTERSECT_5000. |
|---|---|
| CHAMPS | DESCRIPTION |
| PERIMET | Echelle / Périmètre d'analyse |
| SURFPER | Superficie du polygone de zone d'étude considérée (Site, tampon 1 km, tampon 5km) en hectare |
| ID_SITE | Identifiant du site d'étude |
| NOM_SIT | Nom du site d'étude |
| ID_RESV | Identifiant unique du réservoir de biodiversité |
| NOM_RES | Nom du réservoir de biodiversité |
| MILMAJ_ | Sous-trame principale à laquelle le réservoir est rattaché, selon la nomenclature nationale |
| SURFINT_H | Surface intersectée en ha |
| SURFINT_M | Surface intersectée en m² |
| **RES01_LISTE_RESERVOIRS_INTERSECT** | *Tableur : Listes des réservoirs recensés à chaque échelle d'analyse. Chaque échelle d'analyse est traitée dans une feuille Excel distincte* |
| CHAMPS | DESCRIPTION |
| PERIMETRE | Echelle / Périmètre d'analyse |
| SURFPERI_HA | Superficie du polygone de zone d'étude considérée (Site, tampon 1 km, tampon 5km) en hectare |
| ID_SITE | Identifiant unique du site d'étude |
| NOM_SITE | Nom du site d'étude |
| ID_RESV | Identifiant unique du réservoir de biodiversité |
| NOM_RESV | Nom du réservoir de biodiversité |
| MILMAJ_NAT | Sous-trame principale à laquelle le réservoir est rattaché, selon la nomenclature nationale |
| SURFINT_HA | Surface intersectée en ha |
| SURFINT_M | Surface intersectée en m² |
| **RES02_PROP_RESERVOIRS** | *Tableur : Proportions de réservoirs recensés. Chaque échelle d'analyse est traitée dans une feuille Excel distincte* |
| CHAMPS | DESCRIPTION |
| PERIMETRE | Echelle / Périmètre d'analyse |
| SURFPERI_HA | Superficie du polygone de zone d'étude considérée (Site, tampon 1 km, tampon 5km) en hectare |
| ID_SITE | Identifiant unique du site d'étude |
| NOM_SITE | Nom du site d'étude |
| SURFINT_HA | Surface du périmètre d'étude considéré dans le SRCE en réservoir de biodiversité (en ha) |
| PROP_AREA | Proportion du périmètre d'étude considéré dans le SRCE en réservoir de biodiversité | Indicateur Réservoirs |
| **RES02_PROP_RESERVOIRS_MILIEU** | *Tableur : Proportions de réservoirs recensés par sous-trame. Chaque échelle d'analyse est traitée dans une feuille Excel distincte* |
| CHAMPS | DESCRIPTION |
| PERIMETRE | Echelle / Périmètre d'analyse |
| SURFPERI_HA | Superficie du polygone de zone d'étude considérée (Site, tampon 1 km, tampon 5km) en hectare |
| ID_SITE | Identifiant unique du site d'étude |
| NOM_SITE | Nom du site d'étude |
| MILMAJ_NAT | Sous-trame principale à laquelle le réservoir est rattaché, selon la nomenclature nationale |
| SURFINT_HA | Surface du périmètre d'étude considéré dans le SRCE en réservoir de biodiversité (en ha) pour la sous-trame concernée |
| PROP_AREA | Proportion du périmètre d'étude considéré dans le SRCE en réservoir de biodiversité pour la sous-trame concernée |

## Thématique : Réseaux écologiques : Corridors

| | |
|---|---|
| **COR01_SRCE_*X*_INTERSECT_*Y*** | *SIG (Fichiers Shape) : Données brutes de croisement des corridors de type X (où X = L pour les corridors linéaires ; S pour les surfaciques) aux différentes échelles d'analyses Y (Y = Site, 1000 m ou 5000 m).*<br>*6 shape en sortie : COR01_SRCE_L_INTERSECT_Site, COR01_SRCE_L_INTERSECT_1000, COR01_SRCE_L_INTERSECT_5000; ainsi que COR01_SRCE_S_INTERSECT_Site, COR01_SRCE_S_INTERSECT_1000, CCOR01_SRCE_S_INTERSECT_5000.* |
| CHAMPS | DESCRIPTION |
| PERIMET | Echelle / Périmètre d'analyse |
| SURFPER | Superficie du polygone de zone d'étude considérée (Site, tampon 1 km, tampon 5km) en hectare |
| ID_SITE | Identifiant unique du site d'étude |
| NOM_SIT | Nom du site d'étude |
| MILMAJ_ | Sous-trame principale à laquelle le corridor est rattaché, selon la nomenclature nationale |
| OBJ_ASS | Objectif de préservation ou restauration attribué à l'élément |
| ID_CORR | Identifiant unique du corridor écologique |
| NOM_COR | Nom du corridor écologique (le cas échéant) |
| LENGTH_ | Seulement dans COR01_SRCE_L : Distance linéaire de corridor intercepté par le périmètre d'analyse (m) |
| SURFINT_H | Seulement dans COR01_SRCE_S : Surface d'intersection entre le site et le périmètre d'étude (ha) |
| SURFINT_M | Seulement dans COR01_SRCE_S : Surface d'intersection entre le site et le périmètre d'étude (m²) |
| **COR01_SRCE_X_LISTE_TRAMES_INTERSECT** | *Tableurs : Liste des corridors de type X (où X = L pour les corridors linéaires ; S pour les surfaciques) présents aux différentes échelles d'analyses. Chaque échelle d'analyse est traitée dans une feuille Excel distincte.*<br>*Deux tableurs : COR01_SRCE_L_LISTE_TRAMES_INTERSECT et COR01_SRCE_S_LISTE_TRAMES_INTERSECT* |
| **CHAMPS** | **DESCRIPTION** |
| PERIMETRE | Echelle / Périmètre d'analyse |
| SURFPERI_HA | Superficie du polygone de zone d'étude considérée (Site, tampon 1 km, tampon 5km) en hectare |
| ID_SITE | Identifiant unique du site d'étude |
| NOM_SITE | Nom du site d'étude |
| MILMAJ_NAT | Sous-trame principale à laquelle le corridor est rattaché, selon la nomenclature nationale |
| OBJ_ASSI | Objectif de préservation ou restauration attribué à l'élément |
| ID_CORR | Identifiant unique du corridor écologique |
| NOM_CORR | Nom du corridor écologique (le cas échéant) |
| **COR02_NOMBRE_TRAMES_INTERSECT** | *Tableur : Nombre de corridors de chaque type de sous-trame (milieu majoritaire) présents aux différentes échelles d'analyses. Chaque échelle d'analyse est traitée dans une feuille Excel distincte.* |
| **CHAMPS** | **DESCRIPTION** |
| ID_SITE | Identifiant unique du site d'étude |
| NOM_SITE | Nom du site d'étude |
| boisé | Nombre de corridors boisés dans le périmètre d'analyse |
| humide | Nombre de corridors humides dans le périmètre d'analyse |
| littoral | Nombre de corridors littoraux dans le périmètre d'analyse |
| multitrame | Nombre de corridors multitrame dans le périmètre d'analyse |
| non classé | Nombre de corridors non classé dans le périmètre d'analyse |
| ouvert | Nombre de corridors ouvert dans le périmètre d'analyse |
| autre | Nombre de corridors autres dans le périmètre d'analyse |
| NB_TRAME | Nombre total de corridors dans le périmètre d'analyse — *Indicateur Corridors* |
| **COR03_PROPORTION_OBJECTIFS_Y** | *Tableurs : Proportion du périmètre d'analyse Y (Site, 1000 m ou 5000 m) concerné par chacun des types de trames (selon l'objectif et le milieu majoritaire associés). Deux tableurs : Les corridors surfaciques (S) et linéaires (L) sont traités dans deux feuilles Excel distinctes.* |
| **CHAMPS** | **DESCRIPTION** |
| ID_SITE | Identifiant unique du site d'étude |
| NOM_SITE | Nom du site d'étude |
| TYPE | Type du corridor (surfacique) |
| 01.multitrame | Proportion du périmètre d'analyse concerné par un corridor multitrame défini comme étant « à préserver » |
| 02.multitrame | Proportion du périmètre d'analyse concerné par un corridor multitrame défini comme étant « à restaurer » |
| 03.multitrame | Proportion du périmètre d'analyse concerné par un corridor multitrame défini comme étant « à préciser » |
| 01.humide | Proportion du périmètre d'analyse concerné par un corridor humide défini comme étant « à préserver » |
| 02.humide | Proportion du périmètre d'analyse concerné par un corridor humide défini comme étant « à restaurer » |
| 03.humide | Proportion du périmètre d'analyse concerné par un corridor humide défini comme étant « à préciser » |
| 01.ouvert | Proportion du périmètre d'analyse concerné par un corridor ouvert défini comme étant « à préserver » |
| 02.ouvert | Proportion du périmètre d'analyse concerné par un corridor ouvert défini comme étant « à restaurer » |
| 03.ouvert | Proportion du périmètre d'analyse concerné par un corridor ouvert défini comme étant « à préciser » |
| 01.boisé | Proportion du périmètre d'analyse concerné par un corridor boisé défini comme étant « à préserver » |
| 02.boisé | Proportion du périmètre d'analyse concerné par un corridor boisé défini comme étant « à restaurer » |
| 03.boisé | Proportion du périmètre d'analyse concerné par un corridor boisé défini comme étant « à préciser » |

### Thématique : Réseaux écologiques : Unité du territoire (non-fragmentation)

| TRSP_FRGM_01 | SIG (Fichier Shape) : Données brutes de croisement des zonages d'inventaire et de protection aux différentes échelles d'analyses (Y= Site, 1000 m ou 5000 m). 3 shapes en sortie : ZON01_INTERSECT_Site, ZON01_INTERSECT_1000m, ZON01_INTERSECT_5000m. | |
|---|---|---|
| CHAMPS | DESCRIPTION | |
| PERIMET | Echelle d'étude = Site par défaut (car pas de croisement 1000 m ni 5000 m) | |
| NOM_SITE, C,100 | Nom du Site | |
| ID, C,24 | Identifiant unique de l'objet dans la BD TOPO | |
| NATURE, C,29 | Attribut permettant de classer un tronçon de route ou de voie ferrée suivant ses caractéristiques physiques : Routes (Bretelle \| Route à 1 chaussée \| Route à 2 chaussées \| Type autoroutier) ; Voies ferrées (Funiculaire ou crémaillère \| LGV \| Métro \| Tramway \| Voie de service \| Voie ferrée principale) | |
| NOM_1_G, C,127 | Nom principal de la rue côté gauche | |
| NOM_1_D, C,127 | Nom principal de la rue côté droit | |
| NOM_2_G, C,127 | Nom secondaire de la rue côté gauche | |
| NOM_2_D, C,127 | Nom secondaire de la rue côté droit | |
| IMPORTANCE, C,1 | Hiérarchisation du réseau routier, sur l'importance des tronçons de route pour le trafic routier : 1 \| 2 \| 3 \| 4 \| 5 \| 6 | |
| POS_SOL, C,14 | Niveau de l'objet par rapport à la surface du sol (valeur négative pour un objet souterrain, nulle pour un objet au sol et positive pour un objet en sursol). Ici toutes les valeurs doivent être 0 | |
| ETAT, C,15 | Etat ou stade d'un objet qui peut être en projet, en construction, en service ou non exploité. | |
| DATE_CREAT, C,19 | Date de création (Routes) | |
| DATE_CREAX, D | Date de création (Voies ferrées) | |
| DATE_MODIX, D | Date de modification (Voies ferrées) | |
| NB_VOIES, N,9, 0 | Nombre total de voies de circulation automobile tracées au sol ou utilisées : Routes (0 \| 1 \| 2 \| 3 \| 4 \| 5 \| 6 \| 7 \| 8 \| 9 \| 10 \| Sans objet) ; Voies ferrées (0 \| 1 \| 2 \| 3 \| 4) | |
| LARGEUR, N,6, 1 | La largeur de la chaussée, d'accotement à accotement, arrondie au demi-mètre (Routes) | |
| LARGEURX, C,80 | Attribut permettant de distinguer les voies ferrées de largeur standard pour la France (1,435 m), des voies ferrées plus larges ou plus étroites. | |
| USAGE_, C,80 | Précise le type de transport auquel la voie ferrée est destinée : Fret \| Sans objet \| Vélo-rail \| Voyageur \| Voyageur et fret | |
| LONG_KM, N,21, 4 | Longueur en km de l'objet intersectée à l'échelle d'étude (site-1000m-5000m) | →Information à compiler pour le calcul de : **Indicateur Unité territoire (Non-fragmentation)** |

## Thématique : Zonages de biodiversité (dont zonages patrimoniaux)

| | |
|---|---|
| **ZON01-_INTERSECT_Y** | SIG (Fichier Shape) : Données brutes de croisement des zonages d'inventaire et de protection aux différentes échelles d'analyses (Y= Site, 1000 m ou 5000 m).<br>3 shapes en sortie : ZON01_INTERSECT_Site, ZON01_INTERSECT_1000m, ZON01_INTERSECT_5000m.<br>Table récapitulative des tables attributaires des 3 shapes |
| **CHAMPS** | **DESCRIPTION** |
| PERIMET | Échelle/Périmètre d'analyse |
| SURFPER | Surface du périmètre en ha |
| ID_SITE | Identifiant unique du site d'étude |
| NOM_SIT | Nom du site d'étude |
| CATEG_Z | Précise la grande catégorie à laquelle le zonage intersecté appartient (Natura 2000, ZNIEFF ou Espaces protégés) |
| TYPE_ZO | Type précis de zonage (APB, ZNIEFF1, RNR, etc.) |
| ID_ZONA | Identifiant unique du zonage |
| NOM_ZON | Nom du zonage |
| SURFINT_H | Surface en hectare (ha) du zonage inclus dans la zone d'étude |
| SURFINT_M | Surface en mètre carré (m²) du zonage inclus dans la zone d'étude |
| DISTANC | Distance la plus courte entre le zonage et la zone d'étude mesurée bord à bord |
| CATEG_U | Catégorie UICN de l'espace protégé (classement en 7 niveaux) concerné |
| PROC_CR | Catégorie de protection du zonage |
| URL_SOU | URL vers le document de référence rattacher au zonage |
| SCORE_PATRIMO | Valeur du score de patrimonialité (1-2-3) attribué au zonage |
| **ZON01-LISTE_ZONAGES_INTERSECT** | Tableur : Liste et surface de chacun des zonages recensés à chaque échelle d'analyse. Chaque échelle d'analyse est traitée dans une feuille Excel distincte. |
| **CHAMPS** | **DESCRIPTION** |
| PERIMETRE | Échelle/Périmètre d'analyse |
| SURFPERI_HA | Superficie du polygone de zone d'étude considérée (Site, tampon 1 km, tampon 5km) en hectare |
| ID_SITE | Identifiant unique du site d'étude |
| NOM_SITE | Nom du site d'étude |
| CATEG_ZONAGE | Précise la grande catégorie à laquelle le zonage intersecté appartient (Natura 2000, ZNIEFF ou Espaces protégés) |
| TYPE_ZONAGE | Type précis de zonage (APB, ZNIEFF1, RNR, etc.) |
| ID_ZONAGE | Identifiant unique du zonage |
| NOM_ZONAGE | Nom du zonage |
| SURFINT_HA | Surface en hectare (ha) du zonage inclus dans la zone d'étude |
| SURFINT_M | Surface en mètre carré (m²) du zonage inclus dans la zone d'étude |
| DISTANCE | Distance la plus courte entre le zonage et la zone d'étude mesurée bord à bord |
| CATEG_UICN | Catégorie UICN de l'espace protégé (classement en 7 niveaux) concerné |
| PROC_CREATION | Catégorie de protection du zonage |
| URL_SOURCE | URL vers le document de référence rattacher au zonage |
| **ZON02-PROPORTION_ZONAGE_TYPE_(COLONNE ou LIGNE)** | Tableur : Nombre, surface et proportion de la zone d'étude concernée par chaque type de zonage. Chaque échelle d'analyse est traitée dans une feuille Excel distincte. |

| CHAMPS | DESCRIPTION | |
|---|---|---|
| PERIMETRE | Échelle/Périmètre d'analyse | |
| SURFPERI_HA | Superficie du polygone de zone d'étude considérée (site, tampon 1 km, tampon 5km) en hectare | |
| ID_SITE | Identifiant unique du site d'étude | |
| NOM_SITE | Nom du site d'étude | |
| SURF_*X*_HA | Surface de la zone d'étude concernée par un zonage de type X en ha. X = aux zonages APB, ASPIM, BIOS, BPM, CDL, CEN, OSPAR, PNC, PNM, PNR, PNZA, RAMSAR, RBD, RBI, RIPN, RNC, RNCFS, RNN, RNR, SIC, ZPS, ZNIEFF (1, 2, 1 MER, 2 MER) | |
| PROP_*X* | Proportion de la zone d'étude concernée parn zonage de type X | |
| NB_*X* | Nombre de zonages de type X dans la zone d'étude | |
| **ZON03-PROPORTION_ZONAGE_CATEGORIE** | *Tableur : Surface et proportion de chaque périmètre d'étude concernée par les différentes grandes catégories de zonage (NATURA 2000, ZNIEFF et Espaces protégées). Chaque échelle d'analyse est traitée dans une feuille Excel distincte.* | |
| CHAMPS | DESCRIPTION | |
| PERIMETRE | Échelle/Périmètre d'analyse | |
| SURFPERI_HA | Superficie du polygone de zone d'étude considérée (Site, tampon 1 km, tampon 5km) en hectare | |
| ID_SITE | Identifiant unique du site d'étude | |
| NOM_SITE | Nom du site d'étude | |
| SURF_EP_HA | Surface totale de la zone d'étude en espaces protégés (en ha) | |
| PROP_EP | Proportion totale de la zone d'étude en espaces protégés | |
| SURF_N2000_HA | Surface totale de la zone d'étude en N2000 (en ha) | |
| PROP_N2000 | Proportion totale de la zone d'étude en N2000 | |
| SURF_ZNIEFF_HA | Surface totale de la zone d'étude en ZNIEFF (en ha) | |
| PROP_ZNIEFF | Proportion totale de la zone d'étude en ZNIEFF | |
| SURF_ALL | Surface totale de la zone d'étude en zonage d'inventaire ou de protection (en ha) | |
| PROP_ALL | Proportion totale de la zone d'étude en zonage d'inventaire ou de protection | |
| NB_ALL | Nombre total de zonages différents dans la zone d'étude | |
| NB_PATRIM | Nombre de zonages à haute patrimonialité dans la zone d'étude (score 3) | |
| SURF_PATRIMO | Surface de la zone d'étude concernée par l'ensemble des zonages à haute patrimonialité (score 3) en ha | |
| PROP_PATRIMO | Proportion de la zone d'étude concernée par l'ensemble des zonages à haute patrimonialité (score 3) | Indicateur Zonages patrimoniaux |

## Thématique : Espèces à enjeux de conservation (espèces patrimoniales) dans la maille

| ESP_SCAP19_INTERSECT | *SIG (Fichier Shape) : La couche présente le contour des sites avec un remplissage des mailles intersectés (remplissage entier si site intersecté par une seule maille, ou remplissage divisé si plusieurs mailles intersectées). Chaque ligne de la table attributaire est un site (plusieurs lignes pour un même site si celui-ci est intersecté par plusieurs mailles).* |
|---|---|
| CHAMPS | DESCRIPTION |
| PERIMET | Echelle d'étude = Site par défaut (car pas de croisement 1000 m ni 5000 m) |
| SURFPER | Surface du site (en ha ?) |
| ID_SITE | Identifiant unique du site d'étude |
| NOM_SIT | Nom du site d'étude |
| PROP_PE | % de la zone d'étude présente dans ladite maille. Valeur varie de 0 à 0.999 pour les sites présents sur plusieurs mailles, ou = 1 pour les périmètres qui sont sur une maille unique |

| CD_SIG_MAI | Code de référence de la Maille 10x10km Lambert 93 concernée par le site d'étude | |
|---|---|---|
| NbGrInfSeuil | Nombre de groupes taxonomiques inférieurs à leurs seuils de suffisance pour la connaissance (nombre de groupes méconnus). | |
| PourcInfSeuil | Valeur (%) de méconnaissance naturaliste de la maille correspondant au pourcentage de groupes méconnus par rapport au total des 27 groupes étudié. | **Descripteur Méconnaissance** |
| Cat | La catégorie de valeur de méconnaissance (de 1 à 4) correspondant aux 4 intervals de pourcentages définis pour la carte diffusée sur l'ONB (1=11-22 %, 2=26-48 %, 3=52-74 %, 4=78-100 %). | |
| SCAP_NAT | Nombre des taxons de référence pour la SCAP nationale de 2015 dans la maille | |
| DH_DO | Nombre des taxons de la directive Habitats-Faune-Flore ou de la directive Oiseaux (sauf exceptions des taxons considérés non nicheurs par l'expert) dans la maille | |
| CR_EN_V | Nombre des taxons vertébrés ou de la fonge classé CR EN ou VU dans une liste rouge (Listes Rouges Mondiale, Européenne ou France métropolitaine) dans la maille | |
| END | Nombre des taxons vertébrés, invertébré ou de la fonge considérés comme endémiques ou subendémiques dans le référentiel TaxRef (V12) dans la maille | |
| Znieff_NT | Nombre des taxons vertébrés ou de la fonge déterminants ZNIEFF et classés NT dans une liste rouge (Listes Rouges Mondiale, Européenne ou France métropolitaine) dans la maille | |
| End_NT_Z | Nombre des taxons invertébrés remplissant au moins 2 des 3 conditions suivantes : Endémique selon TaxRef (V12), classé NT dans une liste rouge (Listes Rouges Mondiale, Européenne ou France métropolitaine) ou déterminant ZNIEFF dans la maille | |
| CR_EN | Nombre des taxons de la flore classé CR ou EN dans une liste rouge (Listes Rouges Mondiale, Européenne ou France métropole) dans la maille | |
| Richess | Richesse totale des espèces à enjeux de conservation (nbre total d'espèces de la SCAP19) dans la maille. Attention, ce n'est pas une addition des nombres des colonnes précédentes, mais le résultat d'un calcul permettant d'éviter les doubles comptages | **Indicateur Espèces patrimo** |
| ESP_SCAP19_INTERSECT_MAIMAJ | *SIG (Fichier Shape) : La couche présente le contour des sites avec un remplissage des mailles intersectés (remplissage entier même si le site est intersecté par plusieurs mailles intersectées). Dans le cas de plusieurs mailles sur un même site, c'est uniquement la maille majoritaire (% surfacique recouvert) qui est retenue. Chaque ligne de la table attributaire est un site.* | |
| **CHAMPS** | **DESCRIPTION** | |
| *Mêmes champs que sur le tableau de la couche ESP_SCAP19_INTERSECT* | | |
| ESP_SCAP19_MAI10 | *SIG (Fichier Shape) : Données brutes de croisement des mailles 10kmx10km de l'INPN (avec les richesses d'espèces) à l'échelle d'analyse du Site uniquement. La couche présente les mailles (contour des carreaux) intersectant les sites. Chaque ligne de la table attributaire est une maille* | |
| **CHAMPS** | **DESCRIPTION** | |
| CD_SIG_ | Code de référence de la Maille 10x10km Lambert 93 | |
| NbGrInfSeuil | Cf. descriptions des champs sur le tableau de la couche ESP_SCAP19_INTERSECT | |
| PourcInfSeuil | | |
| Cat ; SCAP_NAT ; DH_DO ; CR_EN_V ; END ; Znieff_NT ; End_NT_Z ; CR_EN ; Richess | Cf. descriptions des champs sur le tableau de la couche ESP_SCAP19_INTERSECT | |
| SIT_INT | Identifiant unique du site intersecté par la maille, ou liste de sites intersectés par la maille et séparés par des virgules | |
| RICH_ESP_SCAP19 | *Tableur : Données brutes de croisement des mailles 10kmx10km de l'INPN (avec les richesses d'espèces) à l'échelle d'analyse du Site uniquement. Chaque ligne de la table attributaire est un site (plusieurs lignes pour un même site si celui-ci est intersecté par plusieurs mailles).* | |
| **CHAMPS** | **DESCRIPTION** | |
| *Mêmes champs que sur le tableau de la couche ESP_SCAP19_INTERSECT* | | |

# Annexe 6 :
# Méthodes de discrétisation et algorithme de Jenks

A) Le tableau ci-dessous présente les différentes méthodes de discrétisation existantes, les distributions des valeurs adaptées à chaque méthode, ainsi que les avantages et inconvénients de chaque méthode (Béguin *et al.,* 2017), (Magrit, 2017), (Jegou *et al.,* s. d.) :

| Méthode | Définition | Forme distribution | Avantages ⊕ | Désavantages ⊖ |
|---|---|---|---|---|
| Equidistance = amplitude égale = intervalles égaux | L'écart entre les classes est constant. Il suffit de diviser la dynamique de la série (maximum – minimum) par le nombre de classes souhaité. | Uniforme Symétrique | Facile à réaliser | Qualité du regroupement tributaire de la distribution des valeurs : risque de produire des classes vides si distribution asymétrique ou avec fortes discontinuités |
| Progression arithmétique | l'amplitude des classes augmente selon une progression arithmét à raison de R (étendue de la série) divisée par l'addition des classes. R = (max – min)/(1 + 2 + 3… + x nbre clas.) Amplitude classes : cl.1 (min à min+R), cl.2 (min+R à min+2R)… | Asymétrique | | Conçue pour distributions asymétriques avec beaucoup de valeurs faibles et peu de valeurs fortes. Non adaptée pour distribution normale |
| Équi-fréquence = équi-population = quantiles | Toutes les classes ont, dans la mesure du possible, le même nombre d'individus. Tri par ordre croissant des valeurs et calcul du nombre idéal de valeurs par classe (effectif/nombre classes). Affecter les individus selon leur rang aux classes successives. Bornes des classes définies par valeurs extrêmes de chaque classe. | Toutes, surtout : Asymétrique Plurimodale Uniforme | Diminue le poids des valeurs extrême : équilibre les classes. Produit la carte visuellement la plus équilibrée. Bien pour cartes comparatives. | La carte masque les fortes discontinuités. Peut regrouper des individus assez éloignés. Peut créer des classes vides sous QGIS. |
| Méthode de Jenks | Se base sur le principe de ressemblance/dissemblance en calculant la distance paramétrique entre toutes les valeurs de la série. Minimise la variance intraclasse et maximise la variance interclasse : Pour chaque classe les valeurs sont les plus proches et les classes sont les plus éloignées possibles | Toutes, surtout : Asymétrique Plurimodale | Respecte l'allure de la série. Certainement la plus adaptée des méthodes statistiques. Ne crée pas de classes vides sous QGIS. | Ne permet pas de rendre comparables plusieurs cartes d'une même série. Nécessite l'emploi d'un logiciel statistique. |
| Manuel = seuils observés | L'opérateur découpe la série visualisée graphiquement (histogramme de fréquence) selon les discontinuités (seuils). Le nombre de classes résultant est un compromis entre l'allure de la série (seuils, discontinuités) et le nombre de classes initialement projeté. | Asymétrique Plurimodale | Produit des cartes équilibrées graphiquement. Conseillée pour distributions dont les zones ont leur dispersion caractéristique. | Oblige l'opérateur à analyser graphiquement la série (histogramme de fréquence). Significativité des discontinuités observées dépend de la longueur de la série. Très aléatoire : à partir des mêmes données, ≠ choix |
| Classes standardisées | Classes déterminées selon une fraction d'écart-type (indicateur de l'amplitude des classes) par rapport à la moyenne | Symétrique Normale | Taillée pour les comparaisons | Adaptée si distribution normale. Sinon il faut normaliser ou transformer |
| Moyennes emboîtées | Moyenne divise la série en deux classes. Puis leurs moyennes respectives les divisent à leur tour en deux et ainsi de suite. | | | Nombre de classes est toujours un multiple de 2 ce qui peut s'avérer contraignant. |

Les cartes suivantes illustrent ces méthodes de discrétisation par un exemple sur la densité de la population (habitants/km²) en France par région (données du recensement 1990). Source : Jegou *et al.,* s. d.

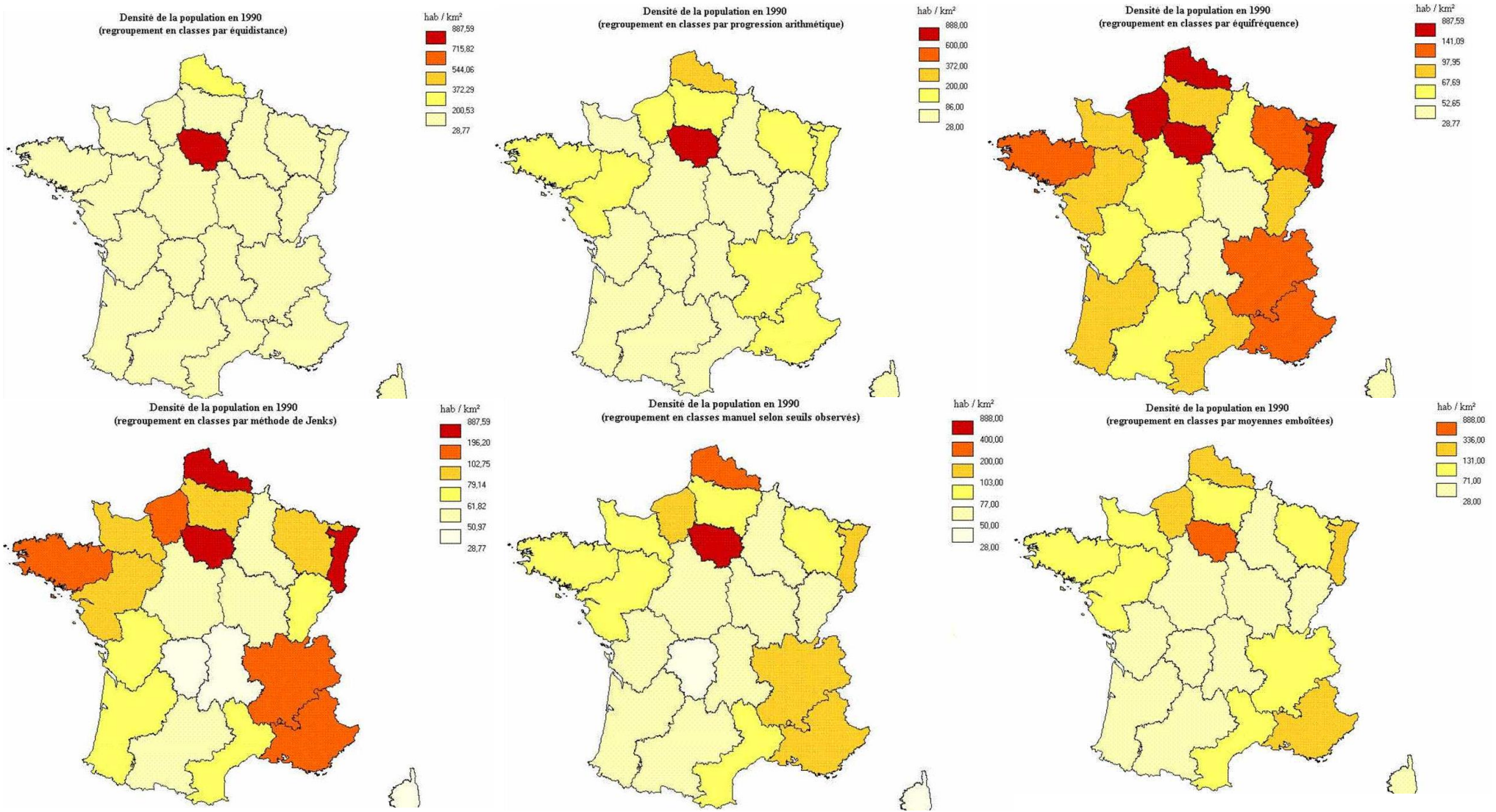

Les tests de CARPO sur deux groupes de sites a mis en évidence que, pour la majorité d'indicateurs, les séries présentent une **distribution de valeurs de type asymétrique (négative ou positive) et plurimodale**, adaptées à une discrétisation par **équi-fréquence ou méthode de Jenks. Cette dernière s'est avérée plus appropriée** en raison de sa puissance statistique, la fidélité à la distribution de valeurs et aux vraies discontinuités, ainsi qu'à l'absence de problème de classes vides.

B)

> ## La méthode de Jenks
>
> ### Introduction
>
> La méthode de Jenks, également connue sous le nom de Ruptures naturelles (Jenks Natural Breaks ou Jenks Optimization method en anglais) est un algorithme utilisé pour classer des entités à l'aide des ruptures naturelles dans une série de valeurs de données. Elle se base sur le principe de Goodness of Variance Fit (GVF), et a été initialement utilisée dans la cartographie choroplèthe (certes principalement utilisées pour afficher des variables surfaciques quantitatives). La méthode a été publiée dans Jenks, George F. 1967. "The Data Model Concept in Statistical Mapping", International Yearbook of Cartography 7: 186-190.
>
> Cette méthode de classification statistique aide à déterminer la meilleure façon de discrétiser les données : elle partitionne les valeurs en classes à l'aide d'un algorithme qui calcule des regroupements de valeurs en fonction de la distribution des données. Cet algorithme réduit la variance au sein des groupes et maximise la variance entre les groupes. Elle crée donc des classes avec des bornes de classes qui sont identifiées parmi celles qui regroupent le mieux des valeurs similaires et optimisent les différences entre les classes. Les entités sont réparties en classes dont les limites sont définies aux endroits où se trouvent de grandes différences dans les valeurs de données (ESRI, 2019).
>
> ### Fonctionnement de l'algorithme
>
> La démarche consiste, de façon schématique, à (Mazurek, 2003) :
> 1. Ordonner les données par valeur croissante
> 2. Chercher tous les groupes possibles pouvant former *k* classes.
> 3. Pour chaque configuration possible pour former de classes, calculer la variance à l'intérieur du groupe, et entre les groupes.
> 4. Comparer l'ensemble des valeurs et prendre la configuration qui minimise la variance dans le groupe et maximisa la variance entre les groupes.
>
> Cependant, le fonctionnement de l'algorithme est plus complexe et **nécessite l'emploi d'un logiciel statistique (QGIS et ArGIS** peuvent traiter cet algorithme). En effet, la méthode se base sur un **processus itératif**. Autrement dit, les calculs doivent être répétés en utilisant différentes ruptures dans l'ensemble de données pour déterminer quel ensemble de ruptures présente la plus petite variance de la classe. L'utilisation de l'algorithme peut être illustrée en quatre étapes (Smith et al., 2007) :
> 1. L'utilisateur sélectionne l'attribut, *x* (par exemple les résultats sur le caractère naturel du groupe de sites), à classer et spécifie le nombre de classes requises, *k*. Dans le cas de CARPO ce sont 3 classes (k = 3).
> 2. Le logiciel ordonne la série de valeurs et, un ensemble de k-1 valeurs aléatoires ou uniformes est généré dans la plage [min {x}, max {x}]. Celles-ci sont utilisées comme bornes de classe initiales. Les divisions des groupes initiales peuvent donc être arbitraires, et seront ensuite corrigées.
> 3. Les **moyennes** (μ) de la série et de chaque classe initiale sont calculées, et la somme des **variances** des membres de la classe par rapport aux valeurs moyennes est calculée. Pour cela :
>    1. Calculer la somme des variances entre chaque valeur et la moyenne de la série de valeurs (**SDAM** = sum of squared deviations from the array mean)

2. Calculer la somme des variances entre chaque valeur et la moyenne de la classe (**SDCM** = sum of squarred deviations from class means).
4. Après avoir inspecté chaque SDCM, une décision est alors prise de déplacer une unité d'une classe avec un SDCM plus grand vers une classe adjacente avec un SDCM inférieur. En fait, les valeurs individuelles de chaque classe seront attribuées aux classes adjacentes en ajustant les bornes de classe afin de vérifier que le **SDCM soit le plus réduit possible**. Les variances entre les valeurs et la moyenne de la classe sont à nouveau recalculées et le processus est répété jusqu'à ce que le SDCM atteigne une valeur minimale.

Ce processus itératif se termine quand l'optimisation de la SDCM tombe en dessous d'un niveau seuil, c'est-à-dire lorsque la variance intra-classe est la plus petite possible et que la variance inter-classes est la plus grande possible. La SDAM est une constante et ne change pas. Bien qu'une véritable optimisation ne soit pas assurée, l'ensemble du processus peut être éventuellement répété à partir de l'étape 1 ou 2 et les valeurs SDCM comparées.

Finalement, la qualité de l'ajustement de la variance (GVF = Goodness of Variance Fit) est calculée :
**GVF = (SDAM - SDCM) / SDAM.**
La GVF varie de 0 (pire ajustement) à 1 (ajustement parfait).
Le processus itératif mentionné se termine donc quand la GVF ne peut plus être incrémenté.

## Exemple de calcul

L'exemple ci-dessous (Ahmad, 2019) illustre le fonctionnement de l'algorithme avec les étapes mentionnées précédemment.
La série de quatre valeurs [4, 5, 9, 10] va être discrétisée en 2 classes (k=2) tout en suivant la méthode de Jenks :

I. Calculer la moyenne ($\mu$) de la série de valeurs :
```
série = [4, 5, 9, 10]
µ = 4 + 5 + 9 + 10)/4 = 7
```

II. Calculer la somme des variances entre chaque valeur et la moyenne de la série (**SDAM**).
```
SDAM = (4-7)² + (5-7)² + (9-7)² + (10-7)² = 9 + 4 + 4 + 9 = 26
```

III. Déterminer toutes les combinaisons de plages de valeurs (classes) possibles, et pour chacune d'entre elles : calculer la somme des variances entre chaque valeur et la moyenne de la classe (**SDCM**), pour ensuite trouver la plus petite SDCM parmi les options résultantes.
```
Pour la variante des classes : [4] et [5,9,10]
SDCM = (4-4)² +(5-8)² +(9-8)² +(10-8)²  = 0 + 9 + 1 + 4 = 14

Pour la variante des classes [4,5] et [9,10]
SDCM = (4-4.5)² +(5-4.5)² +(9-9.5)² +(10-9.5)² = 0.25 + 0.25 + 0.25 + 0.25 = 1

Pour la variante des classes [4,5,9] et[10]
SDCM = (4-6)² +(5-6)² +(9-6)² +(10-10)² = 4 + 1 + 9 + 0 = 14
```

> *La deuxième variante, **[4,5] et [9,10],** est celle qui possède le **SDCM le plus réduit,** ce qui implique une variance minimale.*

IV. Enfin, calculer la (**GVF**) de chaque combinaison possible. Cette variable est définie comme (SDAM - SCDM) / SDAM, et elle varie de 1 (ajustement parfait) à 0 (pire ajustement).
    **GVF** pour **[4,5] [9,10]** = (26 - 1)/26 = 25/26 = **0.96**
    **GVF** pour les deux autres combinaisons = (26 - 14)/26 = 12/26 = **0.46**

Le **GVF pour [4,5] [9,10] est le plus élevé** indiquant que **cette combinaison est la meilleure** pour la définition de deux classes pour la série [4, 5, 9, 10].

Cet algorithme s'alourdit avec l'augmentation de données. Par exemple, pour une discrétisation de 254 valeurs en 6 classes, il existe 8 301 429 675 combinaisons de classes possibles ; d'où la nécessité d'un logiciel informatique pour ces calculs.

**Annexe 7 :**
**Aide à la définition des seuils et classes de niveau via QGIS et exemple sur un groupe de sites**

La définition des seuils et des classes de niveaux de potentialité écologique se base sur une **analyse de la distribution de valeurs de la série du groupe de sites**. Ceci implique que les bornes de chacune des trois classes sont définies sur les résultats des indicateurs CARPO obtenus dans chaque groupe de sites à évaluer, et non pas sur la base de seuils nationaux applicables à tout groupe.

Cette analyse de la distribution des résultats sur la série des sites nécessite l'utilisation **d'un logiciel de traitement de SIG comme QGIS et ArcGIS**, afin de réaliser la définition des bornes, d'obtenir d'histogrammes de distribution de valeurs et de visualiser les résultats. Les indications suivantes illustrent **les étapes à suivre sur QGIS** (version 3.10.3)**, pour analyser la distribution de valeurs et calculer facilement les classes de niveau sous la méthode de Jenks.**

### Etape 1 : Création d'une couche synthétique

Suite aux croisements et aux calculs via l'outil automatisé ou aux opérations manuelles, un ensemble de sorties sont produites. Elles se structurent autour d'un ensemble de tableurs et de fichiers cartographiques (shapefiles) contenant les données brutes. Ces fichiers présentent, entre autres, des colonnes avec les résultats de chaque indicateur CARPO. Les résultats de chaque indicateur se retrouvent donc dans un fichier dédié séparé. **Il n'y a actuellement pas de sortie réunissant sous un seul fichier, l'ensemble des résultats de chacun des 8 indicateurs CARPO. Il est donc essentiel de créer une couche centralisatrice de synthèse,** qui constituera la base de la discrétisation.

Pour cela, il sera nécessaire de **prendre chacune des trois couches contentant les périmètres des sites aux trois échelles** (Site – Voisinage 1000m – Paysage 5000m) **et de les compléter avec les résultats des indicateurs CARPO,** afin de créer cette couche centralisatrice de synthèse. Dans ce cas, il est possible d'ajouter des champs à partir de la table attributaire, afin de créer les 8 colonnes des indicateurs CARPO et d'ajouter les résultats des sites à partir des copier-coller. L'**Annexe 5** présente les intitulés des fichiers et des champs contenant les données à reprendre et à intégrer dans la couche de synthèse. Ces intégrations de données de synthèse doivent permettre de créer une couche complète dont la table attributaire contiendra des champs semblables à ceux sur la figure suivante :

**Etape 2 : Calcul automatique des seuils et classes par QGIS**

A partir de chacune des trois couches centralisatrices de synthèse, il sera possible de lancer le calcul des seuils (fourchettes de valeurs) et donc des 3 classes de niveau, selon la méthode de Jenks, via une fonctionnalité de QGIS. Pour cela, il faudra ouvrir la fenêtre des **propriétés de la couche** de synthèse, puis les paramètres de **Style.** Les réglages suivants permettront de réaliser la discrétisation sous Jenks et de définir les classes dans cette fenêtre :

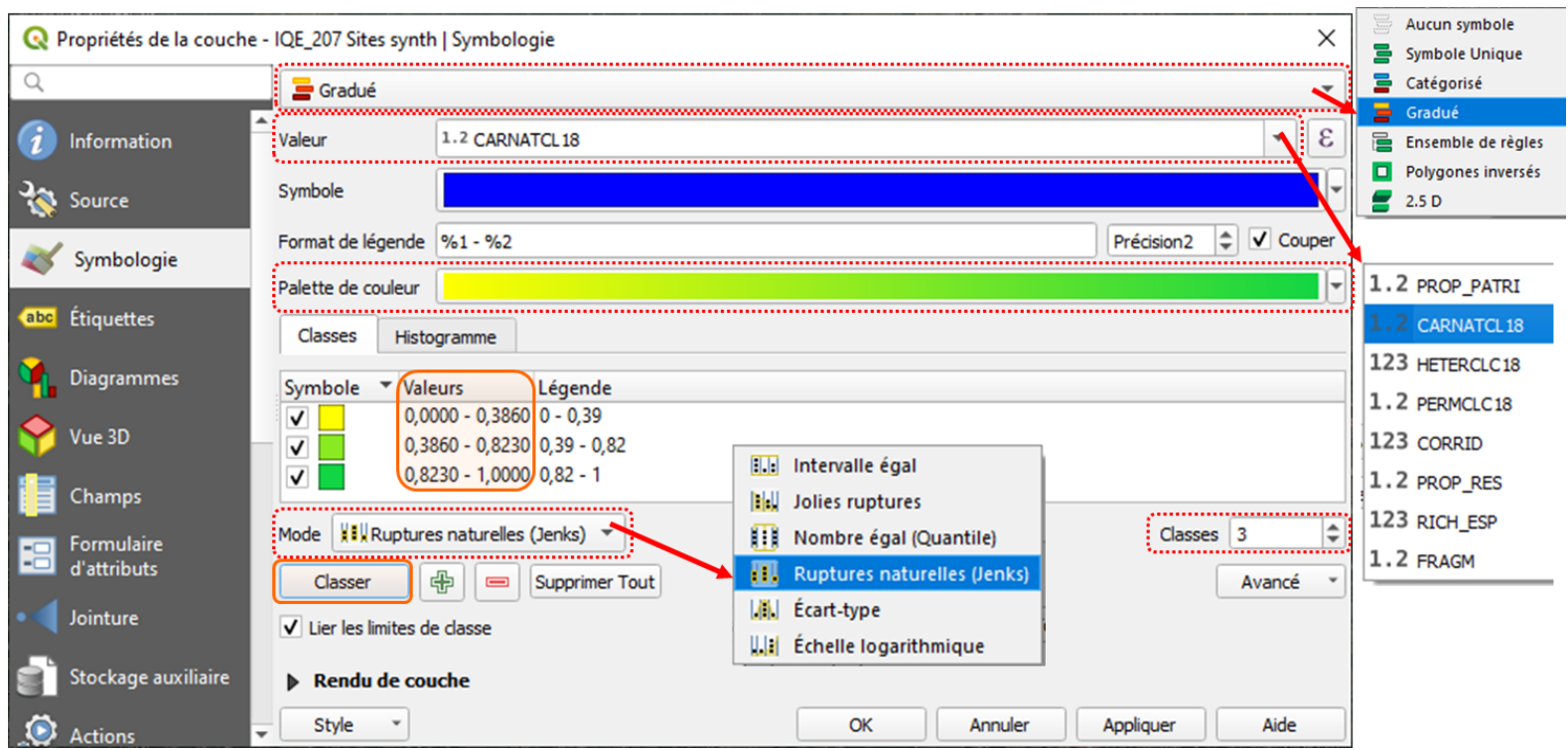

1. Le choix du rendu "**Symbole Gradué**" permet d'afficher toutes les entités, en utilisant un symbole de couche dont la couleur reflétera la classe de niveau associé
2. **Valeur** : choisir **l'attribut qui sera discrétisé**, autrement dit, choisir (à tours de rôle) l'indicateur CARPO dont les résultats seront analysés (le nom du champ dans la table attributaire de la couche de synthèse). Dans cet exemple c'est le champ «CARNATCL18» (pour «Caractère naturel sous CLC18») qui sera discrétisé.
3. **Palette de couleur** : Choisir les couleurs relatifs à chacun des 3 classes de niveau de potentialité (moyen, fort, très fort)
4. **Mode** : Choisir **Ruptures naturelles (Jenks)** comme méthode de discrétisation (regroupement en classes) à appliquer
5. **Classes** : Régler le nombre en 3 (pour les trois classes désirées : moyen, fort, très fort)

Une fois ces paramètres réglés, un clic sur **Classer**, permettra de calculer et d'afficher dans l'onglet **Classes**, l'ensemble des fourchettes de valeurs (seuils) qui définissent les bornes de chacune des trois classes. Ces sont donc les plages indiqués dans le champ **Valeurs** qui établissent les **bornes des classes**.

**Etape 3 : Visualiser l'histogramme de distribution des valeurs**

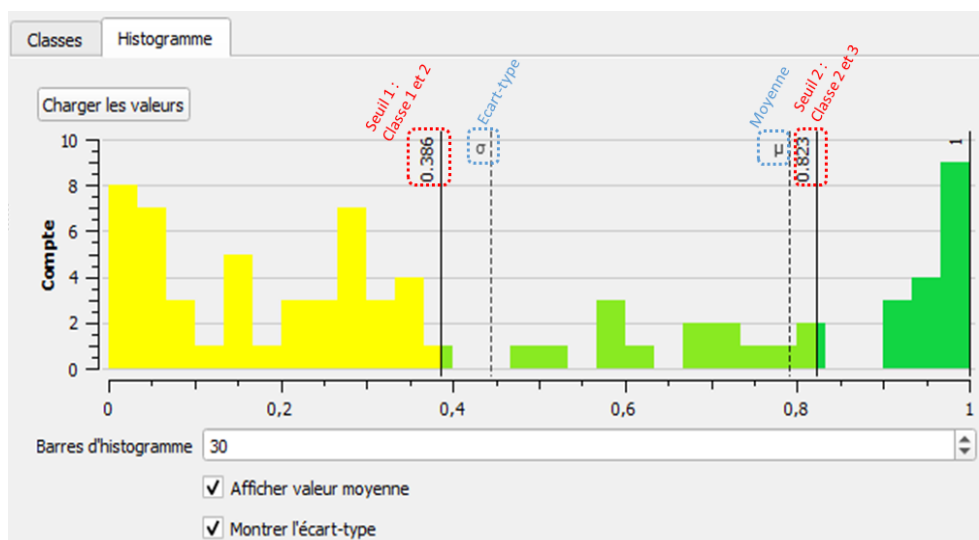

Il est également possible de visualiser la distribution de valeurs de chaque indicateur sous forme d'histogramme. Pour cela il faudra cliquer sur l'onglet **Histogramme** (à côté de l'onglet Classes) puis sur le bouton **Charger les valeurs**.

L'histogramme est automatiquement créé. Les couleurs associées correspondent aux valeurs appartenant à chacune des trois classes. L'axe des abscisses contient les valeurs de l'indicateur en question, et l'axe des ordonnées indique le nombre de sites associé à chaque valeur de l'indicateur. Cette figure indique chacun des **deux seuils** qui ont servi à déterminer les bornes des trois classes : le premier seuil correspond à la borne supérieure de la classe 1 et à la borne inférieure de la classe 2 ; le second correspond à la borne supérieure de la classe 2 et à la borne inférieure de la classe 3.

Il est possible de **visualiser la moyenne et l'écart type** en cochant ces options dans les paramètres.

**Etape 4 : Adaptation des seuils**

Bien que la borne supérieure d'une classe semble être identique à la borne inférieure de la classe suivante, elles ne le sont pas. **Les classes sont exclusives entre elles**. Autrement dit, toutes les valeurs identiques se retrouveront toujours dans une seule classe. Ainsi, **pour la première classe, les deux bornes sont incluses**, alors que pour **les deux classes suivantes, la borne inférieure est exclue et la borne supérieure incluse**.

Au moment d'enregistrer les seuils des classes de niveau sur un document séparé, il suffira de **faire une adaptation à la main : augmenter le dernier chiffre décimal (+1) afin d'avoir des bornes exclusives**.

Il est également possible d'**afficher le nombre de sites associés à chaque classe** ; il suffit de faire un clic droit sur la couche de synthèse et sélectionner l'option **Montrer le décompte d'entités**. Cette adaptation des bornes et cet affichage du nombre d'entités permettra d'établir un tableau de fourchettes de valeurs de classes et de les enregistrer dans la grille (Annexe 8).

Les **étapes 2 à 4 sont à réaliser à tour de rôle pour chacun des 8 indicateurs CARPO et à chacune des 3 échelles d'étude**. Il suffit de modifier l'attribut (indicateur à discrétiser) dans la fenêtre du Style.

**Exemple :**

L'exemple suivant illustre les fourchettes de valeurs et les classes de niveaux établis, suite à l'analyse de la distribution de valeurs de six indicateurs sur un **groupe de 207** sites et le calcul automatique des seuils, selon la méthode de Jenks, via QGIS. Les distributions présentent une forme asymétrique ou plurimodale adapté à la méthode de Jenks. Il faut noter également que la méthode de Jenks ne produit pas de classes à effectifs égaux.

Ce calcul et cette synthèse des seuils par indicateur et par échelle d'étude, permet de compléter la grille d'évaluation.

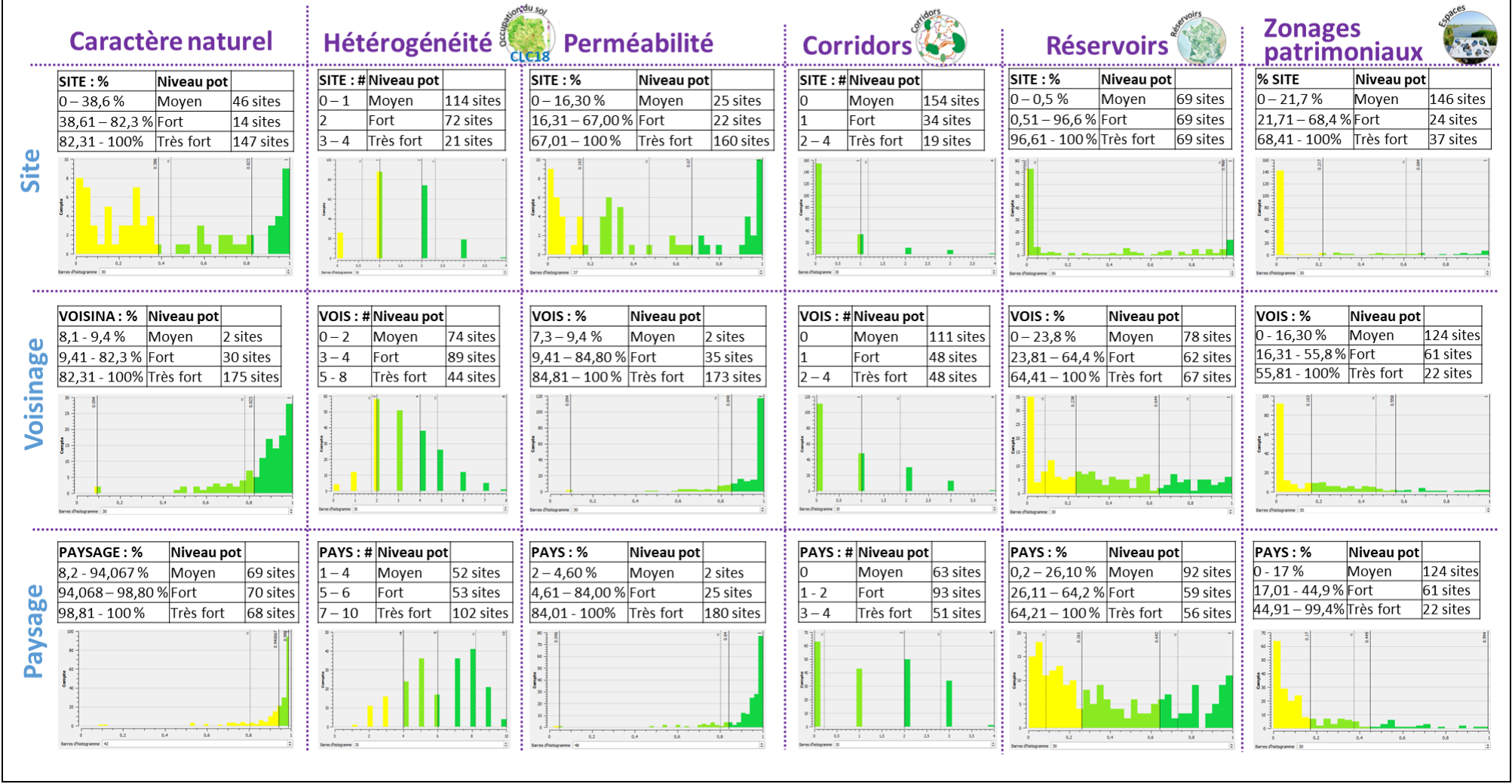

# Annexe 8 :
## Grille d'évaluation du niveau de potentialités écologiques d'un site d'implantation

*Le tableau suivant présente la **grille d'évaluation vierge**. Elle doit être renseignée pour chaque site d'implantation à évaluer et pour chacune des trois échelles à évaluer (**1 tableau par site par échelle**), suite à l'obtention des résultats du calcul des différents indicateurs et des calculs des seuils des classes de niveau de potentialité via un logiciel dédié (QGIS/ArcGIS), sur la base de la distribution des valeurs de la série.*

- *Les **champs en rose** (résultat de l'indicateur et seuils des classes) **doivent être impérativement renseignés**. Les **champs bleus** (moyennes et nombre de sites) **peuvent être renseignés** pour fournir une référence.*
- *La règle décisionnelle pour le cumul des niveaux de potentialité (pour les indicateurs au sein d'une thématique) est présentée dans l'Annexe 9. Cette règle décisionnelle permet de choisir un critère de décision pour le cumul qui doit être noté dans ce tableau dans la colonne « critère décisionnel cumul ».*

| Thématique | Sous thématique | Indicateurs | Résultat | | | Fourchette décisionnelle | | | Niveau potentialité / échelle | Niveau potentialité retenu<br>Préciser échelle d'analyse retenue : ________ | Critère décisionnel cumul indicateurs | Niveau Potentialité thématique |
|---|---|---|---|---|---|---|---|---|---|---|---|---|
| | | | Echelle | Moyenne | Résultat | Seuils | Nbre sites | Niveau | | | | |
| Occupation du sol | Caractère naturel | % surface de la zone d'étude occupée par des milieux considérés naturels et semi-naturels | Site | μ = ___ | ___% | ___ - ___<br>___ - ___<br>___ - ___ | ___ sites<br>___ sites<br>___ sites | Moyen<br>Fort<br>Très fort | □ Moyennes<br>□ Fortes<br>□ Très fortes | □ Moyennes<br>□ Fortes<br>□ Très fortes | ________ | □ Moyennes<br>□ Fortes<br>□ Très fortes |
| | | | Voisinage | μ = ___ | ___% | ___ - ___<br>___ - ___<br>___ - ___ | ___ sites<br>___ sites<br>___ sites | Moyen<br>Fort<br>Très fort | □ Moyennes<br>□ Fortes<br>□ Très fortes | | | |
| | | | Paysage | μ = ___ | ___% | ___ - ___<br>___ - ___<br>___ - ___ | ___ sites<br>___ sites<br>___ sites | Moyen<br>Fort<br>Très fort | □ Moyennes<br>□ Fortes<br>□ Très fortes | | | |
| | Hétérogénéité | Nombre de milieux naturels différents dans la zone d'étude | Site | μ = ___ | ___ | ___ - ___<br>___ - ___<br>___ - ___ | ___ sites<br>___ sites<br>___ sites | Moyen<br>Fort<br>Très fort | □ Moyennes<br>□ Fortes<br>□ Très fortes | □ Moyennes<br>□ Fortes<br>□ Très fortes | | |
| | | | Voisinage | μ = ___ | ___ | ___ - ___<br>___ - ___<br>___ - ___ | ___ sites<br>___ sites<br>___ sites | Moyen<br>Fort<br>Très fort | □ Moyennes<br>□ Fortes<br>□ Très fortes | | | |
| | | | Paysage | μ = ___ | ___ | ___ - ___<br>___ - ___<br>___ - ___ | ___ sites<br>___ sites<br>___ sites | Moyen<br>Fort<br>Très fort | □ Moyennes<br>□ Fortes<br>□ Très fortes | | | |
| | Perméabilité | % surface de la zone d'étude occupée par des modes d'occupation du sol considérés perméables | Site | μ = ___ | ___% | ___ - ___<br>___ - ___<br>___ - ___ | ___ sites<br>___ sites<br>___ sites | Moyen<br>Fort<br>Très fort | □ Moyennes<br>□ Fortes<br>□ Très fortes | □ Moyennes<br>□ Fortes<br>□ Très fortes | | |
| | | | Voisinage | μ = ___ | ___% | ___ - ___<br>___ - ___<br>___ - ___ | ___ sites<br>___ sites<br>___ sites | Moyen<br>Fort<br>Très fort | □ Moyennes<br>□ Fortes<br>□ Très fortes | | | |
| | | | Paysage | μ = ___ | ___% | ___ - ___<br>___ - ___<br>___ - ___ | ___ sites<br>___ sites<br>___ sites | Moyen<br>Fort<br>Très fort | □ Moyennes<br>□ Fortes<br>□ Très fortes | | | |
| Réseaux écologiques | Unité du territoire (non-fragmentation) | Densité du linéaire des réseaux de transport (Rapport : nbre linéaires / surface échelle d'étude) | Site | μ = ___ | ___/ha | ___ - ___<br>___ - ___<br>___ - ___ | ___ sites<br>___ sites<br>___ sites | Moyen<br>Fort<br>Très fort | □ Moyennes<br>□ Fortes<br>□ Très fortes | □ Moyennes<br>□ Fortes | ________ | |
| | | | Voisinage | μ = ___ | | ___ - ___ | ___ sites | Moyen | □ Moyennes | | | |
| | | | | | | ___ - ___ | ___ sites | Fort | □ Fortes<br>□ Très fortes | □ Très fortes | | □ Moyennes<br>□ Fortes<br>□ Très fortes |
| | | | | | /ha | ___ - ___ | ___ sites | Très fort | | | | |
| | | | Paysage | µ = ___ | ___/ha | ___ - ___ | ___ sites | Moyen | □ Moyennes<br>□ Fortes<br>□ Très fortes | | | |
| | | | | | | ___ - ___ | ___ sites | Fort | | | | |
| | | | | | | ___ - ___ | ___ sites | Très fort | | | | |
| | | | Site | µ = ___ | ___ | ___ - ___ | ___ sites | Moyen | □ Moyennes<br>□ Fortes<br>□ Très fortes | | | |
| | | | | | | ___ - ___ | ___ sites | Fort | | | | |
| | | | | | | ___ - ___ | ___ sites | Très fort | | | | |
| | Corridors | Nombre de sous-trames différentes (milieux majoritaires) traversant la zone d'étude | Voisinage | µ = ___ | ___ | ___ - ___ | ___ sites | Moyen | □ Moyennes<br>□ Fortes<br>□ Très fortes | □ Moyennes<br>□ Fortes<br>□ Très fortes | | |
| | | | | | | ___ - ___ | ___ sites | Fort | | | | |
| | | | | | | ___ - ___ | ___ sites | Très fort | | | | |
| | | | Paysage | µ = ___ | ___ | ___ - ___ | ___ sites | Moyen | □ Moyennes<br>□ Fortes<br>□ Très fortes | | | |
| | | | | | | ___ - ___ | ___ sites | Fort | | | | |
| | | | | | | ___ - ___ | ___ sites | Très fort | | | | |
| | | | Site | µ = ___ | ___% | ___ - ___ | ___ sites | Moyen | □ Moyennes<br>□ Fortes<br>□ Très fortes | | | |
| | | | | | | ___ - ___ | ___ sites | Fort | | | | |
| | | | | | | ___ - ___ | ___ sites | Très fort | | | | |
| | Réservoirs | % surface de la zone d'étude occupée par des réservoirs | Voisinage | µ = ___ | ___% | ___ - ___ | ___ sites | Moyen | □ Moyennes<br>□ Fortes<br>□ Très fortes | □ Moyennes<br>□ Fortes<br>□ Très fortes | | |
| | | | | | | ___ - ___ | ___ sites | Fort | | | | |
| | | | | | | ___ - ___ | ___ sites | Très fort | | | | |
| | | | Paysage | µ = ___ | ___% | ___ - ___ | ___ sites | Moyen | □ Moyennes<br>□ Fortes<br>□ Très fortes | | | |
| | | | | | | ___ - ___ | ___ sites | Fort | | | | |
| | | | | | | ___ - ___ | ___ sites | Très fort | | | | |
| | | | Site | µ = ___ | ___% | ___ - ___ | ___ sites | Moyen | □ Moyennes<br>□ Fortes<br>□ Très fortes | | | |
| | | | | | | ___ - ___ | ___ sites | Fort | | | | |
| | | | | | | ___ - ___ | ___ sites | Très fort | | | | |
| Zonages | Patrimonialité | Proportion de zonages patrimoniaux dans la zone d'étude | Voisinage | µ = ___ | ___% | ___ - ___ | ___ sites | Moyen | □ Moyennes<br>□ Fortes<br>□ Très fortes | □ Moyennes<br>□ Fortes<br>□ Très fortes | | |
| | | | | | | ___ - ___ | ___ sites | Fort | | | | |
| | | | | | | ___ - ___ | ___ sites | Très fort | | | | |
| | | | Paysage | µ = ___ | ___% | ___ - ___ | ___ sites | Moyen | □ Moyennes<br>□ Fortes<br>□ Très fortes | | | |
| | | | | | | ___ - ___ | ___ sites | Fort | | | | |
| | | | | | | ___ - ___ | ___ sites | Très fort | | | | |
| Espèces | Patrimonialité | Nombre d'espèces à enjeux de conservation dans la maille de l'INPN | Maille | µ = ___ | ___ | ___ - ___ | ___ sites | Moyen | | □ Moyennes<br>□ Fortes<br>□ Très fortes | | |
| | | | | | | ___ - ___ | ___ sites | Fort | | | | |
| | | | | | | ___ - ___ | ___ sites | Très fort | | | | |

| Thématique | Sous thématique | Descripteur | Valeur/Attribut |
|---|---|---|---|
| Caractéristiques techniques | Dimensions | Surface du site d'implantation | _____ ha |
| Eco zones | Eco zones | Région biogéographique | □ Atlantique  □ Continentale  □ Méditerranéenne  □ Alpine |
| Espèces | Irremplaçabilité | Valeur de l'ICBG (Indice de Contribution à la Biodiversité Globalisée) de la maille (%) | ____ % |
| Espèces | Méconnaissance | Valeur du taux méconnaissance (%) pour les taxons classiques dans la maille | ____ % |

# Exemple

*Les trois tableaux suivants fournissent un exemple d'évaluation des potentialités écologiques d'un site réel de 46,85 ha situé dans la Creuse et appartenant au groupe des 43 sites utilisé pour les tests. Ces tableaux (un par échelle d'étude) illustrent la démarche du renseignement des valeurs (résultat, moyenne, seuils, nombre de sites), de la définition du niveau de potentialité et du cumul des niveaux. Ils indiquent également la portion du tableau qui est commune à toutes les grilles d'une évaluation multi échelle, ainsi que la portion qu'il faudra modifier lorsque l'échelle d'évaluation change.*

### Exemple d'évaluation à l'échelle du Site avec seuils définis selon Jenks (QGIS) pour le Site ABC

| Thématique | Sous thématique | Indicateurs | Résultat | | | Fourchette décisionnelle | | | Niveau potentialité / échelle | Niveau potentialité échelle retenue (Préciser échelle d'analyse retenue : Site) | Critère décisionnel cumul indicateurs | Niveau Potentialité thématique |
|---|---|---|---|---|---|---|---|---|---|---|---|---|
| | | | Echelle | Moyenne | Résultat | Seuils | Nbre sites | Niveau | | | | |
| Occupation du sol CLC18 | Caractère naturel | % surface de la zone d'étude occupée par des milieux considérés naturels et semi-naturels | Site | µ=52,74 % | 10,10% | 0 – 18,90 % / 18,91 – 56,70 % / 56,71 – 100% | 10 sites / 16 sites / 17 sites | Moyen / Fort / Très fort | ■Moyennes □Fortes □Trèsfortes | ■Moyennes □Fortes □Trèsfortes | E) Les autres cas | ■Moyennes ■Fortes □Trèsfortes |
| | | | Voisinage | µ=76,43 % | 91,30% | 6,70 – 43,40 % / 43,41 – 84,70 % / 84,71 – 100% | 9 sites / 10 sites / 24 sites | Moyen / Fort / Très fort | □Moyennes □Fortes ■Trèsfortes | | | |
| | | | Paysage | µ=73,63 % | 85,30% | 20,10 – 29,80 % / 29,81 – 80,80 % / 80,81 – 99,10% | 9 sites / 11 sites / 23 sites | Moyen / Fort / Très fort | □Moyennes □Fortes ■Trèsfortes | | | |
| | Hétérogénéité | Nombre de milieux naturels différents dans la zone d'étude | Site | µ=1,27 | 1 | 0 / 1 – 2 / 3 – 4 | 7 sites / 30 sites / 6 sites | Moyen / Fort / Très fort | □Moyennes ■Fortes □Trèsfortes | □Moyennes ■Fortes □Trèsfortes | | |
| | | | Voisinage | µ=2,41 | 1 | 0 – 2 / 3 – 4 / 5 – 7 | 23 sites / 15 sites / 5 sites | Moyen / Fort / Très fort | ■Moyennes □Fortes □Trèsfortes | | | |
| | | | Paysage | µ=4,51 | 2 | 1 – 3 / 4 – 6 / 7 – 11 | 13 sites / 21 sites / 9 sites | Moyen / Fort / Très fort | ■Moyennes □Fortes □Trèsfortes | | | |
| | Perméabilité | % surface de la zone d'étude occupée par des modes d'occupation du sol considérés perméables | Site | µ=97,99 % | 100% | 30,50% / 30,51 – 86,60 % / 86,61 – 100 % | 1 site / 1 site / 41 sites | Moyen / Fort / Très fort | □Moyennes □Fortes ■Trèsfortes | □Moyennes □Fortes ■Trèsfortes | | |
| | | | Voisinage | µ=96,05% | 100% | 51,70% / 51,71 – 91,30 % / 91,31 – 100 % | 1 site / 6 sites / 36 sites | Moyen / Fort / Très fort | □Moyennes □Fortes ■Trèsfortes | | | |
| | | | Paysage | µ=95,18 % | 98,70% | 66,00% / 66,01 – 88,50 % / 88,51 – 99,80 % | 1 site / 6 sites / 36 sites | Moyen / Fort / Très fort | □Moyennes □Fortes ■Trèsfortes | | | |
| Réseaux écologiques | Unité du territoire (Non-fragmentation) | Densité du linéaire des réseaux de transport (Rapport : nbre linéaires / surface échelle d'étude) | Site | µ=0,5973 km/km² | 3,260 km/km² | 1,2493 – 3,2604 / 0,336 – 1,2492 / 0 – 0,3359 | 5 sites / 14 sites / 24 sites | Moyen / Fort / Très fort | ■Moyennes □Fortes □Trèsfortes | ■Moyennes □Fortes □Trèsfortes | C) 2 M et 1 F | ■Moyennes □Fortes □Trèsfortes |
| | | | Voisinage | µ=2,1329 km/km² | 1,989 km/km² | 3,732 – 8,501 / 2,104 – 3,731 / 0,057 – 2,103 | 1 site / 16 sites / 26 sites | Moyen / Fort / Très fort | □Moyennes □Fortes ■Trèsfortes | | | |
| | | | Paysage | µ=2,2097 km/km² | 2,020 km/km² | 3,225 – 6,225 / 1,707 – 3,224 / 0,772 – 1,706 | 1 site / 29 sites / 13 sites | Moyen / Fort / Très fort | □Moyennes ■Fortes □Trèsfortes | | | |
| | Corridors | Nombre de sous-trames différentes (milieux majoritaires) traversant la zone d'étude | Site | µ=1,06 | 1 | 0 / 1 – 2 / 3 – 4 | 13 sites / 25 sites / 5 sites | Moyen / Fort / Très fort | □Moyennes ■Fortes □Trèsfortes | □Moyennes ■Fortes □Trèsfortes | | |
| | | | Voisinage | µ=1,76 | 1 | 0 / 1 – 2 / 3 – 4 | 8 sites / 22 sites / 13 sites | Moyen / Fort / Très fort | □Moyennes ■Fortes □Trèsfortes | | | |
| | | | Paysage | µ=2,46 | 4 | 0 – 1 / 2 – 3 / 4 – 5 | 13 sites / 15 sites / 15 sites | Moyen / Fort / Très fort | □Moyennes □Fortes ■Trèsfortes | | | |
| | Réservoirs | % surface de la zone d'étude occupée par des réservoirs | Site | µ=34,58 % | 4,40% | 0 – 13,40 % / 13,41 – 52,10 % / 52,11 – 100 % | 22 sites / 8 sites / 13 sites | Moyen / Fort / Très fort | ■Moyennes □Fortes □Trèsfortes | ■Moyennes □Fortes □Trèsfortes | | |
| | | | Voisinage | µ=38,52 % | 20,70% | 0 – 20,70 % / 20,71 – 75,10 % / 75,11 – 100 % | 22 sites / 12 sites / 9 sites | Moyen / Fort / Très fort | ■Moyennes □Fortes □Trèsfortes | | | |
| | | | Paysage | µ=37,53 % | 17,30% | 3 – 17,60 % / 17,61 – 57,70 % / 57,71 – 99,50 % | 20 sites / 11 sites / 12 sites | Moyen / Fort / Très fort | ■Moyennes □Fortes □Trèsfortes | | | |
| Zonages | Patrimonialité | Proportion de zonages patrimoniaux dans la zone d'étude | Site | µ=14,03 % | 0% | 0% / 0,01 – 71,57 % / 71,58 – 99,54 % | 35 sites / 4 sites / 4 sites | Moyen / Fort / Très fort | ■Moyennes □Fortes □Trèsfortes | ■Moyennes □Fortes □Trèsfortes | | |
| | | | Voisinage | µ=11,25 % | 0% | 0 – 5,44 % / 5,45 – 42,76 % / 42,77 – 99,80% | 33 sites / 6 sites / 4 sites | Moyen / Fort / Très fort | ■Moyennes □Fortes □Trèsfortes | | | |
| | | | Paysage | µ=9,92 % | 0,16% | 0 – 10,63 % / 10,64 – 53,64 % / 53,65 – 96,58 % | 31 sites / 11 sites / 1 site | Moyen / Fort / Très fort | ■Moyennes □Fortes □Trèsfortes | | | |
| Espèces | Patrimonialité | Nombre d'espèces à enjeux de conservation dans la maille | Maille | µ=46,16 | 52 | 0 - 38 / 39 – 62 / 63 – 132 | 9 sites / 27 sites / 7 sites | Moyen / Fort / Très fort | | □Moyennes ■Fortes □Trèsfortes | | |

| Thématique | Sous thématique | Descripteur | Valeur/Attribut |
|---|---|---|---|
| Caractéristiques techniques | Dimensions | Surface du site d'implantation | 46,85 ha |
| Eco zones | Eco zones | Région biogéographique | □Atlantique ■Continentale □Méditerranéenne □Alpine |
| Espèces | Irremplaçabilité | Valeur de l'ICBG (Indice de Contribution à la Biodiversité Globalisée) de la maille (%) | 0% |
| | Méconnaissance | Valeur du taux méconnaissance (%) pour les taxons classiques dans la maille | 70% |

| Critère | Si nbre indicateurs | Le niveau global est |
|---|---|---|
| A) | 3 TF | TF |
| B) | 2 TF et (1 F ou 1 M) | |
| C) | 2 M et 1 F | M |
| D) | 3 M | |
| E) | Les autres cas | F |

← Partie commune à toutes les grilles des trois échelles → Partie à personnaliser à chaque échelle

Evaluation cartographique du niveau de potentialités écologiques (CARPO)

# Exemple d'évaluation à l'échelle du Voisinage avec seuils définis selon Jenks (QGIS) pour le Site ABC

| Thématique | Sous thématique | Indicateurs | Résultat | | | Fourchette décisionnelle | | | Niveau potentialité / échelle | Niveau potentialité échelle retenue Préciser échelle d'analyse retenue : 1000 m | Critère décisionnel cumul indicateurs | Niveau Potentialité thématique |
|---|---|---|---|---|---|---|---|---|---|---|---|---|
| | | | Echelle | Moyenne | Résultat | Seuils | Nbre sites | Niveau | | | | |
| Occupation du sol CLC18 | Caractère naturel | % surface de la zone d'étude occupée par des milieux considérés naturels et semi-naturels | Site | μ=52,74 % | 10,10% | 0 – 18,90 % | 10 sites | Moyen | □ Moyennes ■ Fortes □ Très fortes | □ Moyennes → ■ Fortes ■ Très fortes | B) 2 TF et 1 M | □ Moyennes ■ Fortes ■ Très fortes |
| | | | | | | 18,91 – 56,70 % | 16 sites | Fort | | | | |
| | | | | | | 56,71 - 100% | 17 sites | Très fort | | | | |
| | | | Voisinage | μ=76,43 % | 91,30% | 6,70 – 43,40 % | 9 sites | Moyen | □ Moyennes □ Fortes ■ Très fortes | | | |
| | | | | | | 43,41 – 84,70 % | 10 sites | Fort | | | | |
| | | | | | | 84,71 – 100% | 24 sites | Très fort | | | | |
| | | | Paysage | μ=73,63 % | 85,30% | 20,10 – 29,80 % | 9 sites | Moyen | □ Moyennes ■ Fortes □ Très fortes | | | |
| | | | | | | 29,81 – 80,80 % | 11 sites | Fort | | | | |
| | | | | | | 80,81 - 99,10% | 23 sites | Très fort | | | | |
| | Hétérogénéité | Nombre de milieux naturels différents dans la zone d'étude | Site | μ=1,27 | 1 | 0 | 7 sites | Moyen | □ Moyennes ■ Fortes □ Très fortes | ■ Moyennes □ Fortes → □ Très fortes | | |
| | | | | | | 1 – 2 | 30 sites | Fort | | | | |
| | | | | | | 3 – 4 | 6 sites | Très fort | | | | |
| | | | Voisinage | μ=2,41 | 1 | 0 – 2 | 23 sites | Moyen | ■ Moyennes □ Fortes □ Très fortes | | | |
| | | | | | | 3 – 4 | 15 sites | Fort | | | | |
| | | | | | | 5 – 7 | 5 sites | Très fort | | | | |
| | | | Paysage | μ=4,51 | 2 | 1 – 3 | 13 sites | Moyen | ■ Moyennes □ Fortes □ Très fortes | | | |
| | | | | | | 4 – 6 | 21 sites | Fort | | | | |
| | | | | | | 7 – 11 | 9 sites | Très fort | | | | |
| | Perméabilité | % surface de la zone d'étude occupée par des modes d'occupation du sol considérés perméables | Site | μ=97,99 % | 100% | 30,50% | 1 site | Moyen | □ Moyennes □ Fortes ■ Très fortes | □ Moyennes → ■ Fortes ■ Très fortes | | |
| | | | | | | 30,51 – 86,60 % | 1 site | Fort | | | | |
| | | | | | | 86,61 – 100 % | 41 sites | Très fort | | | | |
| | | | Voisinage | μ=96,05 % | 100% | 51,70% | 1 site | Moyen | □ Moyennes □ Fortes ■ Très fortes | | | |
| | | | | | | 51,71 – 91,30 % | 6 sites | Fort | | | | |
| | | | | | | 91,31 – 100 % | 36 sites | Très fort | | | | |
| | | | Paysage | μ=95,18 % | 98,70% | 66,00% | 1 site | Moyen | □ Moyennes □ Fortes ■ Très fortes | | | |
| | | | | | | 66,01 – 88,50 % | 6 sites | Fort | | | | |
| | | | | | | 88,51 – 99,80 % | 36 sites | Très fort | | | | |
| Réseaux écologiques | Unité du territoire (Non-fragmentation) | Densité du linéaire des réseaux de transport (Rapport : nbre linéaires / surface échelle d'étude) | Site | μ=0,5173 km²/km² | 3,260 km/km² | 1,2493 – 3,2604 | 5 sites | Moyen | ■ Moyennes □ Fortes □ Très fortes | □ Moyennes → ■ Fortes ■ Très fortes | E) Les autres cas | ■ Moyennes ■ Fortes □ Très fortes |
| | | | | | | 0,336 – 1,2492 | 14 sites | Fort | | | | |
| | | | | | | 0 – 0,3359 | 24 sites | Très fort | | | | |
| | | | Voisinage | μ=2,1329 km²/km² | 1,989 km/km² | 3,732 – 8,501 | 1 site | Moyen | □ Moyennes ■ Fortes □ Très fortes | | | |
| | | | | | | 2,104 – 3,731 | 16 sites | Fort | | | | |
| | | | | | | 0,057 – 2,103 | 26 sites | Très fort | | | | |
| | | | Paysage | μ=2,2097 km²/km² | 2,020 km/km² | 3,225 – 6,225 | 1 site | Moyen | □ Moyennes ■ Fortes □ Très fortes | | | |
| | | | | | | 1,707 – 3,224 | 29 sites | Fort | | | | |
| | | | | | | 0,772 – 1,706 | 13 sites | Très fort | | | | |
| | Corridors | Nombre de sous-trames différentes (milieux majoritaires) traversant la zone d'étude | Site | μ=1,06 | 1 | 0 | 13 sites | Moyen | □ Moyennes ■ Fortes □ Très fortes | □ Moyennes → ■ Fortes □ Très fortes | | |
| | | | | | | 1 – 2 | 25 sites | Fort | | | | |
| | | | | | | 3 – 4 | 5 sites | Très fort | | | | |
| | | | Voisinage | μ=1,76 | 1 | 0 | 8 sites | Moyen | ■ Moyennes □ Fortes □ Très fortes | | | |
| | | | | | | 1 – 2 | 22 sites | Fort | | | | |
| | | | | | | 3 – 4 | 13 sites | Très fort | | | | |
| | | | Paysage | μ=2,46 | 4 | 0 – 1 | 13 sites | Moyen | □ Moyennes □ Fortes ■ Très fortes | | | |
| | | | | | | 2 – 3 | 15 sites | Fort | | | | |
| | | | | | | 4 – 5 | 15 sites | Très fort | | | | |
| | Réservoirs | % surface de la zone d'étude occupée par des réservoirs | Site | μ=34,58 % | 4,40% | 0 – 13,40 % | 22 sites | Moyen | ■ Moyennes □ Fortes □ Très fortes | ■ Moyennes □ Fortes □ Très fortes | | |
| | | | | | | 13,41 – 52,10 % | 8 sites | Fort | | | | |
| | | | | | | 52,11 – 100 % | 13 sites | Très fort | | | | |
| | | | Voisinage | μ=38,52 % | 20,70% | 0 – 20,70 % | 22 sites | Moyen | ■ Moyennes □ Fortes □ Très fortes | | | |
| | | | | | | 20,71 – 75,10 % | 12 sites | Fort | | | | |
| | | | | | | 75,11 – 100 % | 9 sites | Très fort | | | | |
| | | | Paysage | μ=37,53 % | 17,30% | 3 – 17,60 % | 20 sites | Moyen | ■ Moyennes □ Fortes □ Très fortes | | | |
| | | | | | | 17,61 – 57,70 % | 11 sites | Fort | | | | |
| | | | | | | 57,71 – 99,50 % | 12 sites | Très fort | | | | |
| Zonages | Patrimonialité | Proportion de zonages patrimoniaux dans la zone d'étude | Site | μ=14,03 % | 0% | 0% | 35 sites | Moyen | ■ Moyennes □ Fortes □ Très fortes | ■ Moyennes □ Fortes □ Très fortes | | |
| | | | | | | 0,01 – 71,57 % | 4 sites | Fort | | | | |
| | | | | | | 71,58 – 99,54 % | 4 sites | Très fort | | | | |
| | | | Voisinage | μ=11,25 % | 0% | 0 - 5,44 % | 33 sites | Moyen | ■ Moyennes □ Fortes □ Très fortes | | | |
| | | | | | | 5,45 – 42,76 % | 6 sites | Fort | | | | |
| | | | | | | 42,77 – 99,80 % | 4 sites | Très fort | | | | |
| | | | Paysage | μ=9,92 % | 0,16% | 0 – 10,63 % | 31 sites | Moyen | ■ Moyennes □ Fortes □ Très fortes | | | |
| | | | | | | 10,64 – 53,64 % | 11 sites | Fort | | | | |
| | | | | | | 53,65 – 96,58 % | 1 site | Très fort | | | | |
| Espèces | Patrimonialité | Nombre d'espèces à enjeux de conservation dans la maille | Maille | μ Fr= 46,16 | 52 | 0 - 38 | 9 sites | Moyen | | □ Moyennes ■ Fortes □ Très fortes | | |
| | | | | | | 39 – 62 | 27 sites | Fort | | | | |
| | | | | | | 63 – 132 | 7 sites | Très fort | | | | |

| Thématique | Sous thématique | Descripteur | Valeur/Attribut | Critère | Si nbre indicateurs | Le niveau global est |
|---|---|---|---|---|---|---|
| Caractéristiques techniques | Dimensions | Surface du site d'implantation | 46,85 ha | A) | 3 TF | TF |
| Eco zones | Eco zones | Région biogéographique | □ Atlantique ■ Continentale □ Méditerranéenne □ Alpine | B) | 2 TF et (1 F ou 1 M) | |
| Espèces | Irremplaçabilité | Valeur de l'ICBG (Indice de Contribution à la Biodiversité Globalisée) de la maille (%) | 0% | C) | 2 M et 1 F | M |
| | Méconnaissance | Valeur du taux méconnaissance (%) pour les taxons classiques dans la maille | 70% | D) | 3 M | |
| | | | | E) | Les autres cas | F |

Partie commune à toutes les grilles des trois échelles ←→ Partie à personnaliser à chaque échelle

Evaluation cartographique du niveau de potentialités écologiques (CARPO)

# Exemple d'évaluation à l'échelle du Paysage avec seuils définis selon Jenks (QGIS) pour le Site ABC

| Thématique | Sous thématique | Indicateurs | Résultat | | | Fourchette décisionnelle | | | Niveau potentialité / échelle | Niveau potentialité échelle retenue Préciser échelle d'analyse retenue : 5000m | Critère décisionnel cumul indicateurs | Niveau Potentialité thématique |
|---|---|---|---|---|---|---|---|---|---|---|---|---|
| | | | Echelle | Moyenne | Résultat | Seuils | Nbre sites | Niveau | | | | |
| Occupation du sol CLC18 | Caractère naturel | % surface de la zone d'étude occupée par des milieux considérés naturels et semi-naturels | Site | µ = 52,74 % | 10,10% | 0 – 18,90 % | 10 sites | Moyen | ■ Moyennes □ Fortes □ Très fort | □ Moyennes □ Fortes ■ Très fortes | | |
| | | | | | | 18,91 – 56,70 % | 16 sites | Fort | | | | |
| | | | | | | 56,71 - 100% | 17 sites | Très fort | | | | |
| | | | Voisinage | µ = 76,43 % | 91,30% | 6,70 – 43,40 % | 9 sites | Moyen | □ Moyennes □ Fortes ■ Très fort | | | |
| | | | | | | 43,41 – 84,70 % | 10 sites | Fort | | | | |
| | | | | | | 84,71 – 100% | 24 sites | Très fort | | | | |
| | | | Paysage | µ = 73,63 % | 85,30% | 20,10 – 29,80 % | 9 sites | Moyen | □ Moyennes □ Fortes ■ Très fort → | | | |
| | | | | | | 29,81 – 80,80 % | 11 sites | Fort | | | | |
| | | | | | | 80,81 - 99,10% | 23 sites | Très fort | | | | |
| | Hétérogénéité | Nombre de milieux naturels différents dans la zone d'étude | Site | µ = 1,27 | 1 | 0 | 7 sites | Moyen | □ Moyennes ■ Fortes □ Très fortes | □ Moyennes □ Fortes □ Très fortes | B) 2 TF et 1 M | □ Moyennes □ Fortes ■ Très fortes |
| | | | | | | 1 – 2 | 30 sites | Fort | | | | |
| | | | | | | 3 – 4 | 6 sites | Très fort | | | | |
| | | | Voisinage | µ = 2,41 | 1 | 0 – 2 | 23 sites | Moyen | □ Moyennes ■ Fortes □ Très fortes | | | |
| | | | | | | 3 – 4 | 15 sites | Fort | | | | |
| | | | | | | 5 – 7 | 5 sites | Très fort | | | | |
| | | | Paysage | µ = 4,51 | 2 | 1 – 3 | 13 sites | Moyen | □ Moyennes ■ Fortes □ Très fortes → | | | |
| | | | | | | 4 – 6 | 21 sites | Fort | | | | |
| | | | | | | 7 – 11 | 9 sites | Très fort | | | | |
| | Perméabilité | % surface de la zone d'étude occupée par des modes d'occupation du sol considérés perméables | Site | µ = 97,99 % | 100% | 30,50% | 1 site | Moyen | □ Moyennes □ Fortes ■ Très fort | □ Moyennes □ Fortes ■ Très fortes | | |
| | | | | | | 30,51 – 86,60 % | 1 site | Fort | | | | |
| | | | | | | 86,61 – 100 % | 41 sites | Très fort | | | | |
| | | | Voisinage | µ = 96,05 % | 100% | 51,70% | 1 site | Moyen | □ Moyennes □ Fortes ■ Très fort | | | |
| | | | | | | 51,71 – 91,30 % | 6 sites | Fort | | | | |
| | | | | | | 91,31 – 100 % | 36 sites | Très fort | | | | |
| | | | Paysage | µ = 95,18 % | 98,70% | 66,00% | 1 site | Moyen | □ Moyennes □ Fortes ■ Très fort → | | | |
| | | | | | | 66,01 – 88,50 % | 6 sites | Fort | | | | |
| | | | | | | 88,51 – 99,80 % | 36 sites | Très fort | | | | |
| Réseaux écologiques | Unité du territoire (Non-fragmentation) | Densité du linéaire des réseaux de transport (Rapport : nbre linéaires / surface échelle d'étude) | Site | µ = 0,5173 km/km² | 3,260 km/km² | 1,2493 – 3,2604 | 5 sites | Moyen | ■ Moyennes □ Fortes □ Très fort | □ Moyennes ■ Fortes □ Très fortes | E) Les autres cas | □ Moyennes ■ Fortes □ Très fortes |
| | | | | | | 0,336 – 1,2492 | 14 sites | Fort | | | | |
| | | | | | | 0 – 0,3359 | 24 sites | Très fort | | | | |
| | | | Voisinage | µ = 2,1329 km/km² | 1,989 km/km² | 3,732 – 8,501 | 1 site | Moyen | □ Moyennes ■ Fortes □ Très fort | | | |
| | | | | | | 2,104 – 3,731 | 16 sites | Fort | | | | |
| | | | | | | 0,057 – 2,103 | 26 sites | Très fort | | | | |
| | | | Paysage | µ = 2,2097 km/km² | 2,020 km/km² | 3,225 – 6,225 | 1 site | Moyen | □ Moyennes ■ Fortes □ Très fort → | | | |
| | | | | | | 1,707 – 3,224 | 29 sites | Fort | | | | |
| | | | | | | 0,772 – 1,706 | 13 sites | Très fort | | | | |
| | Corridors | Nombre de sous-trames différentes (milieux majoritaires) traversant la zone d'étude | Site | µ = 1,06 | 1 | 0 | 13 sites | Moyen | □ Moyennes ■ Fortes □ Très fortes | □ Moyennes ■ Fortes □ Très fortes | | |
| | | | | | | 1 – 2 | 25 sites | Fort | | | | |
| | | | | | | 3 – 4 | 5 sites | Très fort | | | | |
| | | | Voisinage | µ = 1,76 | 1 | 0 | 8 sites | Moyen | □ Moyennes ■ Fortes □ Très fortes | | | |
| | | | | | | 1 – 2 | 22 sites | Fort | | | | |
| | | | | | | 3 – 4 | 13 sites | Très fort | | | | |
| | | | Paysage | µ = 2,46 | 4 | 0 – 1 | 13 sites | Moyen | □ Moyennes ■ Fortes □ Très fortes → | | | |
| | | | | | | 2 – 3 | 15 sites | Fort | | | | |
| | | | | | | 4 – 5 | 15 sites | Très fort | | | | |
| | Réservoirs | % surface de la zone d'étude occupée par des réservoirs | Site | µ = 34,58 % | 4,40% | 0 – 13,40 % | 22 sites | Moyen | ■ Moyennes □ Fortes □ Très fortes | ■ Moyennes □ Fortes □ Très fortes | | |
| | | | | | | 13,41 – 52,10 % | 8 sites | Fort | | | | |
| | | | | | | 52,11 – 100 % | 13 sites | Très fort | | | | |
| | | | Voisinage | µ = 38,52 % | 20,70% | 0 – 20,70 % | 22 sites | Moyen | ■ Moyennes □ Fortes □ Très fortes | | | |
| | | | | | | 20,71 – 75,10 % | 12 sites | Fort | | | | |
| | | | | | | 75,11 – 100 % | 9 sites | Très fort | | | | |
| | | | Paysage | µ = 37,53 % | 17,30% | 3 – 17,60 % | 20 sites | Moyen | ■ Moyennes □ Fortes □ Très fortes → | | | |
| | | | | | | 17,61 – 57,70 % | 11 sites | Fort | | | | |
| | | | | | | 57,71 – 99,50 % | 12 sites | Très fort | | | | |
| Zonages | Patrimonialité | Proportion de zonages patrimoniaux dans la zone d'étude | Site | µ = 14,03 % | 0% | 0% | 35 sites | Moyen | ■ Moyennes □ Fortes □ Très fortes | ■ Moyennes □ Fortes □ Très fortes | | |
| | | | | | | 0,01 – 71,57 % | 4 sites | Fort | | | | |
| | | | | | | 71,58 – 99,54 % | 4 sites | Très fort | | | | |
| | | | Voisinage | µ = 11,25 % | 0% | 0 - 5,44 % | 33 sites | Moyen | ■ Moyennes □ Fortes □ Très fortes | | | |
| | | | | | | 5,45 – 42,76 % | 6 sites | Fort | | | | |
| | | | | | | 42,77 – 99,80 % | 4 sites | Très fort | | | | |
| | | | Paysage | µ = 9,92 % | 0,16% | 0 – 10,63 % | 31 sites | Moyen | ■ Moyennes □ Fortes □ Très fortes → | | | |
| | | | | | | 10,64 – 53,64 % | 11 sites | Fort | | | | |
| | | | | | | 53,65 – 96,58 % | 1 site | Très fort | | | | |
| Espèces | Patrimonialité | Nombre d'espèces à enjeux de conservation dans la maille | Maille | µ F= 46,18 | 52 | 0 - 38 | 9 sites | Moyen | | □ Moyennes ■ Fortes □ Très fortes | | |
| | | | | | | 39 – 62 | 27 sites | Fort | | | | |
| | | | | | | 63 – 132 | 7 sites | Très fort | | | | |

| Thématique | Sous thématique | Descripteur | Valeur/Attribut | | Critère | Si nbre indicateurs | Le niveau global est |
|---|---|---|---|---|---|---|---|
| Caractéristiques techniques | Dimensions | Surface du site d'implantation | 46,85 ha | | A) | 3 TF | TF |
| Eco zones | Eco zones | Région biogéographique | □ Atlantique ■ Continentale □ Méditerranéenne □ Alpine | | B) | 2 TF et (1 F ou 1 M) | |
| Espèces | Irremplaçabilité | Valeur de l'ICBG (Indice de Contribution à la Biodiversité Globalisée) de la maille (%) | 0% | | C) | 2 M et 1 F | M |
| | Méconnaissance | Valeur du taux méconnaissance (%) pour les taxons classiques dans la maille | 70% | | D) | 3 M | |
| | | | | | E) | Les autres cas | F |

← Partie commune à toutes les grilles des trois échelles → ← Partie à personnaliser à chaque échelle →

Evaluation cartographique du niveau de potentialités écologiques (CARPO)

## Annexe 9 :
## Calibrage de la règle décisionnelle pour le cumul des niveaux de potentialités

| | Légende |
|---|---|
| M | Potentialité moyenne |
| F | Forte potentialité |
| TF | Très forte potentialité |

| Critère | Si % d'indicateurs : | Niv cumulé |
|---|---|---|
| A) | ≥ 60% de TF | TF |
| B) | ≥ 40% TF et < 25% d'M | TF |
| C) | ≥ 80% d'M | M |
| D) | ≥ 60% d'M et 0% TF | M |
| E) | Les autres cas | F |

Tests pour calibrer les pourcentages seuils

**Cas avec 5 indicateurs**

| Nb indicateurs | | | | | | | | | | | | | | | | | | | | |
|---|---|---|---|---|---|---|---|---|---|---|---|---|---|---|---|---|---|---|---|---|
| TF | 0 | 1 | 0 | 0 | 1 | 2 | 0 | 1 | 2 | 0 | 1 | 1 | 0 | 2 | 2 | 3 | 3 | 3 | 4 | 4 | 5 |
| F  | 0 | 0 | 1 | 2 | 1 | 0 | 3 | 2 | 1 | 4 | 3 | 4 | 5 | 3 | 2 | 0 | 1 | 2 | 0 | 1 | 0 |
| M  | 5 | 4 | 4 | 3 | 3 | 3 | 2 | 2 | 2 | 1 | 1 | 0 | 0 | 0 | 1 | 2 | 1 | 0 | 1 | 0 | 0 |
| Décisions |  | M | M | M | M ? | F | F | F | F | F | F | F | F | F | TF | TF | TF | TF | TF | TF | TF |
| Critère |  | C | C | C | D | E | E | E | E | E | E | E | E | B | B | A | A | A | A | A | A |

| Niv | # cas |
|---|---|
| TF | 8 |
| F | 9 |
| M | 4 |

**Cas avec 4 indicateurs**

| TF | 0 | 0 | 1 | 0 | 1 | 2 | 0 | 1 | 0 | 1 | 2 | 2 | 3 | 3 | 4 |
|---|---|---|---|---|---|---|---|---|---|---|---|---|---|---|---|
| F  | 0 | 1 | 0 | 2 | 1 | 0 | 3 | 2 | 4 | 3 | 2 | 1 | 0 | 1 | 0 |
| M  | 4 | 3 | 3 | 2 | 2 | 2 | 1 | 1 | 0 | 0 | 0 | 1 | 1 | 0 | 0 |
| Décisions |  | M | M | F | F | F | F | F | F | F | F | TF | TF | TF | TF | TF |
| Critère |  | C | D | E | E | E | E | E | E | E | E | B | B | A | A | A |

| Niv | # cas |
|---|---|
| TF | 5 |
| F | 8 |
| M | 2 |

**Cas avec 3 indicateurs**

| TF | 0 | 0 | 1 | 0 | 1 | 0 | 1 | 2 | 2 | 3 |
|---|---|---|---|---|---|---|---|---|---|---|
| F  | 0 | 1 | 0 | 2 | 1 | 3 | 2 | 0 | 1 | 0 |
| M  | 3 | 2 | 2 | 1 | 1 | 0 | 0 | 1 | 0 | 0 |
| Décisions |  | M | M | F | F | F | F | F | TF | TF | TF |
| Critère |  | C | D | E | E | E | E | E | A | A | A |

| Niv | # cas |
|---|---|
| TF | 3 |
| F | 5 |
| M | 2 |

| Critère | Si nbre/% d'indicateurs : | | Niv cumulé |
|---|---|---|---|
| A) | 3 TF | 100% TF | TF |
| B) | 2 TF et (1 F ou 1 M) | ≥ 66% TF et (≤ 34% F ou ≤ 34% M) | TF |
| C) | 2 M et 1 F | ≥ 66% M et ≤ 34 % F | M |
| D) | 3 M | 100% M | M |
| E) | Les autres cas | Les autres cas | F |

**Cas avec 2 indicateurs**

| TF | 0 | 0 | 1 | 0 | 1 | 2 |
|---|---|---|---|---|---|---|
| F  | 0 | 1 | 0 | 2 | 1 | 0 |
| M  | 2 | 1 | 1 | 0 | 0 | 0 |
| Décisions |  | M | F | F | F | TF | TF |
| Critère |  | D | E | E | E | B | A |

| Niv | # cas |
|---|---|
| TF | 2 |
| F | 3 |
| M | 1 |

## Annexe 10 :
## Propositions d'indicateurs à étudier pour une future version de la méthode

| Thématique | Sous-thématique | Indicateur | Base de données et description | Image/source |
|---|---|---|---|---|
| Occupation du sol | Caractère naturel/Artificialisation | Degré d'artificialisation du sol de la commune | Calcul à partir de la couche **Flux d'artificialisation sur la période 2009 – 2017** qui indique l'intensité de l'artificialisation des sols de chaque commune (surface artificialisée en m2). | 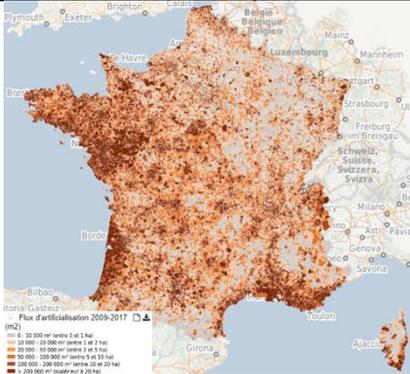 https://artificialisation.biodiversitetousvivants.fr/les-donnees-au-1er-janvier-2017 |
| Occupation du sol | Caractère naturel/Artificialisation | Degré de naturalité (wilderness) | Le projet CartNat de l'UICN a abouti à une cartographie de la naturalité potentielle en France métropolitaine terrestre avec 4 couches spatiales : l'intégrité biophysique de l'occupation du sol, la spontanéité des processus, les continuités spatiales ; et surtout la carte synthétique du gradient de naturalité potentielle à partir des cartes précédentes | 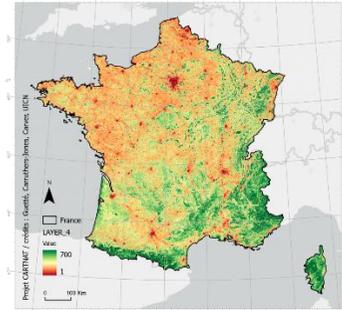 https://uicn.fr/aires-protegees/wilderness/ |
| Occupation du sol | Caractère naturel/Artificialisation | % surface zone d'étude artificialisée | Calcul à partir de la couche **Foncière – Artificialisation des sols** de la surface artificialisée sur la base d'une typologie dichotomique (2 postes) : sols naturels, agricoles forestiers (NAF), et sols artificiels.<br>Nomenclature sur 13 postes (parallèlement à la dichotomie en 2 postes), résolution à l'échelle de la parcelle, mise à jour annuelle. | 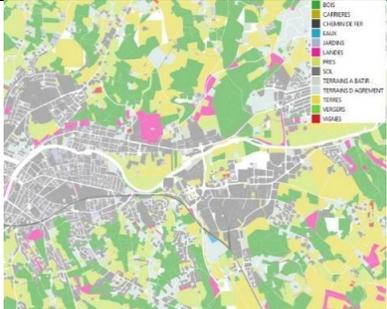 https://artificialisation.biodiversitetousvivants.fr/bases-donnees/les-fichiers-fonciers |

| Occupation du sol | Perméabilité | Valeur de la densité de sols imperméabilisés dans la zone d'étude | Calcul à partir de la couche européenne **CLC HR Imperviousness** qui indique le pourcentage de zone imperméabilisée (2006, 2009, 2012 et 2015). Les couches d'état sont disponibles dans la résolution spatiale d'origine de 20 m et en tant que produits agrégés de 100 m. | 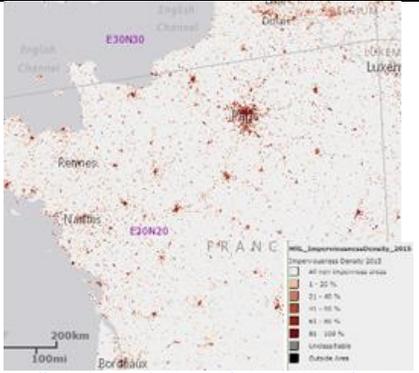<br>https://land.copernicus.eu/pan-european/high-resolution-layers/imperviousness |
|---|---|---|---|---|
| Occupation du sol | Zones humides | % de la zone d'étude concernée par des zones humides | Calcul à partir de la couche **Milieux potentiellement humides en France** qui modélise des enveloppes de zones humides selon trois classes de probabilité (assez forte, forte et très forte). Intérêt : Milieux en régression en lien avec le changement climatique, souvent associés à des espaces et d'espèces patrimoniales. | 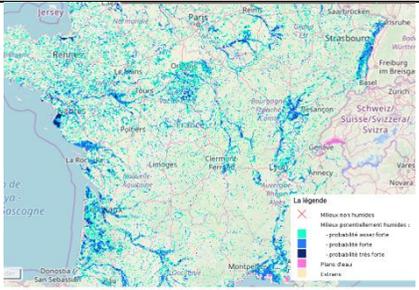<br>http://geowww.agrocampus-ouest.fr/web/?p=1538 |
| Occupation du sol | Hydrographie | Nbre de cours et plans d'eau traversant le site (info sur la qualité ?) | Calcul du nombre d'éléments à partir des couches **BD Carthage Cours d'eau et Plan d'eau Métropole 2016,** présentant les objets hydrographiques suivants : bassins versants, cours d'eau et plans d'eau.<br>Croisement à étudier avec la couche **Naturalité estimée des cours d'eau – ONB** pour les aspects sur la qualité écologique des eaux ? | http://www.sandre.eaufrance.fr/<br>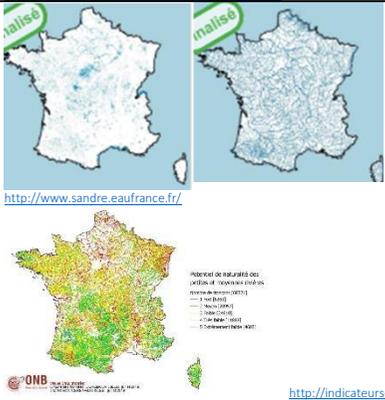<br>http://indicateurs-biodiversite.naturefrance.fr/fr/indicateurs/naturalite-estimee-des-cours-deau |

| Réseaux écologiques | Trame noire | Intensité de la pollution lumineuse | Calcul à partir de la couche **Lightpollutionmmap 2017** de l'ampleur des obstacles aux déplacements des espèces nocturnes (rayonnement en Watts). | 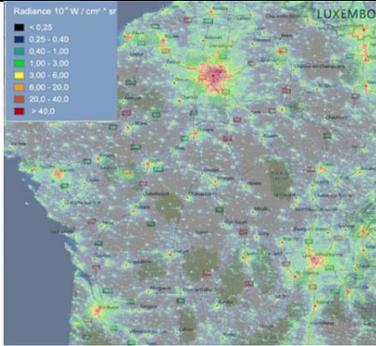<br>https://ngdc.noaa.gov/eog/viirs/download_dnb_composites.html |
|---|---|---|---|---|
| Réseaux écologiques | Unité du territoire (non-fragmentation) | Intensité de la fragmentation d'espaces naturels | Calcul à partir de la couche **Fragmentation – Menaces sur la biodiversité ONB 2019** du degré de fragmentation terrestre dans le territoire d'un site d'implantation. Quelle unité ? Disponibilité de la couche ? | 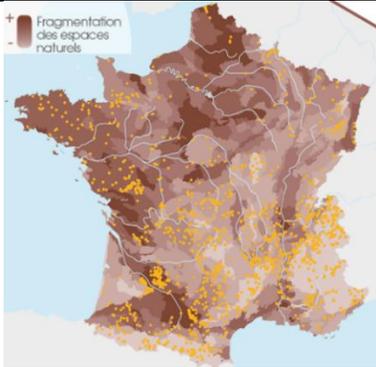<br>http://indicateurs-biodiversite.naturefrance.fr/cartographie-des-pressions |
| Fonctionnalité | Espèces exotiques envahissantes (EEE) | Intensité de l'invasion par des espèces terrestres | Calcul à partir de la couche **EEE – Menaces sur la biodiversité ONB 2019** du degré de la présence d'EEE terrestres dans le territoire d'un site d'implantation. Quelle unité ? Disponibilité de la couche ? | 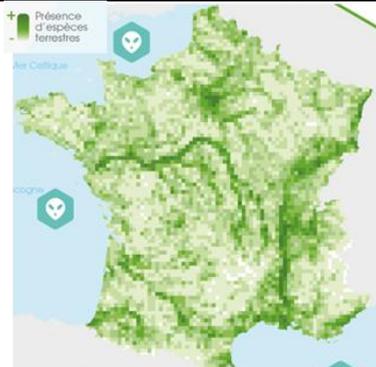<br>http://indicateurs-biodiversite.naturefrance.fr/cartographie-des-pressions |
| Zonages | Enjeux règlementaires | % surface zone d'étude couverte par des aires protégées | Calcul à partir de la couche des zonages INPN déjà intégrée à l'outil automatisé et sur la base des travaux réalisés dans le cadre de ŒIL. | |
| Espèces | Enjeux règlementaires | Nombre d'espèces protégées par maille | Calcul à partir d'une couche à créer (richesse espèces protégées/maille) suite à l'extraction d'espèces protégées recensées dans chaque maille de l'INPN. | |

# RÉSUMÉ

L'Evaluation Cartographique de Potentialités Ecologiques (CARPO) permet de caractériser le contexte écologique d'un groupe de sites à plusieurs échelles (son périmètre et son entourage) et sur la base 4 thématiques : les zonages de biodiversité, l'occupation du sol (les grands types de milieu, le caractère naturel et le dégré de perméabilité des territoires), la connectivité écologique (les réservoirs et corridors de biodiversité, la non-fragmentation du secteur étudié), ainsi que la richesse du territoire en espèces patrimoniales. Elle repose sur une batterie de 8 indicateurs liés à ces thématiques et d'un système de seuils d'évaluation, permettant de définir 3 niveaux des potentialités écologiques (moyennes, fortes et très fortes) pour chaque indicateur et à chaque échelle détude. Le niveau de potentialité reflète l'importance qu'un élément écologique prend sur une zone d'étude, et donc le degré de contribution et de responsabilité d'un site pour le maintien de cet élément favorable pour la biodiversité. Il se traduit aussi par un niveau d'alerte relatif à une biodiversité qui doit être prise en compte dans un site.

Le diagnostic écologique se matérialise à l'aide d'un atlas cartographique réunissant un ensemble de cartes et figures commentées, ainsi que des grilles et des radars d'évaluation des différents indicateurs.

CARPO constitue un outil d'aide à la décision car il permet in fine de comparer les potentialités écologiques à l'intérieur du site avec celles de son entourage, pour définir des pistes d'actions de préservation, de gestion ou de restauration ; ainsi que de comparer l'ampleur des potentialités entre les différents sites du groupe, dans un but de priorisation réfléchie de l'action.

Ce guide présente le cadre méthodologique pour l'application de CARPO sur un groupe de sites.